\documentclass[prb,aps,floatfix,longbibliography,superscriptaddress, nofootinbib]{revtex4-2}
\usepackage{array}
\usepackage{multirow}    

\usepackage{graphicx}
\usepackage{amsmath}
\usepackage{relsize}
\usepackage{amssymb}
\usepackage{gensymb}    
\usepackage{hhline}
\usepackage{bm}
\usepackage{bbm}
\usepackage{cancel}
\usepackage[framed,numbered]{matlab-prettifier}
\usepackage[linesnumbered,ruled]{algorithm2e}
\usepackage{diagbox}   
\usepackage{soul}

\usepackage{babel}
\usepackage{pdfpages}
\usepackage{pgffor}

\usepackage{xr} 
\usepackage{hyperref}            
\makeatletter
\AtBeginDocument{\let\LS@rot\@undefined}
\makeatother

\begin{document}
	
\title{\bf Efficient simulations of Hartree--Fock equations by an accelerated gradient descent method}

 \vskip 1 cm


\author{Y. Ohno}
\affiliation{Department of Mathematics and Statistics, 
	University of Vermont, Burlington, VT 05405, USA}

\author{A. Del Maestro}
\affiliation{Department of Physics and Astronomy, University of Tennessee, Knoxville, TN 37996, USA}
\affiliation{Min H.~Kao Department of Electrical Engineering and Computer Science, University of Tennessee, Knoxville, TN 37996, USA}
\affiliation{Institute for Advanced Materials \& Manufacturing, University of Tennessee, Knoxville, TN 37920, USA}

\author{T.I. Lakoba}
\email[Corresponding author: ]{tlakoba@uvm.edu}
\affiliation{Department of Mathematics and Statistics, 
	University of Vermont, Burlington, VT 05405, USA}

\renewcommand{\baselinestretch}{1.2}
\newcommand{\noi}{\noindent}
\newcommand{\und}{\underline}
\newcommand{\D}{\Delta}
\newcommand{\be}{\begin{equation}}
\newcommand{\ee}{\end{equation}}
\newcommand{\bsube}{\begin{subequations}}
\newcommand{\esube}{\end{subequations}}
\newcommand{\ba}{\begin{array}}
\newcommand{\ea}{\end{array}}
\newcommand{\benum}{\begin{enumerate}}
\newcommand{\eenum}{\end{enumerate}}	
	
\newcommand{\To}{\rightarrow}
\newcommand{\vecx}{{\bf x}}
\newcommand{\Int}{\int_{-\infty}^{\infty}}
\newcommand{\bea}{\begin{eqnarray}}
\newcommand{\eea}{\end{eqnarray}}
\renewcommand{\so}{\Rightarrow}
\newcommand{\dst}{\displaystyle}
\newcommand{\const}{{\rm const}}
\newcommand{\bfSig}{\bm{\Sigma}}
\newcommand{\bfr}{{\bf r}}
\newcommand{\vext}{V_{\rm ext}}
\newcommand{\vint}{V_{\rm int}}
\newcommand{\gslown}{\gamma_{{\rm slow},n}}
\newcommand{\Gslown}{\Gamma_{{\rm slow},n}}
\newcommand{\uslown}{(u_{{\rm slow},j})_n}
\newcommand{\eclip}{E_{\rm clip}}
\newcommand{\dt}{\D t}
\newcommand{\dx}{\D x}
\newcommand{\wD}{\widehat{\D}}
\newcommand{\vD}{\overrightarrow{\D\phi}}
\newcommand{\vu}{\vec{u}}
\newcommand{\wU}{\widehat{U}}
\newcommand{\walpha}{\mathlarger{\mathlarger{\boldsymbol\alpha}}}
\newcommand{\bA}{{\bf A}}
\newcommand{\bB}{{\bf B}}
\newcommand{\bC}{{\bf C}}
\newcommand{\wA}{\widetilde{\bf A}}

\newcommand{\kone}{\bm{\kappa}_1}




\baselineskip 21 pt

\begin{abstract}
	We develop convergence acceleration procedures that enable a gradient descent-type
	iteration method to efficiently simulate Hartree--Fock equations for atoms interacting
	both with each other and with an external potential. Our development focuses on three
	aspects: (i) optimization of a parameter in the preconditioning operator; 
	(ii) adoption of a technique that eliminates the slowest-decaying mode to the case 
	of many equations (describing many atoms); and (iii) a novel extension of the above 
	technique that allows one to eliminate multiple modes simultaneously. We illustrate 
	performance of the numerical method for the 2D model of the first layer of helium atoms 
	above a graphene sheet. We demonstrate that incorporation of aspects (i) and (ii) above
	into the ``plain" gradient descent method accelerates it by at least two orders of
	magnitude, and often by much more. Aspect (iii) --- a multiple-mode elimination --- 
	may bring further improvement to the convergence rate compared to aspect (ii), the
	single-mode elimination. Both single- and multiple-mode elimination techniques are
	shown to significantly outperform the well-known Anderson Acceleration. 
	We believe that our acceleration techniques can also be gainfully employed by other 
	numerical methods, especially those handling hard-core-type interaction potentials. 
\end{abstract}

\maketitle

\vspace*{2cm}


\vskip 1.1 cm

\noi

\bigskip


\newpage

\section{Introduction}

The phases and phase transitions of system of adsorbed bosonic atoms on a substrate at low temperatures 
have received renewed attention due to the possibility of experimentally realizing a spatially modulated superfluid state for the second layer of bosonic $^4$He atoms on corrugated graphite substrates \cite{2016_Nakamura,2017_Nyeki,2021_Choi}. Here, the potential interplay between superfluidity and
broken translational symmetry is argued to arise from a strong competition between these two types of orders. While some zero-temperature numerical simulations \cite{2012_Gordillo, 2020_Gordillo} provide support for
 this scenario, other finite-temperature path integral quantum Monte Carlo results \cite{2008_Corboz} find
only incommensurate solid or superfluid phases.  The origin of continuing uncertainty on the phase diagram
in these few-layer adsorbed helium systems may be a result of the utilization of different adsorption potentials of different degrees of complexity \cite{1973_Steele,2007_Bruch} combined with the complexity of performing full ab initio three-dimensional simulations of this strongly interacting system. 

Seeking to reduce the complexity and enhance throughput of the determination of the adsorbed many-body wavefunction of $^4$He on corrugated substrates, we present a relatively simple and time-efficient 
gradient descent-type iterative method to solve Hartree--Fock (HF) equations that describe particles interacting with each other via hard core-type potentials (i.e., those leading to strong repulsion at
short distances) as well as being influenced by an external adsorption potential.  An early non-optimized version of the algorithm was employed in a recent study by two of the authors \cite{PRB} seeking to determine a low-energy effective model for a single layer of $^4$He atoms adsorbed on a graphene membrane. 
Here we present the details of an improved algorithm, whose convergence is accelerated
by preconditioning and slow-mode elimination, 
	where the following main issues where addressed. First, we show how to optimize the 
	preconditioning operator in the presence of a hard-core-type interatomic potential. 
	Second, we show how the acceleration that relies on elimination of the mode 
	decaying the slowest from one iteration to the next, should be carried out for
	multiple equations. Third, we present a novel extension of the above technique that
	eliminates several slowly decaying modes simultaneously.

We emphasize that our numerical technique is complementary to many existing HF-based 
electronic structure computations, which are designed to handle thousands of particles. Those approaches 
seek the solution of the many-body wavefunction as an expansion over a basis of localized functions \cite{1998_goodHFpaper, 1999_Goedecker_Review, 2010_NumSolHF_molorbs, 2013_nonorthogWannier}.  
In contrast, the approach described in this paper is based on iteratively solving HF equations for smaller numbers of adsorbed particles, expanding them in the (nonlocal) Fourier basis. 
The value of our numerical approach is in the relative simplicity of its implementation compared
to other numerical techniques. 
The novel contribution of this work is in significant acceleration of convergence of the gradient
descent method, as well as in providing understanding of what factors slow down the non-accelerated 
method.

Anderson Acceleration \cite{1965_AA}, also known under the names Pulay Mixing 
and Direct Inversion of the Iteration Subspace, is a widely used 
acceleration technique for gradient descent-type methods
in electronic structure calculations; see, e.g., 
\cite{2011_RedRankExtrap, 2011_WalkerNi, 2019_Nuclear} and references therein. 
We will show that our slow-mode elimination-based acceleration significantly outperforms the 
Anderson Acceleration. We also mention that in a recent
paper \cite{2019_Nuclear}, its authors demonstrated that an acceleration by the Heavy Ball
algorithm, widely used in Optimization, significantly reduces the 
number of iterations of a gradient descent method applied to the so-called Energy Density 
Functional calculations of atomic nuclei. (The latter model is of the HF type, but more complicated.) 
However, optimal performance of this acceleration algorithm relies on having a satisfactory
estimate for the condition number of the linearized iteration operator. While in 
\cite{2019_Nuclear} the authors were able to obtain such an estimate, it is unclear
how to do so in other models, including that considered here. Optimization of performance of our
acceleration techniques does not rely on knowing properties of the linearized iteration operator
(other than it is Hermitian).

A number of other iterative methods for HF equations including quasi-Newton, fixed-point, and 
extrapolation-based methods were reviewed in, e.g., \cite{2011_RedRankExtrap}. Examples
considered there pertained to solving a Hartree-type equation for electrons in single
light atoms, C and Na, and in a single diatomic molecule, CO. The baseline method was
the standard gradient descent method, with the only kind of acceleration used being 
taking a weighted average between two successive values of the potential. In those
examples, the number of iterations of the baseline method 
was under 50, and an extrapolation-based acceleration (the
most efficient one of all types of acceleration considered there) reduced that number by
some factor of two or less. 
In contrast, for the model considered in our study, the number of iterations 
of the baseline method (i.e., of the same non-accelerated method as used in \cite{2011_RedRankExtrap})
is in the (many) tens of thousands. Our acceleration techniques reduce that number
 by at least two orders of magnitude, and often by much more.

The main part of this work is organized as follows. In Sec.~\ref{sec_2} we present the HF equations 
and the general form of the numerical method used to solve them. 
In Sec.~\ref{sec_3}, we begin by discussing specifics of the convergence acceleration technique used by 
the method, including an optimal choice of its parameters. 
This leads us to presenting, in Sec.~\ref{sec_4}, a
non-trivial and novel generalization of the mode elimination technique, which allows one 
to eliminate
{\em multiple} slow modes. In Sec.~\ref{sec_5}, we use our numerical method to obtain results for 
the $^4$He atoms covering a graphene sheet at various filling fractions. 
Among those there are a few patterns which, to our knowledge, have not been 
depicted in the literature. 
We summarize in Sec.~\ref{sec_6}. 
We note that presentation in Secs.~\ref{sec_3} and \ref{sec_4} refers to a
(non-physical) 1D model so as to focus on exposition of the underlying concepts
of the numerical method. The same concepts apply to higher-dimensional
generalizations of the method, and 2D results are presented in Sec.~\ref{sec_5}.
The reader who is interested only in the implementation steps of the
numerical method can find them referenced compactly in Sec.~\ref{sec_6}. 
Appendices contain technical information that could be useful for either
 understanding or implementation of the methods discussed. Supplemental Material
 contains technical details that are less essential for the understanding and 
 implementation than those found in the Appendices.

\section{Hartree--Fock equations and numerical method}
\label{sec_2}

\subsection{Hartree--Fock equations}
\label{sec_2A}

The model for which we will illustrate our numerical method is a system of $^4$He atoms (bosons) 
interacting with each other and with the graphene sheet above which they are located. 
Full determination of the wavefunction of $N$ bosons via Monte Carlo simulations is 
computationally costly. An alternative is to follow an approximate --- HF --- description,
where the full wavefunction $\Psi$ is replaced with an ansatz:
\be
\Psi(\bfr_1,\bfr_2,\ldots,\bfr_N) \, 
\,=\, \frac1{\sqrt{N!}} \, \sum_{{\rm symm},\, q} \;\; \prod_{j=1}^N \phi_{j}(\bfr_q), 
\label{e2_01}
\ee
where $\bfr_q$ are the coordinates of particle $q$, \ $j$ is the label of a quantum 
state  where particle $q$ is found, 
and the summation is over all $N!$ permutations of the $N$ particles.
Ansatz \eqref{e2_01} assumes that the bosons, interacting via hard-core-type potential, 
cannot occupy the same spatial location, and therefore the aforementioned `state' 
amounts to the location. 
The one-particle ``wavefunctions" (in what follows we will drop the
quotes) satisfy the orthonormality condition
\be
\langle \phi_i | \phi_j\rangle \equiv 
\int_{\mathbb{R}^2} d^2\bfr \, \phi^\dagger_i(\bfr)\phi_j(\bfr) = \delta_{i,j}\,,
\label{e2_02}
\ee
and $\dagger$ stands for Hermitian conjugation. 
Thus, wavefunctions of particles occupying different locations are orthogonal. 
It should be clarified that the functions $\phi_{i},\,\phi_j$ are scalars, and the transposition,
implied by the $\dagger$, will be only used in the presentation of the numerical method below
where some of the quantities will be vectors or matrices. 

The wavefunctions $\phi_i$ can be shown to satisfy the HF equations
(see, e.g., \cite{book_Marder, 2005_HFforBosons}):
\be
L_0\phi_j \;\equiv\; L_{00}\phi_j - \sum_i \langle \phi_i |\,L_{00}\phi_j\rangle\,\phi_i \;=\,0,
\label{e2_03}
\ee
where
\be
L_{00}\phi_j \equiv 
\left( -\nabla^2 + \vext(\bfr) \,\right)\phi_j \,  +\, 
\sum_{i\neq j} \langle \phi_i | \,\vint\,|\phi_i \rangle \,\phi_j  
\, + \,
\sum_{i\neq j} \langle \phi_i | \,\vint\,|\phi_j \rangle \,\phi_i,
\label{e2_04}
\ee
$$
 \langle \phi_i | \,\vint\,|\phi_j \rangle \equiv 
 \int_{\mathbb{R}^2} d^2\bfr'  \, 
  \phi^*_i(\bfr') \, \vint(\bfr-\bfr')\,   \phi_j(\bfr')\,.
$$
Equations \eqref{e2_03} arise from minimizing the Hamiltonian functional, 
whose variational derivative is the $L_{00}$-term, subject to the constraints \eqref{e2_02},
which give rise to the sum in \eqref{e2_03}, with 
$\langle \phi_i |\,L_{00}\phi_j\rangle$ being Lagrange multipliers. 
In deriving \eqref{e2_04}, one uses the fact that the particles (helium atoms in our case)
are hard-core bosons, 
which cannot occupy the same site due to strong repulsion via the interaction potential $\vint$.
Also, $\nabla^2$ in \eqref{e2_04} is the Laplacian and $\vext$ is the external potential; the
third and fourth terms describe the mean-field (Hartree) and exchange (Fock) contributions, 
respectively. The spatial variables are nondimensionalized to some scale $d_0$,
usually taken to be a period of $\vext$, and all energies are nondimensionalized to the recoil
energy \ $E_R=h^2/(8md_0)$, where $h$ is the Planck constant and $m$ is the particle mass. 

Potential $\vext$ describes the effect on helium atoms by the hexagonal lattice of
graphene. 
It is obtained from the 3D Steele potential \cite{1973_Steele} as explained in \cite{PRB}.
This $\vext$ has the standard honeycomb shape
with minima over the centers of the graphene cells and maxima over the carbon atoms.
Focusing here on the numerical method rather than on fine aspects of the physical model,
we use the simplest such shape:
\bsube
\label{e2_add_01}
\be
\vext = V_0\, \left( \cos {\bf g}_1{\bf r} + \cos {\bf g}_2{\bf r} + \cos {\bf g}_3{\bf r} 
\right),
\label{e2_add_01a}
\ee
\be
{\bf g}_1 = \frac{2\pi}{d_0}\left[ 1, \frac1{\sqrt{3}} \right], 
\quad
{\bf g}_2 = \frac{2\pi}{d_0}\left[ -1, \frac1{\sqrt{3}}\right], 
\quad
{\bf g}_3 = {\bf g}_1+{\bf g}_2;
\label{e2_add_01b}
\ee
\esube
where the distance between centers of graphene cells is $d_0=2.46$\AA \ and 
$V_0$ is chosen so as to yield $(\vext)_{\max}- (\vext)_{\min}= 2.5E_R\approx 25$K, consistently with the value used in \cite{PRB}.
The recoil energy corresponding to the above $d_0$ is $E_R=9.9$K. 

The helium--helium interaction potential $\vint$ is taken from \cite{1995_Aziz}.
Its numerical modeling in \eqref{e2_04} is discussed in Appendix A.1.

\subsection{Numerical method}
\label{sec_2B}

We will solve Eqs.~\eqref{e2_02}, \eqref{e2_03} by the Accelerated version of the Imaginary-Time
Evolution Method (AITEM) 
given by Eqs.~\eqref{e2_05}--\eqref{e2_07} below \cite{multicomp}:
\bsube
\be
(\phi_j)_{n+1} = (\phi_j)_n -  \dt\,\left[ P^{-1} \,
\big( L^{(0)} \phi_j \big)_n -  \Gslown \; \uslown \,
\right] ,
\label{e2_05a}
\ee
%
where $n$ is the iteration number, $\dt>0$ is an auxiliary parameter called
the ``imaginary time step," 
$\big( L^{(0)} \phi_j \big)_n$ is the modification of \eqref{e2_03} that
takes into account the preconditioning operator $P^{-1}$:
\be
\big( L^{(0)} \phi_j \big)_n  \;\equiv\; (L_{00} \phi_j)_n - (\vec{\phi})_n\,
\langle P^{-1} (\vec{\phi})_n\, | (\vec{\phi})_n \rangle^{-1} \,
\langle P^{-1} \vec{\phi} \,| (L_{00} \phi_j)_n \rangle\,,
\label{e2_05b}
\ee
with
\be
(\vec{\phi})_n \equiv \big( [\phi_1,\phi_2,\ldots,\phi_N] \big)_n\,;
\label{e2_05c}
\ee
\label{e2_05}
\esube
and the positive definite operator $P$ being chosen as 
\be
P = c - \nabla^2
\label{e2_06}
\ee
for some constant $c>0$. 
The specific form of the second term on the r.h.s. of \eqref{e2_05b} is derived
in Supplemental Material based on the general formula presented in \cite{multicomp}.
There we also show that equation \ $L^{(0)}\phi_j = 0$ \ is equivalent to Eq.~\eqref{e2_03}.

The purpose of the $\Gamma$-term in \eqref{e2_05a} is to perform mode elimination (ME),
i.e., eliminate the
slowest-decaying mode from the iterative solution \cite{ME}.  This slowest mode
is approximated by:
\bsube
\be
\uslown = (\phi_j)_n - (\phi_j)_{n-1},
\label{e2_07a}
\ee
and for a single equation (i.e., $N=1$ and solution $\phi_1\equiv\phi$ above), 
the coefficient $\Gslown$ is to be chosen as \cite{ME}:
\be
\Gslown = \gslown\,
\frac{\langle u_{{\rm slow},\,n} | \big( L^{(0)} \phi \big)_n \rangle }
{ \langle u_{{\rm slow},\,n} | P\,u_{{\rm slow},\,n} \rangle }\,,
\qquad\quad
\gslown = 1 - \frac{s}{\alpha_n\dt},
\label{e2_07b}
\ee
where $\alpha_n$ is an approximation of the eigenvalue of the slowest-decaying eigenmode.
For the case of a single equation, one has
\be
\alpha_n = \frac{ \langle u_{{\rm slow},\,n} | L u_{{\rm slow},\,n} \rangle}
{ \langle u_{{\rm slow},\,n} | P u_{{\rm slow},\,n} \rangle}\,.
\label{e2_07c}
\ee
\label{e2_07}
\esube
%
The form of $\Gslown$ and $\alpha_n$ for $N>1$ coupled equations \eqref{e2_03} will be
discussed in Sec.~\ref{sec_3}. 
The parameter $s\in(\alpha_n\dt,1)$ in \eqref{e2_07b} determines how much of the slowest mode is being
subtracted at each iteration, with $s=\alpha_n\dt$ and $s=1$ corresponding to 0 and
100\% of the mode, respectively. An optimal range for $s$ will also be discussed in Secs.~\ref{sec_3}--\ref{sec_5}.
%
%
Operator $L$ in \eqref{e2_07c} is the linearized operator of \eqref{e2_05b}, defined as follows:
If $\vec{\phi}$ satisfies \eqref{e2_03} and $\|\widetilde{\vec{\phi}}\| \ll \|\vec{\phi}\|$, then
in the linear approximation:
\be
L^{(0)}\left( \vec{\phi} + \widetilde{\vec{\phi}} \; \right) \equiv 
L^{(0)}  \vec{\phi} + L  \widetilde{\vec{\phi}} = L_{0}  \vec{\phi} + L  \widetilde{\vec{\phi}}  
= L  \widetilde{\vec{\phi}}\,,
\label{e2_08}
\ee
where in the last step we have used \eqref{e2_03}. 
Per the note after \eqref{e2_06}, $L$ is also the linearized operator of $L_0$. 
Conveniently
for the method, the $L$-term in \eqref{e2_07c} 
does not need to be computed exactly but can instead be approximated by:
\addtocounter{equation}{-2}
\bsube
\addtocounter{equation}{3}
\be
 L\uslown \approx \big( L^{(0)} \phi_j \big)_n - \big( L^{(0)} \phi_j \big)_{n-1}\,,
\label{e2_07d}
\ee
%
\esube
assuming that $(\vec{\phi})_n$ is sufficiently close to the exact solution $\vec{\phi}$. 
(For future reference, 
we have written this relation for the case of multiple coupled equations, i.e., with 
the index $j$.)
The final step of the iteration method is the Gram--Schmidt orthonormalization of $\phi_j$'s
so as to ensure condition \eqref{e2_02} to hold to numerical precision.

Seven remarks about the above numerical method are in order. First, the main contributions 
of this work, listed in the order in which they are discussed in Sec.~\ref{sec_3},
are: \ (i) Finding a range of values of parameter $c$ in \eqref{e2_06} that 
significantly accelerates convergence of the iterations; \ 
(ii) \ Extension of the ME technique, i.e., the
form of the $\Gamma$-term in \eqref{e2_05a} and of the expressions in
 \eqref{e2_07b} and  \eqref{e2_07c}, from a single
equation to a system; \ and \ 
(iii) Extension of the same technique that allows one to eliminate {\em multiple}
slow modes simultaneously.
Thus, there are two ways in which the textbook ITEM is accelerated by our AITEM
\eqref{e2_05a}: via the preconditioning by operator $P$ and via ME by the $\Gamma$-term.

Second, a necessary condition for iterations \eqref{e2_05a} to converge is that the
eigenvalues of the linearized operator $L$ be positive. This cannot be guaranteed without
knowing the solutions $\phi_j$. However, it was shown in \cite{multicomp} that if $\phi_j$ are
dynamically stable (usually, ground-state) stationary solutions of time-dependent evolution
equations \ $i(\phi_j)_t = L_0\phi_j$, where $L_0\phi_j$ is 
a Schr\"odinger-type operator, as in \eqref{e2_03}, 
and `$i$' is the imaginary unit, then
iterations \eqref{e2_05a} without the $\Gamma$-term and for a sufficiently small $\dt$, do
converge. In this case they will also converge {\em with} a properly designed $\Gamma$-term,
since it does not change the sign of eigenvalues of the linearized operator $L$ but
effectively changes only the magnitude of the slowest-decaying eigenmode
of $L$ \cite{ME}. 

Third, parameter $\dt$ 
needs to be chosen to satisfy a relation:
\addtocounter{equation}{1}
\be
\dt < 2/\max \lambda_{P^{-1}L}
\label{e2_09}
\ee
so as to ensure convergence of the iterations; see, e.g., \cite{AITEM}.
The denominator in \eqref{e2_09} is the maximum eigenvalue of operator $P^{-1}L$.
Since this eigenvalue is usually not known, $\dt$ is chosen empirically from a small number
of trial simulations. More will be said about this in Sec.~\ref{sec_3}. 

Fourth, for periodic boundary conditions, which we use in this study, the Laplacian operator
in \eqref{e2_04} and \eqref{e2_06} is efficiently computed using the Fast Fourier Transform. 
An efficient computation of the $\vint$-terms is discussed in Appendix A.1.

Fifth, the exchange interaction between helium atoms 
(the Fock term in \eqref{e2_04}) is much weaker
than the mean-field interaction (the Hartree term).
Therefore, one does not need to compute the Fock term until the iteration
error becomes comparable in size to it. 
This can be monitored by computing the exchange interaction
between any one pair of neighboring atoms every so many (say, $O(10)$) iterations
and comparing it to the mean-field interaction between the same two atoms. 
Moreover, the Fock term is also more expensive to compute, as its computation
scales with the number of helium atoms as $O(N^2)$, whereas the computation of 
the Hartree term can be implemented to scale as $O(N)$ (see Appendix A.1).
Therefore, we used a well-known method from molecular dynamics (see, e.g.,
\cite{1992_Molly, 1999_Molly}) whereby this small but expensive term is computed
only every $n_{\rm Fock}\gg 1$ iterations, in between which one uses its latest
computed value. Furthermore, we verified for a number of selected cases 
(for high filling fractions) that using any value $n_{\rm Fock}$ from 1 to infinity
(i.e., not including the Fock term at all) makes very little difference for the 
final solution: it changed the central peak of the wave functions 
by a negligible amount and the much weaker side peaks
(see, e.g., Fig.~7 in \cite{PRB}), by some 10--20\%. 


Sixth, the Conjugate Gradient Method (CGM) is known to be the fastest generic 
fixed point-type method for problems with Hermitian positive definite operators, as far 
as the iteration count is concerned. An extension of the CGM to nonlinear wave
equations with quadratic constraints, such as \eqref{e2_02}, was presented in \cite{2009_CGM}. 
We tried to employ this method instead of the simpler AITEM \eqref{e2_05}--\eqref{e2_07},
but did not succeed. First and most importantly, the CGM iterations diverged. While we did not 
understand the reason for that divergence, we hypothesized that it could be intrinsic to how the
iteration method ``chooses" increments of the solution for the next iteration. Indeed, method 
\eqref{e2_05}--\eqref{e2_07}, which is a gradient decent-type method, chooses those increments 
so as to,
essentially, minimize energies of the helium atoms. This prevents such increments that lead to a
(significant) increase of the $\vint$-terms and thus keeps the overlap of neighboring $\phi_j$'s to a
minimum. On the other hand, increments chosen by the CGM are
designed to explore a wider range of directions \cite{1994_nopainCGM}. As a result, they can potentially
cause neighboring $\phi_j$'s to overlap to such an extent that the nearly singular behavior of 
$\vint(|{\bfr}|)$ near $\bfr=0$ would lead to a sharp increase in the $\vint$-terms and could
thus cause divergence of iterations. \ Second, while the CGM requires 
a factor of two to three fewer iterations to converge
than the ME-accelerated method \eqref{e2_05}--\eqref{e2_07} \cite{2009_CGM}, the cost of each
iteration may be more than twice greater for the CGM. This is due to two factors: \ 
(i) \ The CGM computes both $L^{(0)}\phi_j$ and $L$ acting on some auxiliary quantity 
(versus one $L^{(0)}\phi_j$ for the AITEM), as well as computes two constraint-enforcing
terms (those with angle brackets in \eqref{e2_05b}), per iteration.  (ii) \ Moreover, 
the linearized operator $L$ must be computed explicitly, contrary to that in \eqref{e2_07d}.
In examples considered in \cite{2009_CGM}, nonlinear terms had much simpler form
than the $\vint$-terms in \eqref{e2_04} and evaluation of the corresponding terms in $L$ was not
the most computationally expensive part of an iteration. In contrast, not only are the $\vint$-terms in
\eqref{e2_04} the most expensive ones to evaluate, but also their linearization is quite
cumbersome to code. Thus, even if CGM iterations had converged, their computational time might not
had been less than that of the simpler method \eqref{e2_05}--\eqref{e2_07}, and the coding
of the CGM is significantly more complex.

Finally, we chose to use a fixed point-type method over a Newton-type one, 
considered, e.g., in \cite{2011_RedRankExtrap}, because 
for the former methods, it is straightforward to compute spatial
derivatives using the time-efficient and spectrally accurate Discrete Fourier transform,
whereas it is not possible to do this time-efficiently for Newton-type methods.

\subsection{Initial condition for the iterations}
\label{sec_2C}

%
%

\begin{figure}[!ht]
	\vspace*{0.6cm}
	{\centering
		\includegraphics[width=8cm,angle=0]{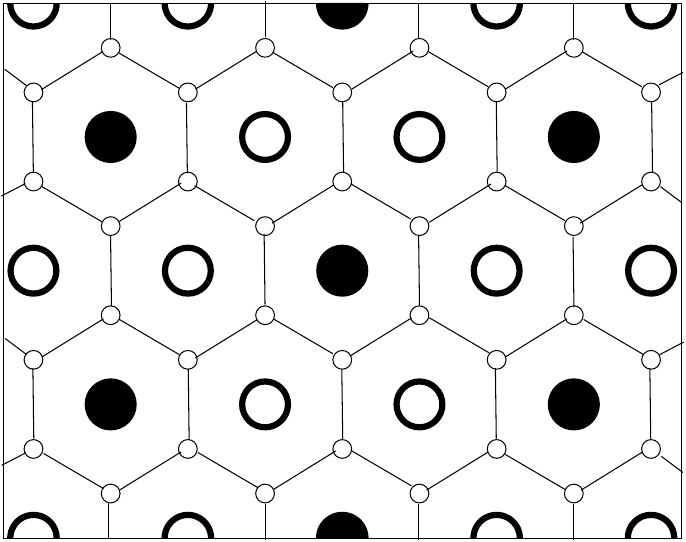}
	}
	\vspace{0.5cm}
	\caption{Schematics of helium atoms over a hexagonal
		graphene lattice. Small (large) circles
		denote carbon (helium) atoms. Thin lines are used to improve visibility 
		of the lattice. Filled circles show the $1/3$-filling
		of the lattice by helium. 
	}
	\label{fig_1}
\end{figure}

The ratio of the number $N$ of helium atoms and the number $N_{\rm cells}$
of graphene cells defines the filling fraction, $f\!f$, 
of the lattice by the atoms. Examples of commensurate $f\!f=1$ and
$f\!f=1/3$ are shown in \ref{fig_1}; examples of incommensurate filling fractions, 
where helium atoms are
{\em not} located over the centers of the graphene cells, will be presented in Sec.~\ref{sec_5}.  
The lattice has periodic boundary conditions to ensure consistency with the 
discrete Fourier transform used to compute the $\nabla^2$- and nonlocal terms in \eqref{e2_04}; see
Appendix A.  Due to this periodicity, non-unit filling fractions
 impose restrictions on the number of 
periods of the lattice. For commensurate filling fractions, 
these restrictions can be found by inspection.
For example, for  $f\!f=1/3$, one needs to have the number of cells in the $x$-direction, 
$N_{{\rm cells},\;x}$, be a multiple of 3. 
(Note that in Fig.~\ref{fig_1}, this restriction is violated
intentionally in order to illustrate that without it, $f\!f=1/3$ could not
be realized with periodic boundary conditions.)
For other filling fractions to result in periodic patterns, 
restrictions on $N_{{\rm cells},\;x}$ and $N_{{\rm cells},\;y}$
need to be found by trial and error on a case-by-case basis;
see Sec.~\ref{sec_5} for examples. 
In addition, for{\em all} filling fractions, 
one needs to have $N_{{\rm cells},\;y}$ to be a multiple of 2
in order to fit a hexagonal lattice to a rectangular computational window while respecting
periodic boundary conditions.

%

%

For the helium-over-graphene system considered here, 
fillings fractions with high $f\!f$ are only relevant for a
strictly 2D system.  In 3D simulations, or in experiments, 
a second layer of helium atoms begins to form above the first one for $f\!f\approx 0.62$ 
(equivalently, for area coverage of about 0.118 \AA$^{-2}$) \cite{1998_Whitlock,2012_Gordillo, 2012_Kwon} over either graphite or graphene.
To prevent this from occurring and thereby achieve higher values of $f\!f$ in 
3D simulations, one would restrict the motion of helium atoms away from the surface by 
placing a ``cap" on the simulation box, as was done in Ref.~\onlinecite{PRB}.
In the 2D model considered here, we can meaningfully study filling fractions in the
entire range $f\!f$ between 0 and 1.


To create an initial condition $(\phi_j)_0$ for unit-filling simulations, one places identical
copies of a 2D Gaussian function (with width of about a quarter of the lattice 
 side) at the center of each graphene cell. 
 For a non-unit filling $p/q$, we 
experimented with two alternative initial placement procedures of the aforementioned Gaussian. 
These details are described in Appendix A.3. 
The initial placement of helium atoms does not need to be (and in most cases is not) 
close to that obtained at the outcome of the
simulation, as the equations themselves will ``dictate" the equilibrium
 placement of the atoms.

\section{Choosing parameters of the numerical method}
\label{sec_3}

In Sec.~\ref{sec_3A}, we will discuss an optimal choice of parameter $c$ in the preconditioning
operator $P$ \eqref{e2_06}.
In Sec.~\ref{sec_3B} we will discuss the form of the slow 
mode-eliminating term (the $\Gamma$-term) in \eqref{e2_05a} for the case of multiple
coupled equations. In Sec.~\ref{sec_3C} we will demonstrate that ME accelerates the ``original"
AITEM (i.e., method \eqref{e2_05} {\em without} the $\Gamma$-term) by well over an 
order of magnitude. 
However, we will also reveal a problem of convergence of this ME-accelerated AITEM,
which we will seek to overcome by presenting a novel, multiple-mode eliminating
technique in Sec.~\ref{sec_4}. 

The discussions in Secs.~\ref{sec_3A},\ref{sec_3B} rely on the following well-established fact. Consider
a fixed point iteration method whose linearized form is
\be 
\vec{q}_{n+1} = \vec{q}_n - \dt A\, \vec{q}_n,
\label{e3_01}
\ee
where $\vec{q}$ is the iterated quantity and $A$ is a Hermitian positive definite matrix;
compare with \eqref{e2_05a}. 
Quantity $\vec{q}$ can be expanded over the set of eigenvectors $\vec{a}_i$ of $A$:
\ $\vec{q} = \sum_i b_i \, \vec{a}_i$,
where the expansion coefficients in subsequent iterations satisfy:
\be
(b_i)_{n+1} = (1 - \dt \lambda_{A,i}) (b_i)_n,
\label{e3_add_01}
\ee
with $\lambda_{A,i}$ being eigenvalues of $A$. 
Based on \eqref{e3_add_01}, convergence rate of \eqref{e3_01} for an 
optimal choice of $\dt$ can be shown to be (see, e.g., \cite{AITEM} or
 references in \cite{2009_CGM}):
\be
\frac{\|\vec{q}_n\| }{\|\vec{q}_{n+1}\| } = 
\frac{1 + (\lambda_A)_{\min}/(\lambda_A)_{\max}}
     {1 - (\lambda_A)_{\min}/(\lambda_A)_{\max}},
\label{e3_02}
\ee
Thus, to ensure the fastest convergence of
\eqref{e3_01} with a (nearly) optimal $\dt$, one needs to modify $A$ so as to minimize 
${\rm cond} A \equiv (\lambda_A)_{\max}/(\lambda_A)_{\min} > 1$.

In the context of iteration method \eqref{e2_05}, the role of $A$ is played by 
$P^{-1}L$, where $L$ is the linearized operator introduced 
in Sec.~\ref{sec_2B}.\footnote{
Here we adopt a simplified point of view
that the $\Gamma$-term merely reduces the amplitude of $L$'s eigenmode with 
$(\lambda_A)_{\min}$ \cite{ME}. In reality, the
situation is more complicated, as evidenced by the numerics presented in Sec.~\ref{sec_3C},
but this would require a separate study. Thus, we focus on 
the eigenvalues (and eigenfunctions) of $P^{-1}L$.
} 
It is not feasible to find $\lambda_{P^{-1}L}$ 
for the 2D problem \eqref{e2_03}--\eqref{e2_05}. 
Therefore, for the purposes of convergence analysis of the AITEM, 
we considered a 1D and Hartree-only reduction of the instead,
for which we were able to find those quantities numerically for a small number of 
atoms $N$. Those numerics used the idea of plane-wave expansion of the eigenfunctions 
of a Shr\"odinger-type operator (see, e.g., \cite{book_PC_Jianke}), but were more technically
complex due to the presence of the second term on the r.h.s. of \eqref{e2_05b} and the
nonlocal nonlinear terms in \eqref{e2_04}. Since details of those numerical implementations
are not central to this work, we omit them and below will present only the setup (in this
paragraph) and the results (in Secs.~\ref{sec_3A},\ref{sec_3B}). 
The Hartree-only reduction implies dropping the Fock term in
\eqref{e2_04} and using only $N$ normalization conditions \eqref{e2_02}
with $i=j$:
\bsube
\label{e3_03}
\be
L_{0}^{\rm 1D,H}\phi_j \,\equiv\, 
L_{00}^{\rm 1D,H}\phi_j -  \langle \phi_j |\,L_{00}^{\rm 1D,H}\phi_j\rangle\,\phi_j \;=\,0,
\qquad 
\langle \phi_j |\phi_j\rangle = 1,
\label{e3_03a}
\ee
\be
L_{00}^{\rm 1D,H}\phi_j \equiv 
\left( -\partial_x^2 + \vext(\bfr) \,\right)\phi_j \,  +\, 
\sum_{i\neq j} \langle \phi_i | \,\vint\,|\phi_i \rangle \,\phi_j  \,.
\label{e3_03b}
\ee
\esube
The discrete version of $P^{-1}L$ is a $NM_x\times NM_x$ matrix,
where $M_x$ is the number of grid points. 
For the results below, we chose, as representative values:
\be
\vext=-2\cos(2x), \qquad 
E_{\rm clip}=100\; \mbox{(see Appendix A)}
\label{e3_04}
\ee
in nondimensional units.

For the purpose of illustrating the behavior of the AITEM \eqref{e2_05}, 
all examples considered in this section will be for the 1D, Hartree-only, model
\eqref{e3_03}, \eqref{e3_04}. 
The behavior of the AITEM for the 2D HF model \eqref{e2_02}--\eqref{e2_04}
are qualitatively the same. The results for it will be presented in 
Sec.~\ref{sec_5}.


\subsection{Optimal value of $c$ in (\ref{e2_06})}
\label{sec_3A}

The role of $P^{-1}$ is to reduce large eigenvalues of $L$ associated with high
 wavenumbers $k$ (see, e.g., \cite{AITEM}); the same preconditioning is widely used 
 in computational condensed matter and nuclear physics: see, e.g., 
\cite{2019_Nuclear} and references therein). While the concept of preconditioning
\eqref{e2_06} is well-known, the question of an optimal $c$ value has not been,
to our knowledge, systematically explored. We are aware of only one such study,
\cite{AITEM}, where an optimal $c$ value was found for a different class of 
problems and led to a different result than the one found below. 

Large eigenvalues of $L$ occur due to the Laplacian in \eqref{e2_04} whose 
contribution to the large-$k$ eigenvalues 
(with corresponding eigenfunctions containing fast oscillations)
dominates that of the $\vext$- and $\vint$-terms. Those
eigenvalues of $P^{-1}L$ can be roughly estimated to be:
\be
\lambda_{P^{-1}L}(|k|\gg1)\sim (k^2+O(\vext)+O(\vint\mbox{-term}))/(k^2+c).
\label{e3_05}
\ee
An expression for the `$\vint$-term' is given in Appendix B.  
When the Laplacian dominates the contributions of the $\vext$- and $\vint$-terms
(and the value of $c$), the above estimate is $O(1)$. If we now, as an initial
motivation, assume that this estimate also holds, in the order of magnitude sense, 
for {\em all} eigenvalues, then we get: \ 
$\lambda_{P^{-1}L}(|k|\lesssim 1)\sim (O(\vext)+O(\vint\mbox{-term}))/c$. 
This suggests that if \ $|\vint\mbox{-term}|\gg 1$, then one needs to have
$c= O(\vint\mbox{-term})$ in order to have $\lambda_{P^{-1}L}(\,|k| =O(1)\,) = O(1)$ also
(here we have used that $\vext=O(1))$. 
We stress that the above order-of-magnitude estimate provides only an initial motivation
for the choice of $c$. 
Moreover, it is not possible to estimate the $\vint$-term without actually
computing eigenfunctions of $P^{-1}L$ (see Appendix B), and that  
computation requires one to assume a specific value of $c$ in $P$.

\begin{figure}[!ht]
	\begin{minipage}{7.1cm}
		\includegraphics[width=7.1cm,angle=0]{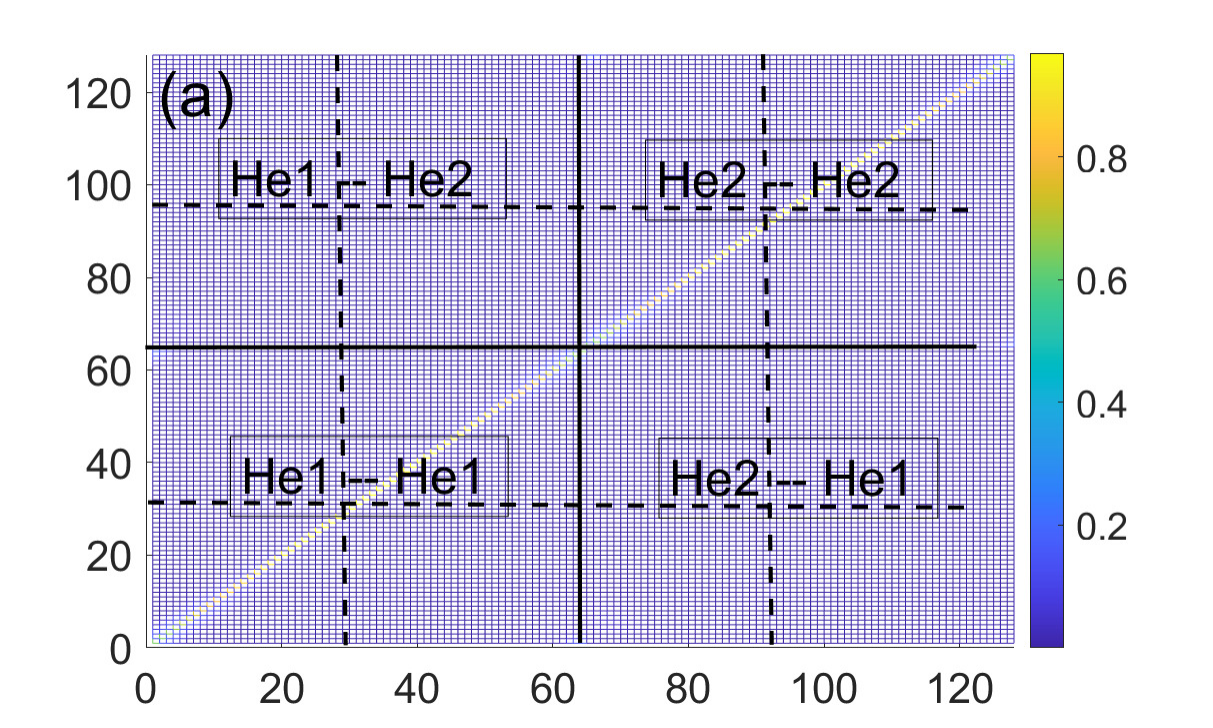}
		\vspace{1cm}
	\end{minipage}
	\hspace{2cm}
	\begin{minipage}{7.1cm}
		\vspace*{-0.3cm}
		\includegraphics[width=7.1cm,angle=0]{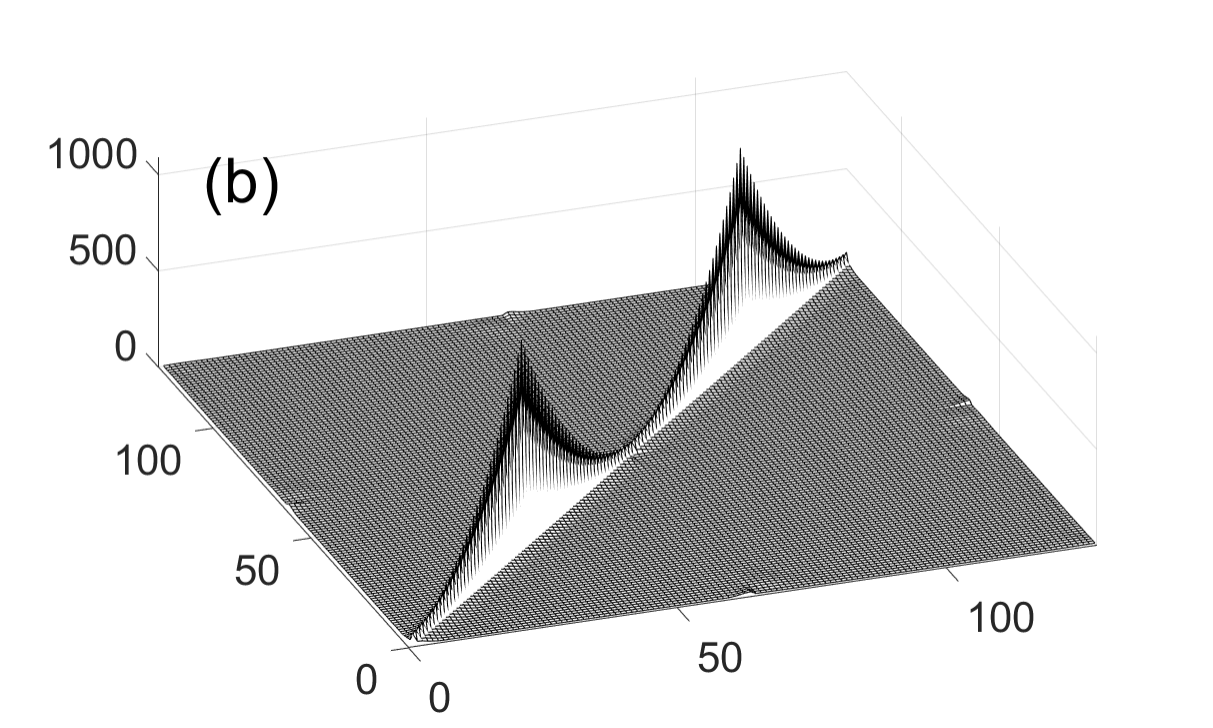}
		\vspace{1cm}
	\end{minipage}

\vskip -0.5 cm

	\begin{minipage}{7.1cm}
		\includegraphics[width=7.1cm,angle=0]{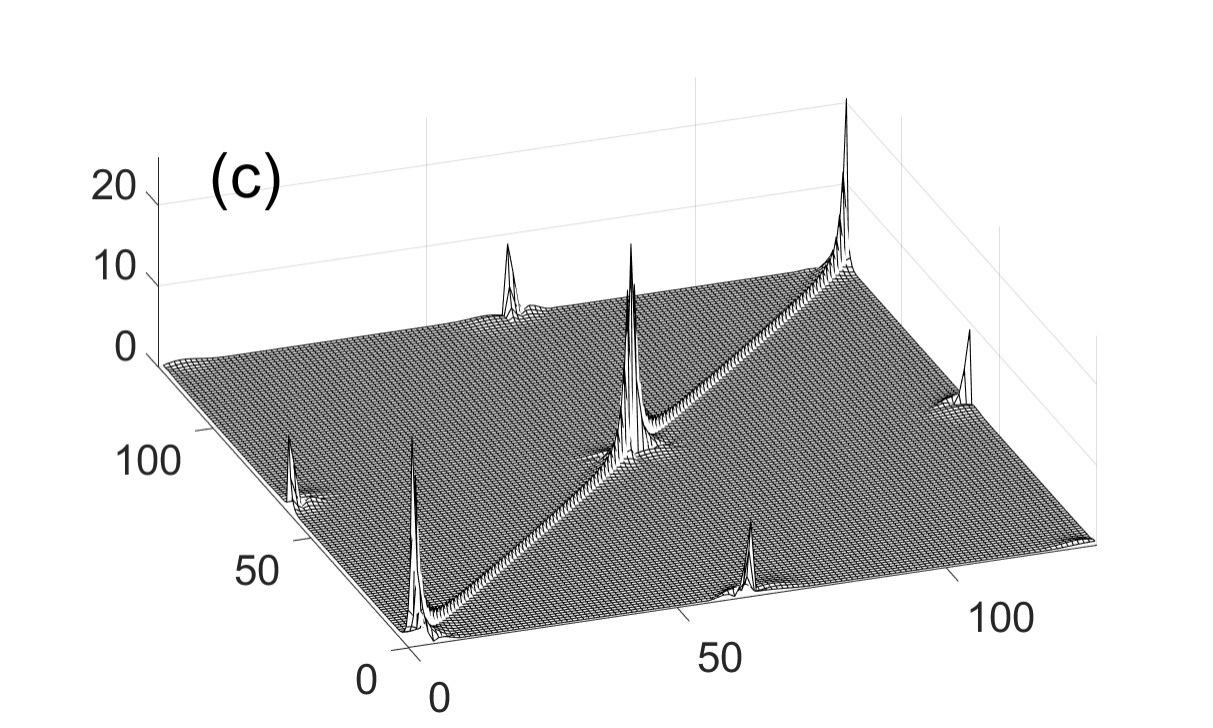}
		\vspace{0.1cm}
	\end{minipage}
	\hspace{2cm}
	\begin{minipage}{7.1cm}
		\vspace*{-0.3cm}
		\includegraphics[width=7.1cm,angle=0]{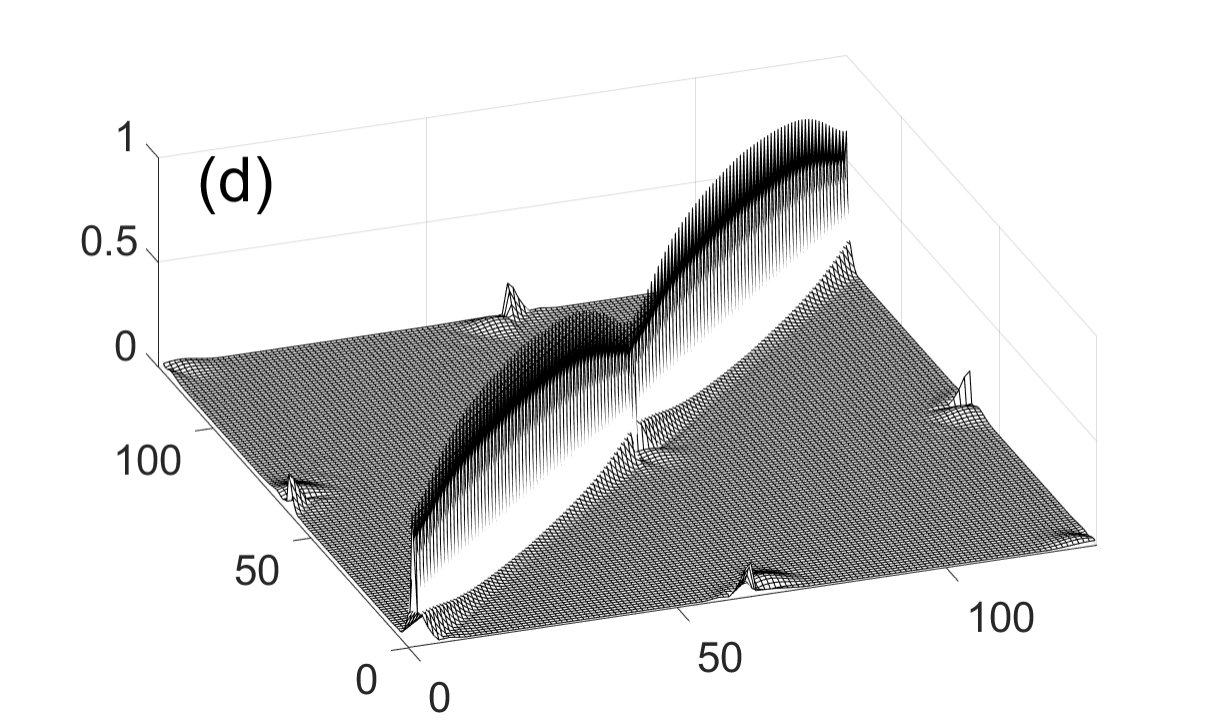}
		\vspace{0.1cm}
	\end{minipage}
	\caption{Absolute values of entries of $P^{-1}L$ for $N=2$ helium atoms
		in two adjacent wells of $\vext$ \eqref{e3_04}; $M_x=64$. \ 
		(a) \ Top view of panel (d), to illustrate the matrix structure:
		solid lines separate the blocks pertaining to atom $i$'s eigenfunction
		contributing to atom $j$'s linearized equation \eqref{e3_03a}, as marked; 
		dashed lines separate the two parts of the wavenumber vector, as per
		\eqref{A_03}. Panels (b,c,d) show the entries for $P=I$ and $P$ given
		by \eqref{e2_06} with $c=2$ and $c=100$, respectively. 
	}
	\label{fig_2}
\end{figure}

To break this circular logic, we instead examined
the size of entries of the $N M_x \times N M_x$ matrix $P^{-1}L$. Representative 
results are shown in Fig.~\ref{fig_2}.
This matrix appears to be diagonally dominant (if not in the
strict mathematical sense, then in appearance). 
The Gerschgorin Circles theorem implies that for a diagonally
dominant matrix $A$, a {\em necessary} (but not sufficient) condition
for ${\rm cond} A$ to be $O(1)$ is
that the entries along the diagonal be uniform. 
From Fig.~\ref{fig_2}, one sees that this occurs only when 
\be
c = O(E_{\rm clip}).
\label{e3_06}
\ee
 This implies that
for an optimal acceleration with the preconditioner $P^{-1}$,
one needs to choose $c$ according to \eqref{e3_06}.

To confirm this, we computed  ${\rm cond} (P^{-1}L)$ 
for the examples shown in Fig.~\ref{fig_2}. When $P=I$ (no preconditioning),
$\lambda_{\min}\approx 0.61$, $\lambda_{\max}\approx 1100$, 
${\rm cond} (P^{-1}L) \approx 1800$.  
For $P$ with $c=2$ (a representative value of order $O(\vext)$,
as previously found in \cite{AITEM} in the context of
nonlinear wave equations),
$\lambda_{\min}\approx 0.030$, $\lambda_{\max}\approx 35$, 
${\rm cond} (P^{-1}L) \approx 1150$. 
For $P$ with $c=100 = E_{\rm clip}$,
$\lambda_{\min}\approx 0.052$, $\lambda_{\max}\approx 0.98$, 
${\rm cond} (P^{-1}L) \approx 19$.

\begin{figure}[!ht]
	\vspace*{0.5cm}
    \begin{minipage}{7.5cm}
	\includegraphics[width=7.5cm,angle=0]{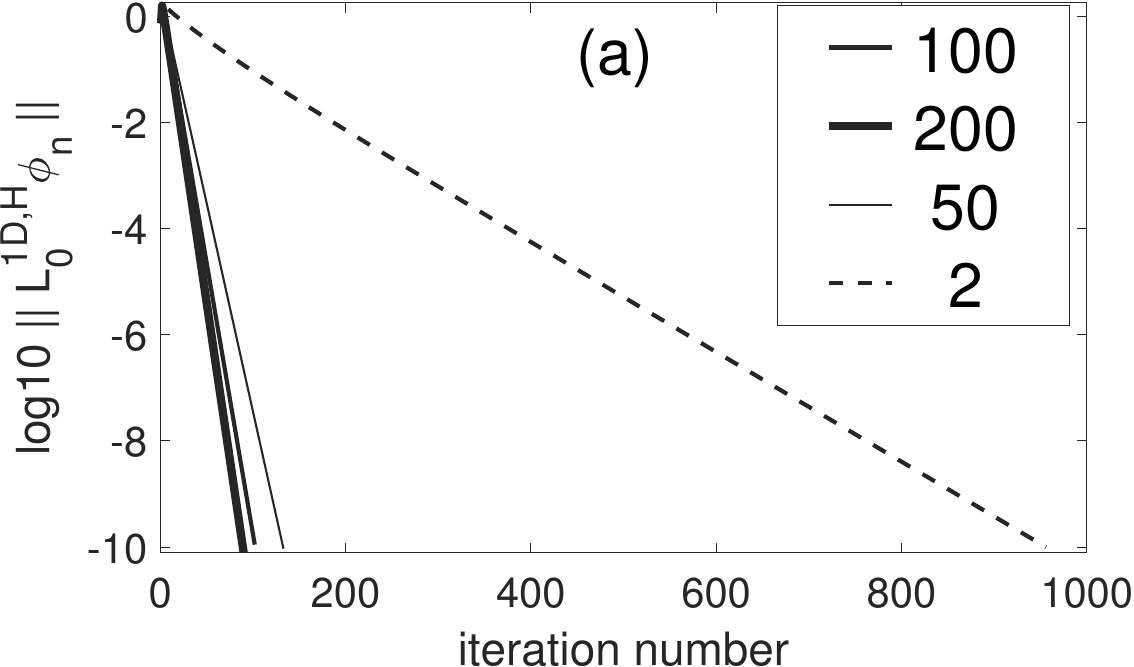}
	\end{minipage}
\hspace{1cm}
    \begin{minipage}{7.5cm}
	\includegraphics[width=7.5cm,angle=0]{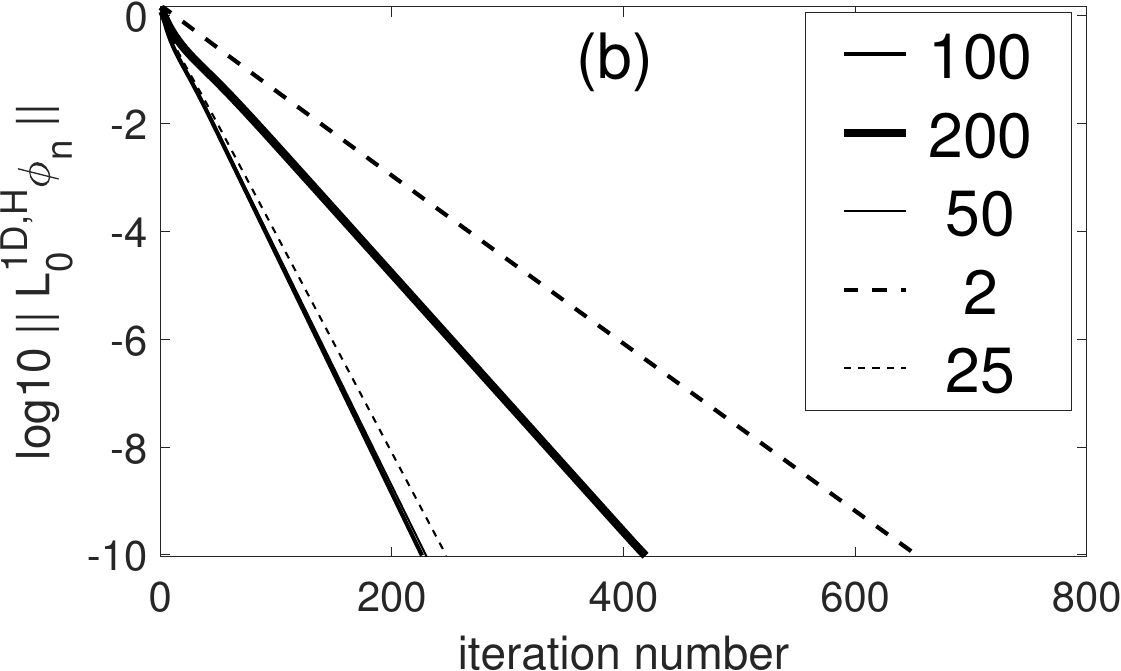}
    \end{minipage}
\vspace*{0.3cm}
	\caption{Evolution of the error in iterated Eqs.~\eqref{e3_03} for 
		 $N=20$ (a) and $N=10$ (b)
		 equidistant helium atoms at $\vext$'s 20 minima.
		 Here and below the error norm is defined as:
		 \ $\| L_0^{\rm 1D,H}\phi_n\| \equiv 
		 \sqrt{ \int dx \sum_{j=1}^N | L_0^{\rm 1D,H}(\phi_j(x))_n |^2 }\,/N$.
		  See main text for more details.
		The values of $c$ are listed in the legend.
		In (b), the thin and medium solid lines are 
		indistinguishable.
		Also, in (b) unlike in (a), {\em all} $\dt$ values were
		optimized by trial and error, and the corresponding values
		are $\dt=2.02, 2.20, 1.02, 0.044, 0.52$ for 
		$c=100, 200, 50, 2, 25$, respectively.
	}
	\label{fig_3}
\end{figure}

Figure \ref{fig_3}(a) illustrates that for high filling fractions, i.e., 
$f\!f\lesssim 1$, 
 $c$ can be in a rather wide range \eqref{e3_06}
and still provide superior performance over the case where $c=O(1)$.
Method \eqref{e2_05} without the $\Gamma$-term was used to iterate Eqs.~\eqref{e3_03} 
for $N=20$ helium atoms initially {\em placed exactly at the centers}
of $N$ adjacent wells of $\vext$:
\be
(\phi_j)_0 = \exp\left[ -(x-x_j)^2/w_0^2 \right], 
\label{e3_07}
\ee
where $x_j$ are the locations of $\vext$'s minima and the width $w_0=0.2\pi$. 
The exact value of the width is not important as long as the initial Gaussians do not
overlap; the importance of the italicized clause in the previous sentence and the
absence of the need to include the $\Gamma$-term in \eqref{e2_05} in this case
will be explained
in Sec.~\ref{sec_3B}. The values of parameter $\dt$ for $c=50,100,200$ were {\em not} optimized.
We merely checked that $\dt_{100}=0.6$ worked for $c=100$ and then used $\dt=0.5\dt_{100}$
and $\dt=2\dt_{100}$ for $c=50$ and $c=200$, respectively. This was motivated by 
the rough estimate \eqref{e3_05}, which along with \eqref{e2_09} suggested that 
$\dt$ increases with $c$ (again, this is just a very approximate estimate). 
The point of not optimizing $\dt$ was to emphasize that method \eqref{e2_05}
with a preconditioner $P$ where $c$ satisfies \eqref{e3_06}, does not require
fine-tuning of $\dt$ to yield a much faster convergence than when $c=O(1)$.
On the contrary, for $c=2$ in Fig.~\ref{fig_3}(a), 
we used a (nearly) optimal $\dt=0.025$, which we had found
by trial and error. The main conclusion is that for sufficiently high filling fractions
(with $f\!f=1$ considered above as an example), a value of $c$ satisfying \eqref{e3_06}
improves the convergence rate by at least on order of magnitude and likely more, given
that we did not optimize $\dt$ for those $c$ values.

For low filling fractions, we have found that an optimal value of $c$ is still in the
range \eqref{e3_06}. However, the sensitivity of the convergence rate on $c$ is weaker
than for $f\!f=1$, considered above. Figure \ref{fig_3}(b) shows the error evolution
for $f\!f=1/2$, which is the largest commensurate filling with $f\!f<1$ in 1D, with $\dt$
being optimized for each value of $c$. The conclusions for such low
(see below)  filling fractions
are: \ (i) The convergence rate for the optimal $c$ is only some factor of three better
than for a suboptimal $c=O(1)$; \ and \ (ii) \ It may be advisable to choose $c$ in the
 lower part of range \eqref{e3_06}, contrary to the case of higher filling fractions.

We conclude this subsection with two comments. First, $f\!f=1/2$ in 1D corresponds to
$f\!f=(1/2)^2=1/4$ in 2D, which is lower than the commensurate filling with $f\!f=1/3$ in 2D.
All filling fractions that we consider in Sec.~\ref{sec_5} are greater than 1/3. Based on
Fig.~\ref{fig_3}, we then set
\be
c = E_{\rm clip}
\label{e3_08}
\ee
in all simulations presented in this work and do {\em not} optimize
$\dt$, as we focus on other aspects of convergence acceleration. \ Second, the relatively
low iteration count, in the low hundreds, suggested by Fig.~\ref{fig_3}, is due to the 
initial guesses \eqref{e3_07} for the wavefunctions $\phi_j$ being centered {\em precisely}
at the minima of $\vext$. When this restriction is lifted (which can be avoided
only for a small minority of filling fractions such as 1, $1/3$, and a small number of others), 
the iteration count goes up
by well over an order of magnitude and sometimes much more. This, as well as the ME-based
acceleration technique required to overcome this problem, is considered in the next
subsection. 


\subsection{Form of the $\Gamma$-term in (\ref{e2_05a})}
\label{sec_3B}

In this subsection we will motivate and present the form of the $\Gamma$-term in
\eqref{e2_05a} for the case of multiple equations, as in \eqref{e2_03}. 
Recall that in \eqref{e2_07} in Sec.~\ref{sec_2A} that term was presented for a single equation,
as introduced in \cite{ME}.

\begin{figure}[!ht]
	\centering
	\includegraphics[width=9cm,angle=0]{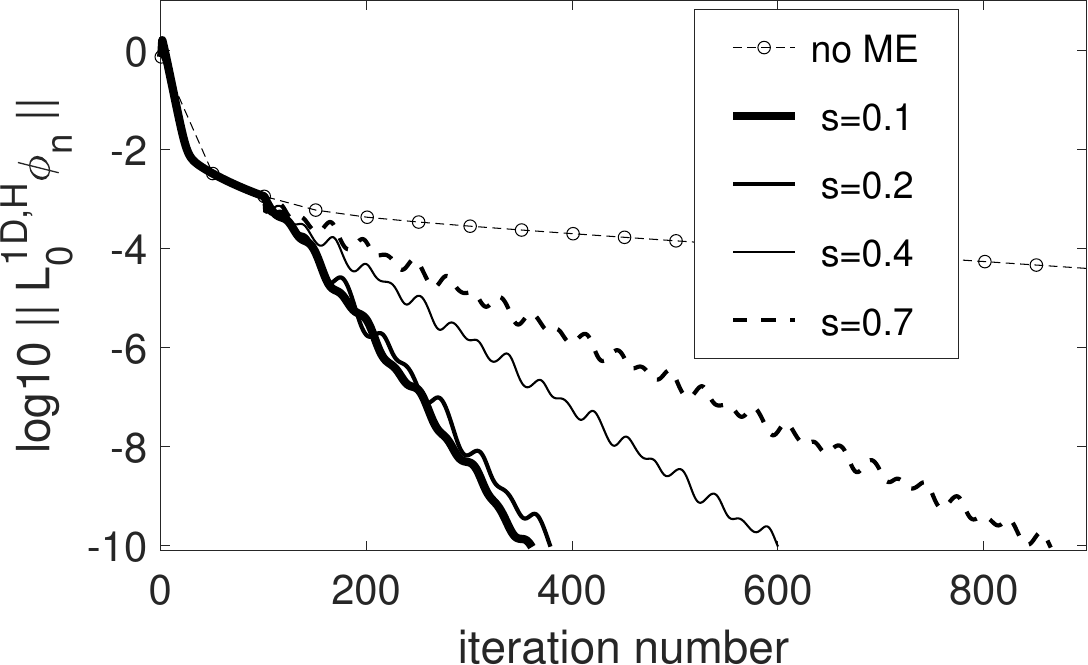}
	\caption{Error evolution for the same setup as in Fig.~\ref{fig_3}(a)
		with $c=100$, $\dt=0.6$, but when in the initial condition \eqref{e3_07} 
		the atoms are
		shifted by small random amounts from the minima of $\vext$. 
		The line with circular symbols shows the behavior of iteration method
		\eqref{e2_05} without the $\Gamma$-term (no slow mode elimination);
		the other lines show the behavior of the same method with the $\Gamma$-term
		for different values of parameter $s$.  In those cases, ME was started at the
		100th iteration. Starting it at another iteration between 50th and 200th did not
		change the results significantly.
	}
	\label{fig_4}
\end{figure}

The dashed line with circles in Fig.~\ref{fig_4} 
shows the error evolution for the same setup as in Fig.~\ref{fig_3}(a),
but when in the initial condition \eqref{e3_07} the atoms are
shifted slightly (by $\sim 1$\% of $\vext$'s period) by random amount from the minima
of their respective potential wells. The iterations converge to the same accuracy as in
Fig.~\ref{fig_3} in 4930 iterations, i.e., about 50 times slower than in the case
when initial conditions are centered exactly at the minima of the wells. 
The culprit, then, must be a mode, or modes, of the linearized operator $P^{-1}L$
which are related to shifts of the atoms from their equilibria (i.e., the wells' minima)
and which must have much smaller eigenvalues than the other modes, as per \eqref{e3_add_01}.
It is those modes that will need to be eliminated by an appropriately chosen $\Gamma$-term
in \eqref{e2_05a}. The question is then: \ Should one eliminate those ``shift" modes one
per atom, or are there certain combinations of the shifts that compose the slow 
(i.e., slowly decaying) modes? In the latter case, it would be those combinations that 
need to be eliminated. 

\begin{figure}[!ht]
	\centering
	\includegraphics[width=10cm,angle=0]{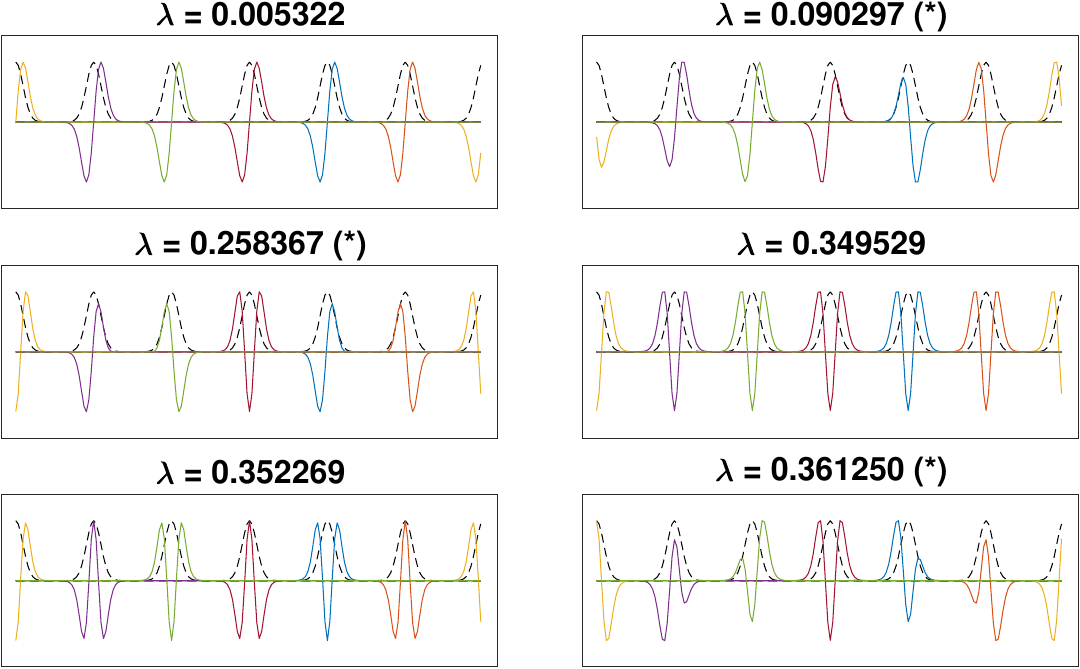}
	\vspace{0.5cm}
	\caption{Eigenfunctions of $P^{-1}L$ with $c=100$ for 
		$N=6$ helium atoms placed in $N$ adjacent wells of $\vext$,
		corresponding to
		its first few smallest eigenvalues.
		Thin dashed lines show the locations of the helium atoms.
		Vertical scale is in arbitrary units.
		The modes with asterisks next to the $\lambda$ value are doubly repeated.
		In those cases, the second eigenfunction (not shown) is obtained from the
		first by a shift by one period of $\vext$.
	}
	\label{fig_5}
\end{figure}

The answer to that question is inspired by Fig.~\ref{fig_5}. A unique mode with the lowest
eigenvalue corresponds to a {\em common shift} of all helium atoms. (To see that component 
$j$ of this eigenfunction corresponds to a shift of atom $j$, think about the shape of
the derivative of wavefunction $\phi_j$, which is qualitatively similar to \eqref{e3_07},
with respect to its center $x_j$.) The next eigenfunction, corresponding to two groups of
the atoms shifting out of sync, has a much larger (by almost 20 times) eigenvalue. 
Higher modes in that figure correspond to ``breathing" (width--amplitude changes) of
$\phi_j$'s and its combinations with shifts. A discussion about why the common shift
corresponds to the lowest-$\lambda$ eigenfunction for the system in question, is found
in Appendix C.1. 

Thus, based on the above, the $\Gamma$-term must eliminate {\em a single mode common to
all $\phi_j$'s}. Then, the following expressions replace their counterparts in 
\eqref{e2_07b}, \eqref{e2_07c}:
\bsube
\label{e3_09}
\be
\Gslown = \gslown\,
\frac{\sum_{j=1}^N \langle  (u_{{\rm slow},\,j})_n | \big( L^{(0)} \phi_j \big)_n \rangle }
{ \sum_{j=1}^{ N \vphantom{\mathlarger{T}} } 
  \langle (u_{{\rm slow},\,j})_n | P\,(u_{{\rm slow},\,j})_n \rangle }
\,\equiv\,
\gslown\, \frac{ \langle \big( \overrightarrow{u_{\rm slow}}\big)_n^T \,|\, 
		         \big( L^{(0)} \vec{\phi}\big)_n^T \rangle }
 { \langle \big( \overrightarrow{u_{\rm slow}}\big)_n^{T \vphantom{\mathlarger{T}} } \,|\, 
           \big( P\, \overrightarrow{u_{\rm slow}} \big)_n^T \rangle } ;
%
\label{e3_09a}
\ee
with $\gslown$ still being given by \eqref{e2_07b}, and where now 
\be
\alpha_n = \frac{ \sum_{j=1}^N 
	              \langle (u_{{\rm slow},\,j})_n | L (u_{{\rm slow},\,j})_n \rangle}
{ \sum_{j=1}^{N \vphantom{\mathlarger{T}} }  
  \langle (u_{{\rm slow},\,j})_n | P (u_{{\rm slow},\,j})_n \rangle}
\,\equiv\,
 \frac{ \langle \big( \overrightarrow{u_{\rm slow}}\big)_n^T \,|\, 
    	\big( L \, \overrightarrow{u_{\rm slow}} \big)_n^T \rangle }
  {	\langle \big( \overrightarrow{u_{\rm slow}}\big)_n^{T \vphantom{\mathlarger{T}}} \,|\, 
    \big( P\, \overrightarrow{u_{\rm slow}} \big)_n^T \rangle }.
\label{e3_09b}
\ee
\esube
Here $\overrightarrow{u_{\rm slow}}$ is a vector defined from its components
 \eqref{e2_07a}  similarly to \eqref{e2_05c}.

The error evolution of such an ME-accelerated AITEM \eqref{e2_05} is 
shown in Fig.~\ref{fig_4} for several representative values of parameter $s$
(see the text after \eqref{e2_07c}). We note that in the case of a single equation
\cite{ME}, a relatively large value $s=0.7$ was advocated as optimal. In contrast,
here, much smaller values $s\in[0.1, 0.2]$ are shown to result in a considerably
faster convergence. 
We hypothesize that the smaller values of $s$ being optimal here
is related to the fact that the (preconditioned) linearized operator $P^{-1}L$ is
still quite stiff, with ${\rm cond}\, (P^{-1}L) = \lambda_{\max}/\lambda_{\min}
\approx 1.8/0.0053 > 300$.

An important note is in order about what mode is eliminated by the ME. Initially,
i.e. right after the ME is started,
it is the mode with the smallest eigenvalue of $P^{-1}L$ 
(or a combination of a few such modes if 
their eigenvalues are close to one another). Indeed, it is that mode which,
in the absence of the $\Gamma$-term in \eqref{e2_05a},
``survives" the longest in the iteration error and hence dominates
$(u_{{\rm slow},\,j})_n$ in \eqref{e2_07a}.
However, after the ME has been applied for several tens of iterations, the
contribution of those lowest-$\lambda$ modes to $(u_{{\rm slow},\,j})_n$
is reduced (by design), and the latter becomes a mix of several low-$\lambda$
modes, with $\alpha_n$ representing some kind of weighed average of the
eigenvalues of those modes at iteration $n$. This mix of modes and, hence, 
$\alpha_n$ vary from one iteration to the next. This will be illustrated
in Sec.~\ref{sec_4B}.

Concluding this subsection, we mention that  for {\em commensurate}
filling fractions smaller
than 1 in 1D, the benefits of ME for AITEM \eqref{e2_05} are slightly less pronounced.
For $f\!f=1/2$, considered in Fig.~\ref{fig_3}(b), the AITEM with $c=100$, $\dt=0.6$ 
and without ME converges in about 1450 iterations when initial conditions are offset
from $\vext$'s minima. ME with $s\in[0.1, 0.3]$\footnote{Note that this optimal range
	is slightly higher than that for $f\! f=1$, which is consistent with our hypothesis
	about it two paragraphs above and the estimate of ${\rm cond}(P^{-1}L)$ 
	found in Appendix C.1.
}
brings the iteration count below 300.
This 5-fold improvement should be compared to the more than 10-fold one for $f\!f=1$,
as shown in Fig.~\ref{fig_4}.  A brief discussion of this is found in Appendix C.1.
We observed a similar relation between iteration counts
for $f\!f=1/3$ and $f\!f=1$ in 2D. It is then surprising that for {\em incommensurate}
$f\!f<1$, the benefits of ME can be much greater. This is demonstrated in the next
subsection.

\subsection{Performance of the ME-accelerated ITEM in 1D}
\label{sec_3C}

Here we will examine performance of AITEM \eqref{e2_05}, \eqref{e3_09} for two
incommensurate filling fractions. 
Our findings will, on the one hand, confirm a very significant
benefit of ME for convergence acceleration, but, on the other, highlight a problem,
which will be dealt with in the next section. 

\begin{figure}[!ht]
	\centering
	\includegraphics[width=9cm,angle=0]{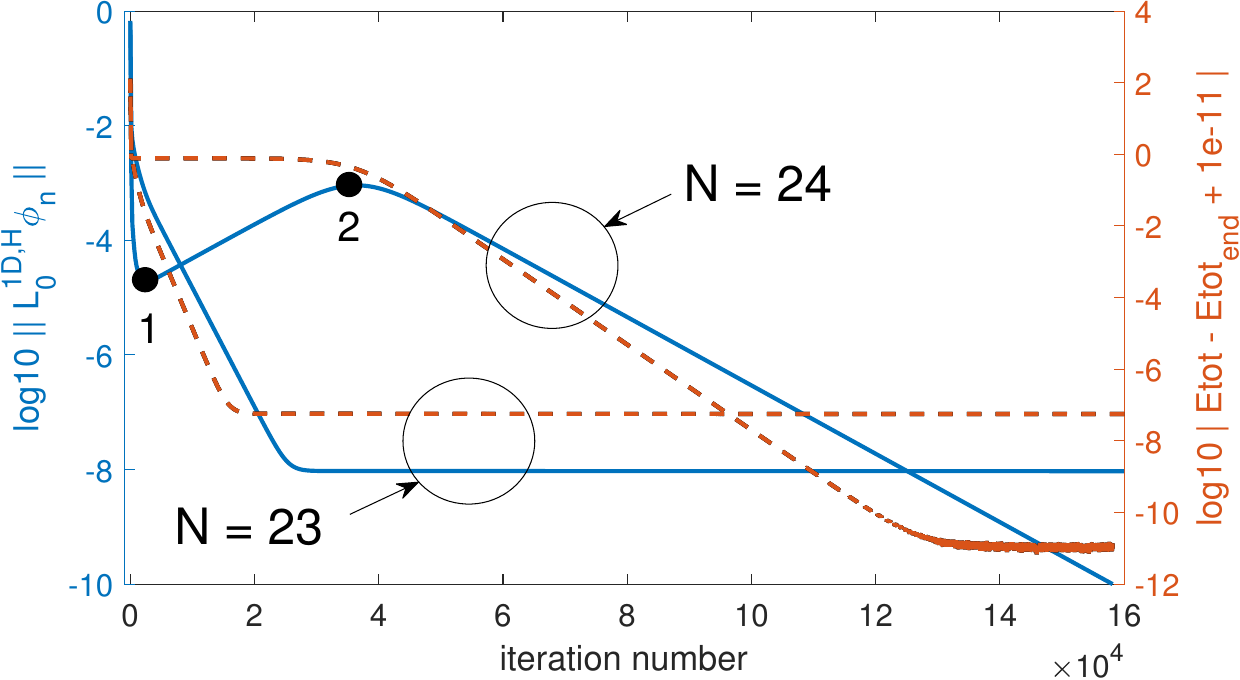}
	\vspace{0.5cm}
	\caption{Error (solid lines) and total energy (dashed lines) evolutions 
		of the AITEM \eqref{e2_05} without the $\Gamma$-term. 
		See main text for more details. 
		In the $y$-label of the right axis, the subscript `end' refers 
		to the last iteration in each of the two cases: \ $\sim 158$
		thousandth iteration
		for $N=24$ and  $\sim 115$ millionth iteration for $N=23$, when the 
		equation error reaches $10^{-10}$. The term $10^{-11}$ is added 
		to the argument of the
		logarithm to avoid running into $\log 0$. 
		A noisy plateau in the energy evolution between iterations 130K and
		160K is a numerical artifact; in the absence of a machine round-off error, 
		the curve would continue going down at a constant slope.
	}
	\label{fig_6}
\end{figure}

Figure \ref{fig_6} shows the evolution of the error and the total energy 
of Eqs.~\eqref{e3_03}:
\be
E_{\rm tot} = \sum_{j=1}^N \int_{\mathbb{R}^2} \, \left( \, 
 \left( \partial_x \phi_j\right)^2 + \frac12 \vext\, \phi_j^2 + 
 \frac12  \langle \phi_j \,| \, 
  \sum_{i\neq j} \langle \phi_i | \,\vint\,|\phi_i \rangle \,\phi_j  \,\rangle 
  \,\right) dx
\label{e3_10}
\ee
obtained by AITEM \eqref{e2_05} {\em without} ME for 30 periods of $\vext$ 
and $N=23$ and
$N=24$ helium atoms. The initial conditions are \eqref{e3_07} with $x_j$ uniformly
placed in the computational window. 
We used the same simulation parameters as those used in Sec.~\ref{sec_3B}
and listed in the caption to Fig.~\ref{fig_4}. 
Not optimizing parameters $c$ and $\dt$ of the AITEM, as well as the ME parameter $s$ mentioned
later, was a deliberate choice made so as to focus on other issues. 

The error in the $N=24$ case ($f\!f =0.8$) 
exhibits a peculiar behavior: After dropping to almost
$10^{-5}$ in a couple thousand iterations, it then slowly climbs up almost 2 orders
of magnitude before decaying again. The total energy \eqref{e3_10} decays monotonically,
with the local maximum of the error corresponding to an inflection point of the 
energy curve. Note that the energies of the intermediate solutions corresponding to
points labeled `1' and `2' are $O(1)$ greater than that of the final solution. 
Moreover, the differences of the solutions $\phi_j$ themselves at these points from
the final solution amount to a common shift of all atoms (similar to the first panel
in Fig.~\ref{fig_5} but with slightly non-uniform amplitudes) of sizes around 0.5 and 0.25, 
respectively (i.e., both also being $O(1)$). 
Interestingly, the difference from the final solution at point `2' is smaller than the 
difference at point `1', even though the error at `2' is greater. In other words, even 
though the error (of the equations)
exhibits a non-monotonic evolution, the iterated solution appears to 
approach the exact solution monotonically. 

From the error evolution, one can estimate (see Appendix C.2) 
\footnote{
	It is not possible to compute eigenvalues of $P^{-1}L$ in this case with the code
	mentioned earlier in this section because for $N \gtrsim 10$, Matlab runs
	out of memory.}
that for the incommensurate filling fraction $f\!f=0.8$ ($N=24$), 
the smallest eigenvalue of the linearized operator $P^{-1}L$ is
 $O\left(10^{-3}\ldots 10^{-4}\right)$, i.e., significantly smaller than that
 for both commensurate filling fractions considered in Sec.~\ref{sec_3B}.
A qualitative explanation of this is the following. For commensurate filling fractions, 
when the atoms undergo a common shift, the energy of each atom increases as it moves away 
from ``its" minimum of $\vext$. However, for an incommensurate filling, a (nearly)
uniform shift causes external potential energy of some atoms increase while those
of other decrease, with the net amount of energy change being close to zero.

The case $N=23$ ($f\!f\approx 0.767$) 
illustrates the phenomenon of $L$ having extremely small eigenvalues
even more dramatically. In Fig.~\ref{fig_6} it may appear that after having
decreased to $O(10^{-8})$, the error stalls, and so does the total energy. 
However, the error does actually decay, monotonically, but does so extremely slowly:
it takes about 115 million iterations (and about 5 days on a PC) to reach $10^{-10}$.
The estimates found in Appendix C.2 suggest that the minimum eigenvalues of 
$P^{-1}L$ and $L$ are $O(10^{-7})$ and $O(10^{-5})$, respectively.

\begin{figure}[!ht]
	\centering
	\includegraphics[width=9cm,angle=0]{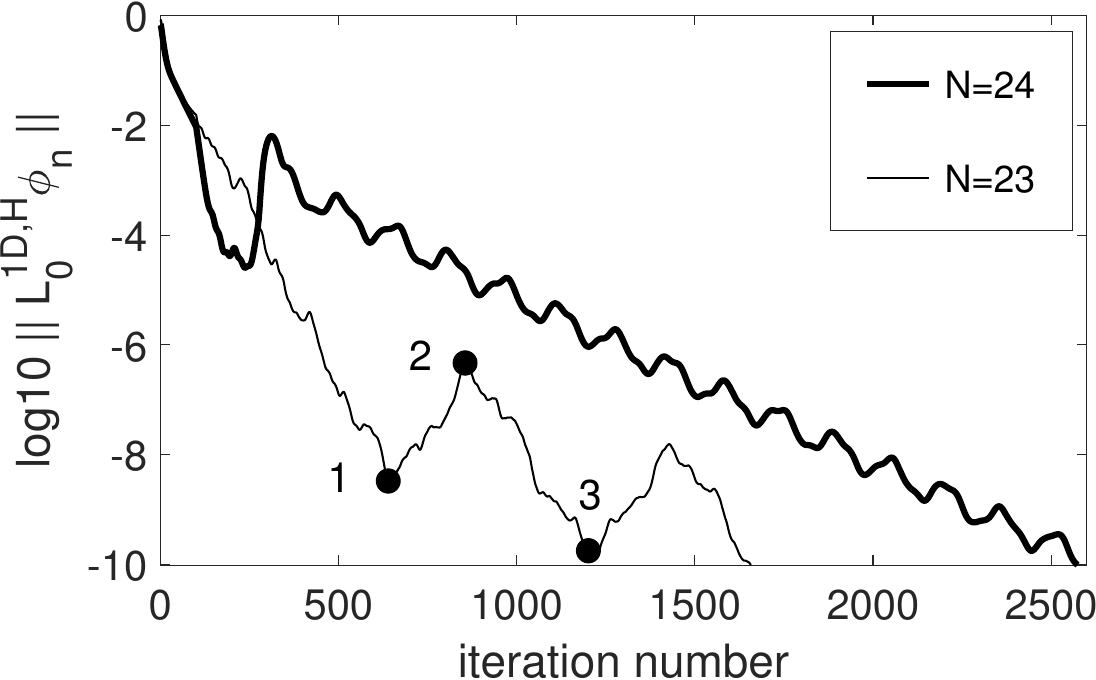}
	\vspace{0.5cm}
	\caption{Error evolution 
		of the AITEM \eqref{e2_05} accelerated by ME.  
		Points labeled `1'--`3' here are not related to points `1', `2' in Fig.~\ref{fig_6};
		they are commented on in Sec.~\ref{sec_6}.
	}
	\label{fig_7}
\end{figure}

Figure \ref{fig_7} shows the evolution of the error for the same two cases as in 
Fig.~\ref{fig_6}, but with the ME using parameter $s=0.1$. That value was used 
merely as a 
representative one advocated for in the previous subsection. In the case $N=23$
the convergence rate is insensitive to $s$ varying in a wide range $s\in[0.05, 0.4]$,
while for $N=24$ it actually improves for $s$ decreasing towards $0.05$.

Figure \ref{fig_7} illustrates two salient points about ME. First,
it improves the convergence rate of the AITEM for incommensurate  filling fractions
even more than for commensurate ones: by over 50 times for $N=24$ and by more than
four orders of magnitude for $N=23$. Second, the error evolution of the AITEM with ME
can be highly non-monotonic, exhibiting oscillations of some two orders of magnitude
(here, for the $N=23$ case). 
Both types of error evolution, of which the one for $N=24$ was found to be more common,
are also the typical scenarios in 2D, as we will discuss in Sec.~\ref{sec_5}.

In order to further 
improve the convergence rate, it is essential to understand an origin of 
error oscillations and devise a way to reduce or eliminate them. 
Especially problematic may seem the large oscillations, where the error first decreases
to a value that appears sufficient for practical purposes (e.g., $10^{-8}$ at point `1'
in Fig.~\ref{fig_7}), but then, unexpectedly, increases for several
hundred iterations before starting to decrease again.
We have been unable to understand the origin of those oscillations; 
this open problem
is to be addressed in future research. We have also tried a number of ad hoc ways
to reduce the oscillations, none of which has led to a significant or consistent
improvement. These attempts are summarized in the Supplemental Material, so that
future researchers would know what has not worked.

One of our attempts, however, led to a novel and non-trivial extension of the ME
technique, with the motivation being the following. 
We mentioned in Sec.~\ref{sec_3B} that the ME \eqref{e3_09} eliminates not the lowest-$\lambda$
eigenmode of $P^{-1}L$ but rather a mix of several low-$\lambda$ modes that compose
\eqref{e2_07a}. 
Each of those modes is multiplied by the same $\Gslown$, which represents
some ``average reduction factor" that is designed to reduce the amplitude of an ``average" mode.
As the analysis in \cite{ME} suggests, while 
the lower-$\lambda$ modes composing \eqref{e2_07a} 
will be suppressed by such an ``average reduction factor," the higher-$\lambda$ modes
in \eqref{e2_07a} 
will be amplified by it. This may be a reason leading to an oscillatory evolution 
of the error. Therefore, one can expect to be able
 to reduce error oscillations if one eliminates
{\em not} a single ``average mode" \eqref{e2_07a} but instead decomposes \eqref{e2_07a}
into several, ``more exact," eigenmodes and eliminates each of them with its proper 
reduction factor $\Gslown$. This multiple-mode elimination technique
 is described in the next section.

\section{Multiple-mode elimination (mME)}
\label{sec_4}

In the first subsection, we will 
derive equations of the mME, which generalize \eqref{e3_09} to the case of
multiple modes being eliminated with their individual reduction factors $\Gamma_{{\rm slow},\,n}$.
In the second subsection, we will illustrate improvement brought about by this method
over the single-mode ME \eqref{e3_09} for the 1D Hartree equations \eqref{e3_03},
as well as discuss certain technical implementation details of both kinds of ME.
In the third subsection we will
 show that both ME techniques are superior to the Anderson Acceleration
for the type of problems considered in this work. 

\subsection{Derivation of the mME equations}
\label{sec_4A}

Here we present a conceptual derivation of the method, which culminates in the Algorithm
presented towards the end of this subsection. 
Its implementation issues are discussed in Appendix D.

The main idea is to extract the modes, which will be eliminated later, from
increments of the solution at several consecutive iterations. To that end, 
consider a $K \times (N\cdot M_xM_y)$ matrix that generalizes \eqref{e2_07a}:
\be
\wD = \left( 
\ba{c} \vD_{n-1} \\ \vdots \\ \vD_{n-K} \ea \right),
\label{e4_01}
\ee
where: \ $K$ is the number of modes one plans on eliminating, $M_x,M_y$ are numbers
of grid points (see Appendix A), and, by analogy with \eqref{e2_05c}, 
\be
\vD_{n-i} = (\vec{\phi})_{n-i+1} - (\vec{\phi})_{n-i}, 
\qquad i = 1,\ldots, K.
\label{e4_02}
\ee
Similarly, define another $K \times (N\cdot M_xM_y)$ matrix:
\be
L\,\wD \equiv \left( 
\ba{c} \D \big( L^{(0)} \vec{\phi}\big)_{n-1} \\ \vdots \\
       \D \big( L^{(0)} \vec{\phi}\big)_{n-K}  \ea \right) \,\approx \,
 \left( \ba{c}  L\, \vD_{n-1} \\ \vdots \\ L\,\vD_{n-K} \ea \right),      
\label{e4_03}
\ee
where
\be
\D \big( L^{(0)} \vec{\phi}\big)_{n-i} = 
 L^{(0)} (\vec{\phi})_{n-i+1} -  L^{(0)} (\vec{\phi})_{n-i}, 
\qquad i = 1,\ldots, K,
\label{e4_04}
\ee
and the approximate equality in \eqref{e4_03} holds 
by analogy with \eqref{e2_07d}. In what follows we will treat this approximate 
equality as exact; we verified that for $(\vec{\phi})_n$ sufficiently close to
the exact solution, it holds with high accuracy.

We now assume that each of $\vD_{n-i}$ is a linear combination of the
eigenmodes $\vu_i,\,i=1,\ldots,K$ of 
the linearized iteration operator $P^{-1}L$.
Inverting, for future convenience, the relation between $\vD_{n-i}$'s 
and $\vu_i$'s, we write it as:
\be
\wU \equiv \left( \ba{c} \vu_1 \\ \vdots \\ \vu_K \ea \right) = 
\bA\, \wD 
\label{e4_05}
\ee
for some $K\times K$ matrix $\bA$.  Eigenmodes of $P^{-1}L$ are to 
satisfy the relation
\be
P^{-1}L \,\wU \equiv 
\left( \ba{c} P^{-1}L \,\vu_1 \\ \vdots \\ P^{-1}L\, \vu_K \ea \right) 
= \walpha \, \wU,
\label{e4_06}
\ee
%
where
\be
\walpha = {\rm diag} \, (\alpha_1, \ldots, \alpha_K).
\label{e4_07}
\ee
%
Entries $\alpha_i$ in \eqref{e4_07}
denote the ``average eigenvalue" corresponding to modes 
$\vu_i$. By slight abuse of notations, the subscripts of $\alpha_i$ pertain
to different modes, whereas previously subscripts of $\alpha_n$ pertained to iterations;
this is not expected to lead to a confusion. 
As at the end of Sec.~\ref{sec_3B}, we stress that 
eigenmodes $\vu_i$ are {\em not} the 
exact eigenfunctions of $L$ but are some mixes of those eigenfunctions, and thus
$\alpha_i$ are not necessarily $L$'s eigenvalues $\lambda_i$, although the former
are expected to approximate the latter. 
More will be said about this in the next subsection.
We will continue to refer to the mixes $\vu_i$ 
of the exact eigenfunctions of $P^{-1}L$ as eigenmodes (or modes), thus making
a distinction between the modes extracted from $\wD$ with our procedure described
below, on the one hand, and the exact eigenfunctions, on the other hand.

The goal of the subsequent derivation is to find matrix $\bA$; then the modes can be
recovered from \eqref{e4_05} and $\alpha_i$'s can be found from \eqref{e4_06}. 

Following \cite{AITEM,multicomp} we will work, instead of $P^{-1}L$, with a
similar operator $P^{-1/2}LP^{-1/2}$ which, unlike $P^{-1}L$, is Hermitian. 
Eigenmodes of $P^{-1/2}LP^{-1/2}$ are $P^{1/2}\vu_i$, and the latter can be chosen
to form an orthonormal set:
\be
P^{1/2}\,\wU \, \big( P^{1/2}\,\wU \big)^T =  {\bf I},
\label{e4_08}
\ee
where ${\bf I}$ is the $K\times K$ identity matrix. 
Here and below we use the fact that solutions of our HF equations are real-valued;
for the case of complex-valued solution the transposition of a matrix would need to 
be replaced by Hermitian conjugation. 
Using \eqref{e4_05} and the fact that
actions of matrix $\bA$ and operator $P$ commute, one rewrites the last equation as
\be
\bA \, P^{1/2}\,\wD \, \big( P^{1/2}\,\wD \big)^T \, \bA^T = {\bf I}.
\label{e4_09}
\ee
If one writes the Cholesky decomposition of
$P^{1/2}\,\wD \, \big( P^{1/2}\,\wD \big)^T$ as
\be
P^{1/2}\,\wD \, \big( P^{1/2}\,\wD \big)^T = \bC\,\bC^T,
\label{e4_10}
\ee
where $\bC$ is a $K\times K$ matrix, then the last two equations yield
\be
\wA \, \wA^T = {\bf I},
\qquad 
\wA \equiv 
\big(\bA \, \bC\big);
\label{e4_11}
\ee
i.e., $\wA$ must be an orthogonal matrix.

Next, one acts on both sides of \eqref{e4_06} with $P$, multiplies them by 
$\wU^T$ on the right, and uses \eqref{e4_08} and then \eqref{e4_05} to obtain:
\be
\bA\, \bB\, \bA^T = \walpha, \qquad
\bB \equiv L\,\wD \,\wD^T.
\label{e4_12}
\ee
Here we have used that actions of $\bA$ and $L$ commute and that
\be
P\,\wU \, \wU^T = 
P^{1/2}\,\wU \, \big( P^{1/2}\,\wU \big)^T,
\label{e4_13}
\ee
which follows from the facts that the matrix product in \eqref{e4_13} is the
discrete version of the inner product, as defined in \eqref{e2_02}, and that
$P$ is a Hermitian operator.
Using \eqref{e4_11}, one rewrites \eqref{e4_12} as:
\be
\bC^{-1} \,\bB\, (\bC^{-1})^T = \wA^T \,\walpha\,\wA,
\label{e4_14}
\ee
which means that $\wA$ is the orthogonal matrix that diagonalizes the Hermitian
matrix $\bC^{-1} \,\bB\, (\bC^{-1})^T$. The Hermitian character of the latter matrix
follows from that of $\bB$, which, in turn, amounts to the equality
\be
L\, \vD_i \; \vD_j^T = L\, \vD_j \; \vD_i^T.
\label{e4_15}
\ee
This last relation follows from the fact that the inner products there are the discrete
approximations of the continuous inner product defined in \eqref{e2_02} and that $L$ is
a Hermitian operator in the space of functions satisfying the linearized constraints
\eqref{e2_02} \cite{AITEM, multicomp}. Indeed, $\vD_i = \tilde{\phi}_{n-i+1}-\tilde{\phi}_{n-i}$
(see \eqref{e2_08}) satisfy those linearized constraints by virtue of individual 
$\tilde{\phi}_{n-i}$'s satisfying them (when the iterated solution is sufficiently close 
to the exact one). This issue is further commented on in Appendix D.1.

The above derivation leads to the following {\em Algorithm of mME}.
\benum
	\item 
	Compute the matrix on the l.h.s. of \eqref{e4_10} and find its Cholesky decomposition,
	given by the r.h.s. of that equation.
	\item 
	Compute matrix $\bB$ as defined in \eqref{e4_12}; then find the diagonalization
	of the matrix on the l.h.s. of \eqref{e4_14}. This gives one matrices
	$\walpha$ and  $\wA$ and then, via \eqref{e4_11}, $\bA$.
	\item 
	Modes $\vu_i$ (or $P^{1/2}\vu_i$) are then computed via \eqref{e4_05}. 
	The corresponding $\alpha_i$ have already been computed at the previous step.
	\item 
	For each mode $i$ we compute its individual $\gamma_{{\rm slow},\,i}$ 
	via \eqref{e2_07b} that uses $\alpha_i$ computed at Step 2. (Note that the
	subscript refers to the mode, not iteration, number.)
	Then compute $\Gamma_{{\rm slow},\,i}$ by \eqref{e3_09a}, where 
	$\overrightarrow{u_{\rm slow}}$ is replaced by the respective $\vu_i$. 
	\item 
	Finally, the entire $\Gamma$-term in \eqref{e2_05a} is replaced by the sum
	of all $\Gamma_{{\rm slow},\,i}\,$-terms. 
\eenum


\subsection{Performance of mME in 1D}
\label{sec_4B}

Here we will discuss how mME performs relative to the single-mode ME 
for the same two cases that were considered
in Sec.~\ref{sec_3C}. Through extensive experimentation, we have found
these results to be typical. 
At the end we will
 address some of the implementation details of both kinds of ME;
other details are found in Appendix D. 
Simulation parameters were the same as those used in Secs.~\ref{sec_3B} and
\ref{sec_3C}.

For the case of $N=23$ atoms in 30 minima of $\vext$, we did not observe any
consistent or significant 
benefit of eliminating multiple modes versus a single one, for any values
of $s$ in the range $[0.1,\, 0.8]$. In fact, in most of our attempts, an increase
of the number of eliminated modes led to an increase in the number of iterations. 
We conclude, therefore, that the large error oscillations seen in Fig.~\ref{fig_7}
must be related to factors other than the presence of multiple slow modes.

Results for the other case, with $N=24$ helium atoms, are shown in Table \ref{tab_1}.
These results show that increasing
the number $K$ of eliminated modes does, on average, tend to decrease the number
of iterations.
However, since both the computational time and computer memory
usage increase with $K$, one should focus on the {\em minimum} $K$ value that 
still provides a sizable reduction of the iteration count. Such a value appears
to be $K=3$ or $4$. This observation will guide our choice of $K$ in 2D simulations,
presented in Sec.~\ref{sec_5}, where the computational cost of experimentation with mME's
adjustable parameters in more prohibitive than that in 1D. 

\begin{table}[h!]
	\vspace*{0.2cm}
	\begin{center}
		\begin{tabular}{ |c|c|c|c|c|c|c|c|}
			\hline 
\diagbox[width=6em]{$\hspace*{1em} s$}{$K \hspace*{1em}$} 
     & \hspace*{0.4cm} 1 \hspace*{0.4cm} & \hspace*{0.4cm} 2 \hspace*{0.4cm}  
     & \hspace*{0.4cm} 3 \hspace*{0.4cm} & \hspace*{0.4cm} 4 \hspace*{0.4cm}  
     & \hspace*{0.4cm} 5 \hspace*{0.4cm} & \hspace*{0.4cm} 6 \hspace*{0.4cm} 
     & \hspace*{0.4cm} 7 \hspace*{0.4cm} \\ \hline 
0.05 & 1500 & 1200 & 1500 & 1100 & 1400 & 1400 & 1200     \\ \hline 
0.1  & 2700 & 1400 & 1700 & 1500 & 1200 & 1300 & 1000   \\ \hline 
0.2  & 4500 & 1600 & 1100 & 1000 & 1200 & 1300 & 1000   \\ \hline 
0.4  & 5600 & 1500 & 1400 & 1300 & 1300 & 1200 & 1100   \\ \hline 
		\end{tabular}                      
	\end{center}
	\caption{Number of iterations (rounded to the nearest hundred)
		required for the error to decrease to $10^{-10}$, versus
		the number of eliminated modes $K$ and parameter $s$,
		for $N=24$ helium atoms in $\vext$ \eqref{e3_04} with 30 minima.
		Other adjustable parameters of (m)ME 
		are as stated at the end of this subsection. 
	}
	\label{tab_1}          
\end{table}

For completeness and comparison with the results reported in the next
subsection, we mention that increasing $K$ past seven does not
consistently reduce the number of iterations, which stays around 1000
up to $K=15$, the largest number that we tested. Yet, increasing
$K$ {\em did} improve the convergence rate when the error got sufficiently
small. What made the {\em total} number of iterations not decrease with 
the increase of $K$ was that the duration of the transient --- i.e.,
when the error would not decay {\em overall}, as between iterations 100 and
 $\sim 300$ in Fig.~\ref{fig_7} for $N=24$, --- tended, on average, 
 to be longer for a greater $K$. In principle, this would open up the
possibility of reducing the total number of iterations by using a smaller
$K$ at the beginning --- for a shorter transient --- and a grater $K$
later --- for a faster convergence rate, --- but we did {\em not} explore
this further as our focus was not on fine-tuning the performance of the
mME.

\begin{figure}[!ht]
	\begin{minipage}{5.2cm}
		\hspace*{0cm} 
		\includegraphics[width=5.2cm,angle=0]{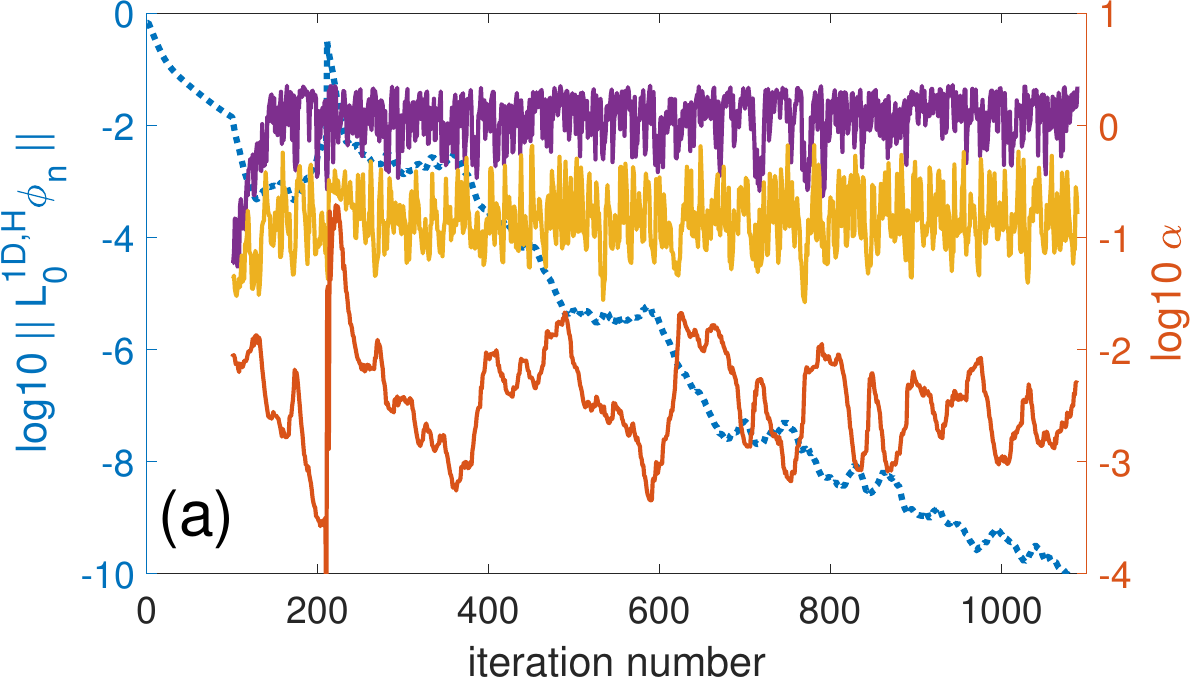}
		\vspace{0.2cm}
	\end{minipage}
	\hspace{0.1cm}
	\begin{minipage}{5.2cm}
		\includegraphics[width=5.2cm,angle=0]{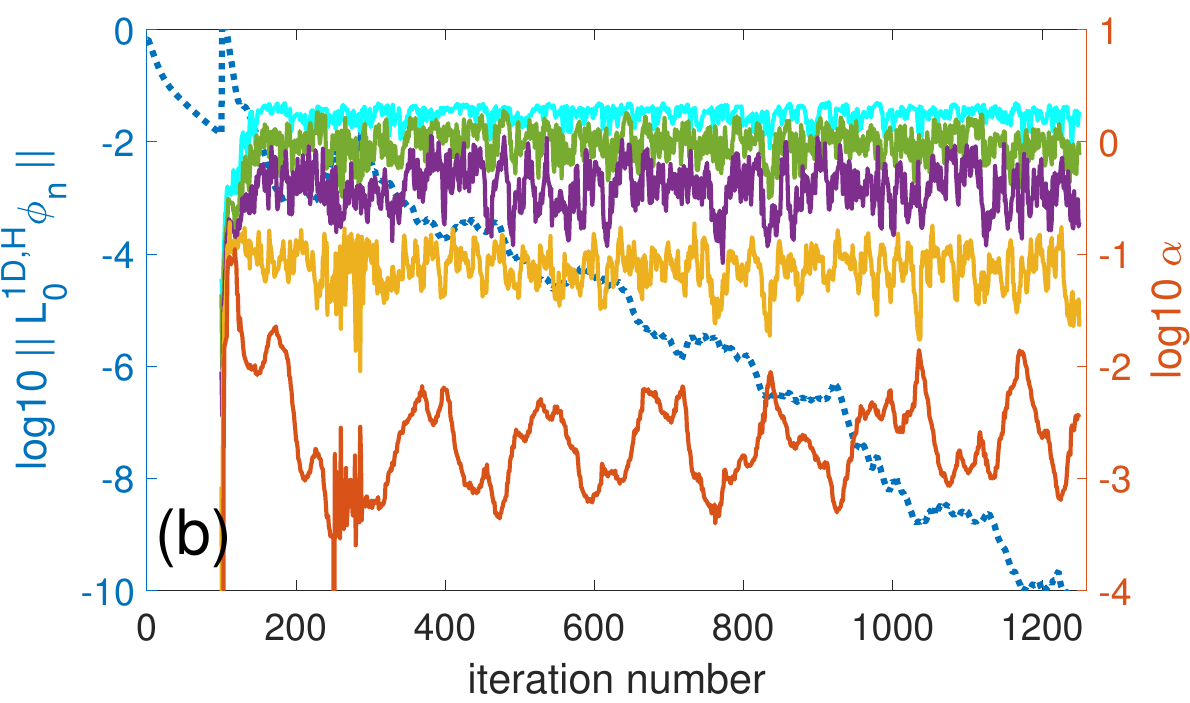}
		\vspace{0.2cm}
	\end{minipage}
    \hspace{0.1cm}
	\begin{minipage}{5.2cm}
	    \includegraphics[width=5.2cm,angle=0]{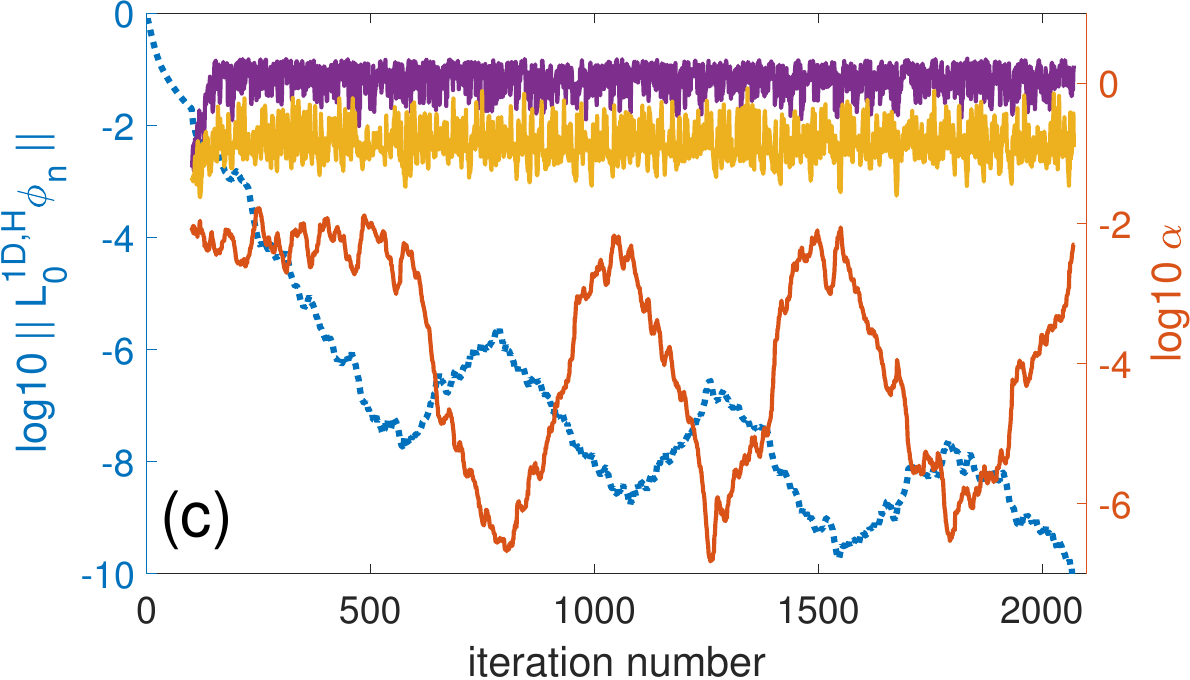}
	    \vspace{0.2cm}
    \end{minipage}
	\caption{Evolutions of the error (blue dotted line, left axis) and 
		$\alpha_i$'s, $i=1,\ldots,K$, 		 computed from
		 \eqref{e4_14} (solid lines, right axis)
		for the cases considered in Sec.~\ref{sec_3C}. Panels (a) and (b): $N=24$;
		panel (c): $N=23$. Numbers of eliminated modes are: $K=3$ (a,c) and 
		$K=5$ (b). In all panels, $s=0.2$. 
		Note that the right axis in (c) has a different scale than that in 
		(a,b). 
	}
	\label{fig_8}
\end{figure}

Now, we again emphasize that the eliminated modes $\vu_i$ are {\em not}
 the exact
eigenfunctions of the linearized iteration operator $P^{-1}L$ and, thus,
the corresponding $\alpha_i$ are {\em not} the exact eigenvalues of $P^{-1}L$.
The composition of the eliminated modes in terms of the eigenfunctions appears
to depend on the number $K$ of the modes that one sets to eliminate, and,
hence, so do the ``dynamic average eigenvalues" $\alpha_i$. This is illustrated
in Fig.~\ref{fig_8}(a,b), where one notices that $\alpha_i$ depend on the
value of $K$ chosen (and, of course, they vary from one iteration to the next). 
Incidentally, the minimum computed values of $\alpha_1$ are seen to agree, by the
order of magnitude, with the estimates of $(\lambda_{P^{-1}L})_{\min}$ presented
in Appendix C.2 for both $N=24$ and $N=23$ cases. 

We conclude this subsection with two remarks about implementation of
the ME. The first remark pertains to both single- and multiple-mode ME.
Parameter(s) $\alpha_n$ ($(\alpha_i)_n$) found by the code at some iterations
can be quite small and even negative. The former (i.e., $\alpha_n$ being
``too small") would make $\gamma_n$, computed by \eqref{e2_07b}, very large.
While the corresponding $\Gamma$-term will reduce the specific mode of the
iterated solution, as intended, it may amplify other modes (unintentionally),
thereby impeding the convergence. 
Similarly, having $\alpha_n<0$ would make $\gamma_n>0$, and this can also
impede convergence via the same mechanism. To prevent this from occurring,
we imposed the condition:
\be
-10^7 \le \gamma_n \le 0,
\label{e4_16}
\ee
where the lower bound was found to be overall optimal across many tens
of tested cases in both 1D and 2D.

The second remark is specific to the mME. As mentioned earlier, the memory
requirements increase with $K$, the number of modes one needs to store. 
Yet, Fig.~\ref{fig_8} shows that only few
of the $\alpha_i$'s can be as low as $O(10^{-2})$ or lower, which per \eqref{e3_add_01}
would make the corresponding modes to converge in as many as several
{\em thousand} iterations. Such modes indeed need to be eliminated by the ME.
However, most other $\alpha_i$'s are at least an order of magnitude 
greater and the corresponding modes would decay to the required tolerance 
of $10^{-10}$  ``on their own," i.e., without being eliminated
by the $\Gamma$-term, in only a few {\em hundred}, or fewer, iterations.
Therefore, to save computational resources, 
we first computed $\alpha_i$'s as eigenvalues
of the l.h.s. of \eqref{e4_14} but then computed and 
eliminated only the modes with
\be
\alpha_i < \alpha_{\max}.
\label{e4_17}
\ee
 For the results presented in this subsection,
we used $\alpha_{\max}=0.2$.

It is important to clarify that while only few $\alpha_i$'s (usually, no 
more than four) ``go" below $\alpha_{\max}$, increasing $K$ still may 
improve the convergence rate, as evidenced by Table \ref{tab_1}. 
This is because the allowing for more modes ``pushes" the smaller of their
eigenvalues down (compare Figs.~\ref{fig_8}(a)  and \ref{fig_8}(b)),
thereby making more of them satisfy \eqref{e4_17}.

%


\subsection{Comparison of ME with Anderson Acceleration (AA)}
\label{sec_4C}

\begin{figure}[!ht]
	\begin{minipage}{5.2cm}
		\hspace*{0cm} 
		\includegraphics[width=5.2cm,angle=0]{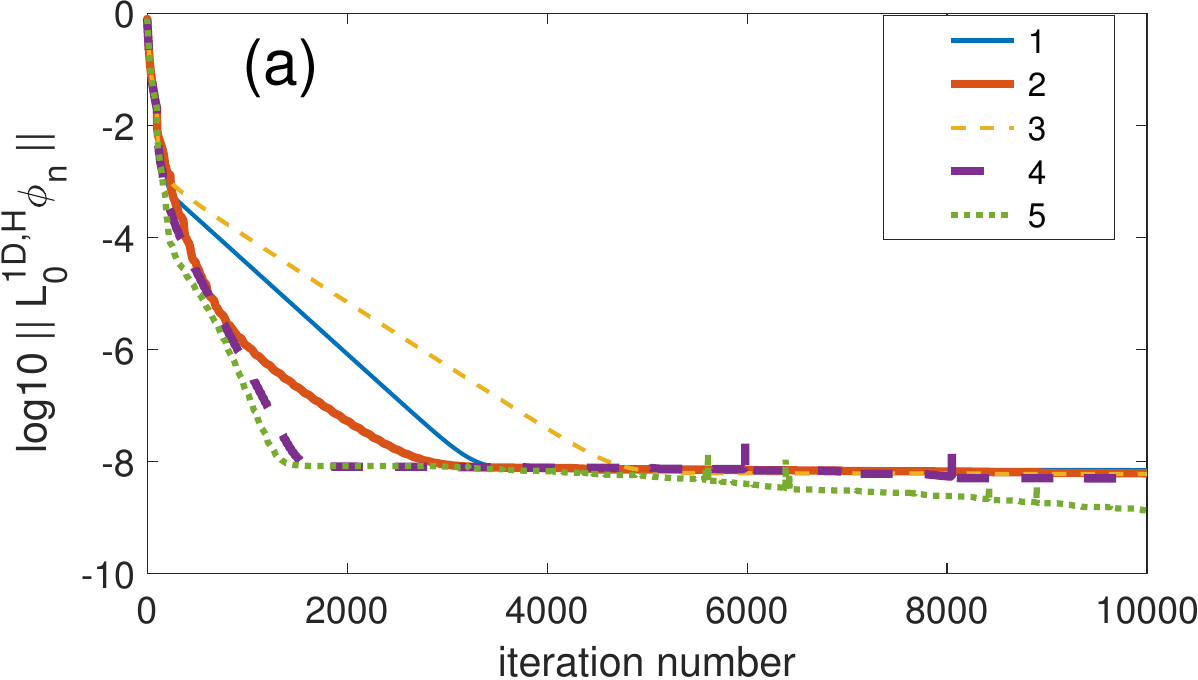}
		\vspace{0.1cm}
	\end{minipage}
	\hspace{0.2cm}
	\begin{minipage}{5.2cm}
		\includegraphics[width=5.2cm,angle=0]{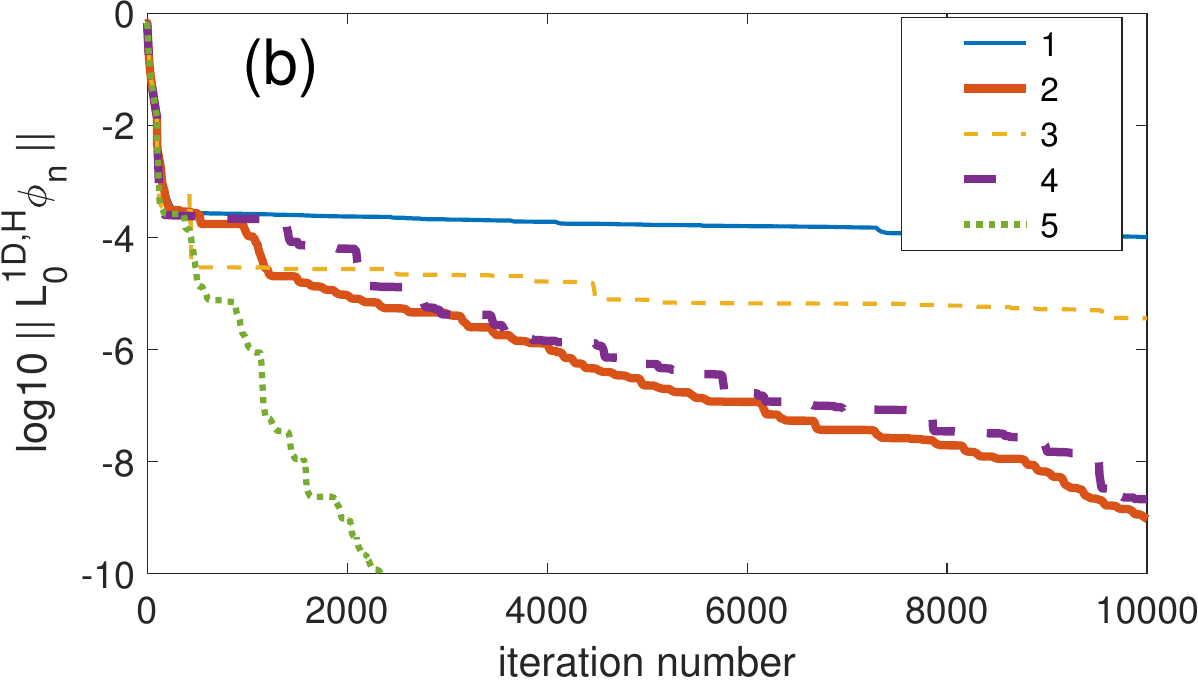}
		\vspace{0.1cm}
	\end{minipage}
    \hspace{0.2cm}
	\begin{minipage}{5.2cm}
		\includegraphics[width=5.2cm,angle=0]{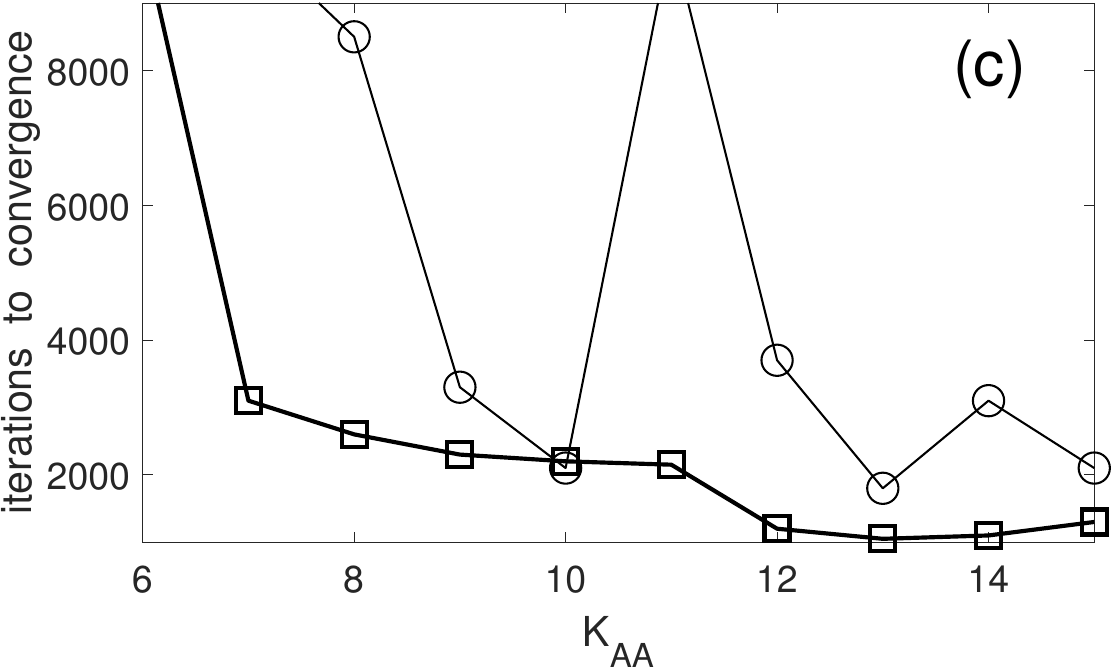}
		\vspace{0.1cm}
	\end{minipage}
	\caption{(a,b): \ Error evolution of the AITEM 
		(with $\dt=0.6$ and $c=100$ in $P$) with the Anderson Acceleration 
		for the two cases considered in Secs.~\ref{sec_3C}, \ref{sec_4C}: $N=23$ (a) 
		and $N=24$ (b). See main text for more details. \ (c): Number of iterations
		to convergence to $10^{-10}$ versus the number of previous iterations 
		(up to 15) whose
		information is used by the Anderson Acceleration: $N=23$ (circles) and 
		$N=24$ (squares); the lines are guides to the eye. 
		Numbers of iterations greater than 9000 indicate that convergence was not
		achieved in 10,000 iterations.
	}
	\label{fig_9}
\end{figure}

Figure \ref{fig_9} demonstrates 
that in those cases where the ME accelerates the AITEM so that it
converges in one to two thousand of iterations, the AITEM with 
the well-known AA \cite{1965_AA} performs considerably worse.
The AA does not eliminate modes, but instead uses the
solution at $K_{AA}$ previous iterations to subtract their certain combination
to minimize the error. The values of $K_{AA}\le 5$ are shown in the legend
of panels (a,b). 
For those lower $K_{AA}$, the best    
result achieved by the AA is about twice as slow as that achieved by the mME
(for $N=24$; refer to Sec.~\ref{sec_4B} regarding $N=23$),
and for more previous iterations/eliminated modes (namely, five) than the best results
shown in Table \ref{tab_1}. 
The benefit of the AA with $K_{AA}\ge 12$ becomes comparable to that brought about 
by the mME with $K$ as low as 3: compare Fig.~\ref{fig_9}(c) with Table \ref{tab_1}.
While the memory requirements associated with a larger number of previous iterations  
stored is not an issue in 1D, it becomes significant in 2D.  
Thus, due to this inferior overall performance by the AA relative to the mME, 
we will not use the former for the
2D simulations reported in the next section.

\smallskip

%

\section{2D patterns obtained with AITEM}
\label{sec_5}

Here we will compare the performance of mME and the original single-mode ME (sME) 
for several representative 2D periodic
patterns that helium atoms can form over graphene. The equations 
simulated are those presented in Sec.~\ref{sec_2A}. The common parameters for all
simulations are: \ $\dt = 0.4$; $\eclip=2000$K leading to $c=200$ in \eqref{e2_06};
ME starts at the 100th iteration; conditions \eqref{e4_16} and
\eqref{e4_17} were imposed, with $\alpha_{\max}=0.1$. 
The values $s$ and $K$ which we used in this section
were motivated by the results of Sec.~\ref{sec_4B}.

\begin{figure}[!ht]
	\vspace*{0.5cm}
	\begin{minipage}{5.2cm}
		\hspace*{0cm} 
		\includegraphics[width=5.2cm,angle=0]{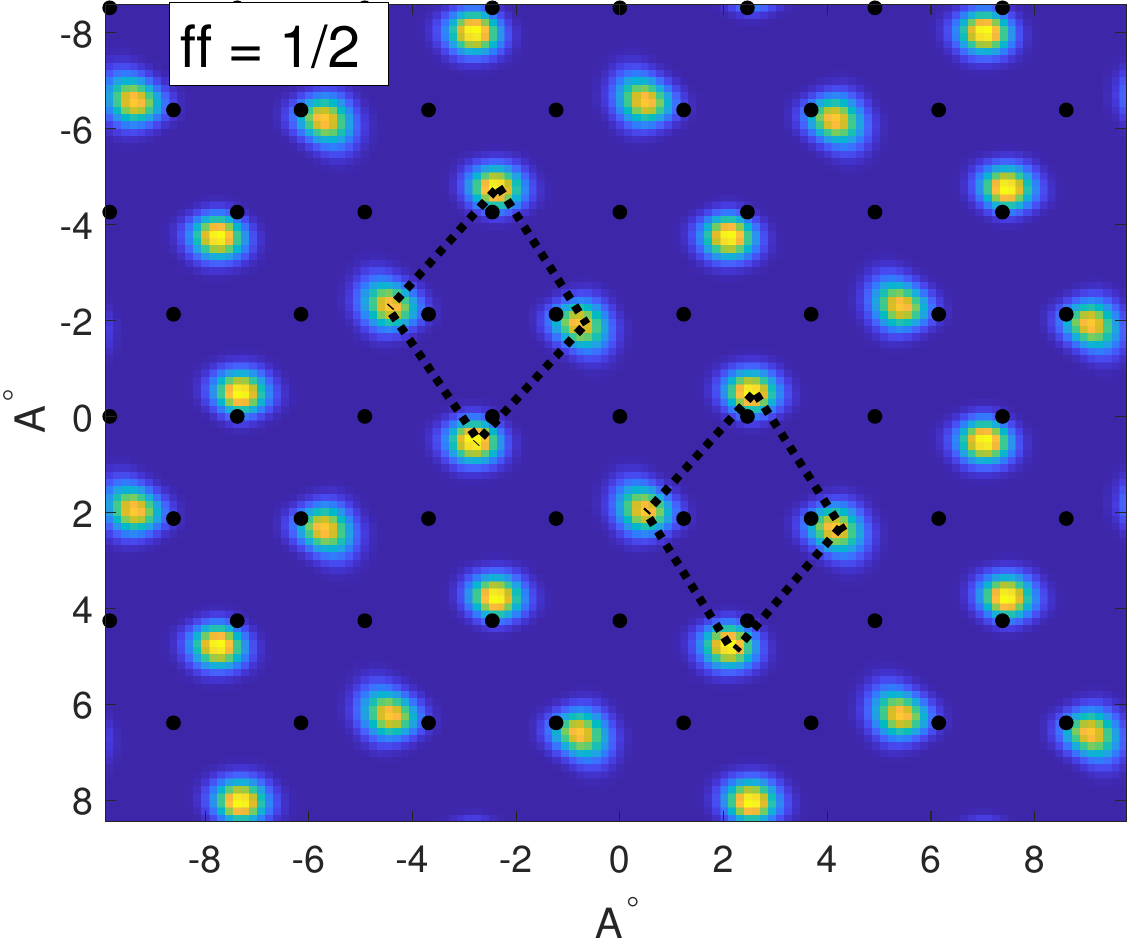}
		\vspace{0.1cm}
	\end{minipage}
	\hspace{0.2cm}
	\begin{minipage}{5.2cm}
		\includegraphics[width=5.2cm,angle=0]{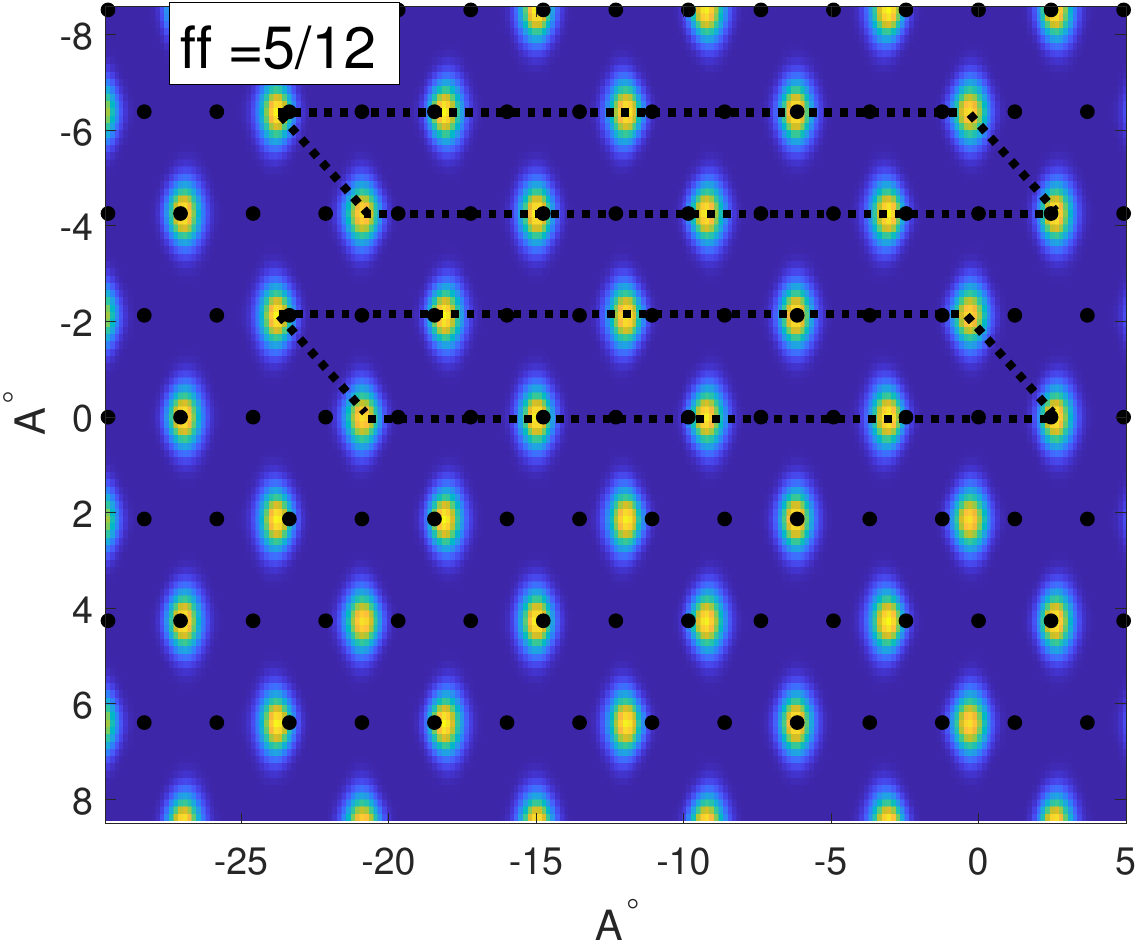}
		\vspace{0.1cm}
	\end{minipage}
	\hspace{0.2cm}
	\begin{minipage}{5.2cm}
		\includegraphics[width=5.2cm,angle=0]{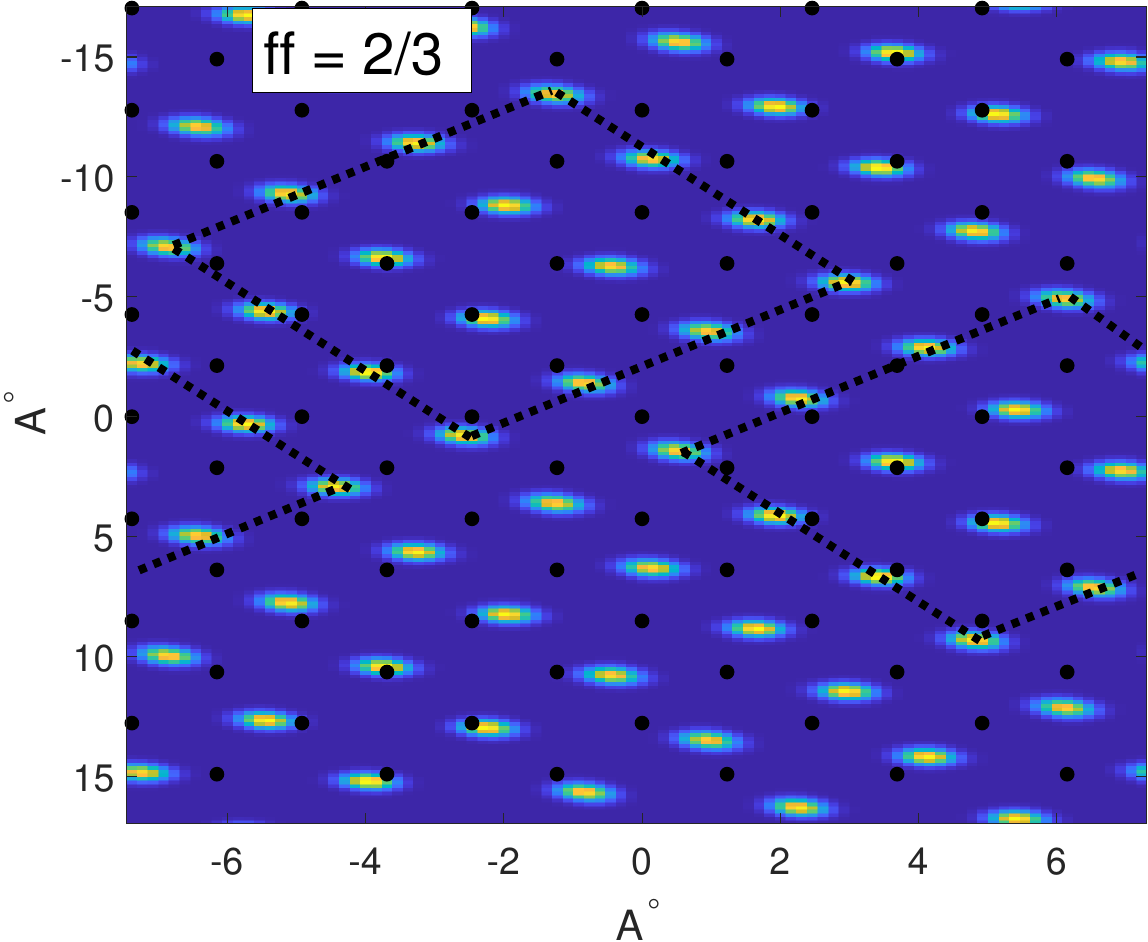}
		\vspace{0.1cm}
	\end{minipage}
\vspace*{-0.5cm}
	\caption{Equilibrium particle densities, i.e., 
		$\sum_{j=1}^N |\phi_j|^2$,  for some of the filling fractions 
		reported in this section. Darkest blue in each panel corresponds to zero,
		while brightest yellowish, to the maximum density. 
		Black dots denote the
		locations of centers of graphene hexagons (i.e., minima of $\vext$).
		Dotted lines show two periods of the pattern.
		The reverse direction/labels of the vertical axis are an artefact
		of Matlab's command {\sf imagesc}. 
	}
	\label{fig_10}
\end{figure}

As in Sec.~\ref{sec_4B}, there were patterns (i.e., values of $f\!f$) 
for which mME performed, on average, better than sME, and those for which it did not.
The results for the former group are shown in Table \ref{tab_2}.
The equilibrium pattern for $f\!f=1/2$ is depicted in Fig.~\ref{fig_10};
the patterns for $f\!f=7/16$ and $7/12$ can be found in the literature
and therefore are shown in Supplemental Material. 
Initial placement of helium atoms was done by the Second Initial placement procedure
(Appendix A.3) for all filling fractions except $f\!f = 1/2$
(and $f\!f=2/3$ mentioned below), where that procedure
led to non-periodic patterns in a large percentage of random initial placement; therefore,
for $f\!f = 1/2$ ((and for $f\!f=2/3$) we used the First procedure. 
The simulation window consisted of $8\times 8$, $8\times 4$, and $12\times 4$
rectangular periods for $f\!f=7/16,\,1/2$, and $7/12$, respectively; recall
(Sec.~\ref{sec_2C}) that one such a period contains two minima of $\vext$.
For each pair ($f\!f$, $s$), Table \ref{tab_2} shows
 the number of iterations for six different initial
conditions for the placement of helium atoms, with the same condition being used to obtain
results for the sME ($K=1$) and mME ($K=3$). While this number of initial conditions is not 
statistically significant, we have observed qualitatively similar performance of the mME
relative to the sME in many exploratory simulations not reported here.

%

\begin{table}[ht]
\begin{tabular}{|c|ccc|ccc|ccc|}
	\hline
	\multirow{2}{*}{} & 
	\multicolumn{3}{c|}{$f\!f=7/16$} & 
	\multicolumn{3}{c|}{$f\!f=1/2$} & 
	\multicolumn{3}{c|}{$f\!f=7/12$} 
	\\ 	\cline{2-10}
	 & $\quad s=0.05\quad$ & $s=0.1\quad$ & $s=0.2\quad$ 
	 & $\quad s=0.05\quad$ & $s=0.1\quad$ & $s=0.2\quad$ 
	 & $\quad s=0.05\quad$ & $s=0.1\quad$ & $s=0.2\quad$ 
	\\   \hline
	$\;K=1$ \quad & 3300 & 3800 & 4300 & 2400 & 3300 & 4400 & 4600 & 4600 & 4000 \\
	$K=3$         & 3200 & 3600 & 3700 & 1600 & 1600 & 1800 & 3300 & 3100 & 3000 \\
	ratio         & 0.97 & 0.95 & 0.86 & 0.67 & 0.48 & 0.41 & 0.72 & 0.67 & 0.67 \\
	\hline
	$K=1$         & 2200 & 3100 & 4700 & 2500 & 3400 & 4300 & 2800 & 3000 & 2700 \\
	$K=3$         & 2100 & 1900 & 4200 & 1600 & 1600 & 1500 & 2100 & 3600 & 2900 \\
	ratio         & 0.95 & 0.61 & 0.89 & 0.64 & 0.47 & 0.35 & 0.75 & 1.2 & 1.1 \\
	\hline
	$K=1$         & 3600 & 4200 & 4800 & 2000 & 2000 & 2300 & 2200 & 12800 & 4100 \\
	$K=3$         & 2500 & 3100 & 3000 & 1500 & 1600 & 1600 & 4600 & 2700 & 2400 \\
	ratio         & 0.69 & 0.74 & 0.63 & 0.75 & 0.80 & 0.70 & 2.1 & 0.21 & 0.68 \\
	\hline
	$K=1$         & 3500 & 4000 & 4000 & 2100 & 2700 & 3100 & 3400 & 9200 & 7100 \\
	$K=3$         & 2600 & 2800 & 2400 & 1500 & 1600 & 1800 & 2600 & 2300 & 2700 \\
	ratio         & 0.74 & 0.70 & 0.60 & 0.71 & 0.59 & 0.58 & 0.76 & 0.25 & 0.38 \\
	\hline
	$K=1$         & 2800 & 3700 & 3900 & 2200 & 2300 & 2300 & 2300 & 2200 & 2600 \\
	$K=3$         & 2100 & 2200 & 2900 & 1700 & 1700 & 1900 & 2500 & 2400 & 2600 \\
	ratio         & 0.75 & 0.59 & 0.74 & 0.77 & 0.74 & 0.83 & 1.1 & 1.1 & 1.0 \\
	\hline
	$K=1$         & 3100 & 3800 & 4400 & 1700 & 1800 & 1800 & 2600 & 3100 & 3900 \\
	$K=3$         & 3000 & 2300 & 2300 & 1300 & 1500 & 1400 & 2500 & 2500 & 2600 \\
	ratio         & 0.97 & 0.61 & 0.52 & 0.76 & 0.83 & 0.78 & 0.96 & 0.81 & 0.67 \\
	\hline
\end{tabular}
\caption{Number of iterations (rounded to the nearest hundred)
	required for the error to decrease to $10^{-10}$ for three different periodic
	patterns, as obtained with sME ($K=1$) and mME ($K=3$). A triplet of vertically 
	aligned numbers enclosed between two horizontal lines corresponds to one 
	random realization of the initial condition (see Appendix A.3); thus,
	for each $f\!f$, the Table  presents results for six random initial conditions 
	for each $s$ value. Within the same set of horizontal lines, the initial conditions 
	are different for different $f\!f$ 	(naturally) and may be different for the same
	 $f\!f$ and different $s$. The latter is because some initial conditions could 
	 converge to non-periodic patterns for some of the values of $s$. 
	The ratios in the third line of each row are those of the two numbers directly 
	above, shown for ease of comparison of performances of sME and mME. 
}
\label{tab_2}
\end{table}

This performance can be summarized as follows: While the mME is not guaranteed
to {\em always} reduce the number of iterations, it does so
{\em on average} by about 30\%. Perhaps more importantly, the use of the mME leads to a
more narrow {\em range} for the iteration numbers across different initial conditions than
the sME. Also, for any $f\!f$, using $s=0.05$ leads, again {\em on average}, to fewer
iterations than the larger values of $s$.

On the other hand, for $f\!f=5/12$ and $2/3$, whose equilibrium patterns are shown
 in Fig.~\ref{fig_10}, we found that mME does not reduce the number
of iterations and may even, on average, increase it compared to the sME. This is analogous
to the case $N=23$ considered in Sec.~\ref{sec_4B}. Just as there, we found that a hallmark
of this case is the high oscillations of the error, as in Fig.~\ref{fig_8}(c). In contrast,
for the cases where the mME helped to reduce the number of iterations, the error oscillations
were much smaller, similarly to those in Figs.~\ref{fig_8}(a,b). The evolution of $\alpha_i$
in 2D were found to also be similar to the respective cases shown in Fig.~\ref{fig_8}.
For completeness, we note that the simulation window consisted of $24\times 4$
and $6\times 8$ rectangular periods for $f\!f=5/12$ and $2/3$, respectively.

\section{Conclusions and Discussion}
\label{sec_6}

In this paper we have presented the details of an Accelerated Imaginary Time Evolution Method
(AITEM) for Hartree--Fock (HF) equations.
Its essential steps are given in 
\eqref{e2_05} with the $\Gamma$-term (for a single eliminated
mode) given by \eqref{e3_09}. The optimal value of parameter $c$ in the
preconditioner \eqref{e2_06} is given by relation \eqref{e3_06}, with $E_{\rm clip}$ 
defined in Appendix A.1. We stress that this relation is expected to hold for
{\em any} hard-core-type (i.e., strongly repulsive at short distances) potentials,
which will need to be clipped when implemented in an ITEM or another 
relaxation-type method. Other implementation details of the AITEM \eqref{e2_05}
are found in Appendices A.1 and A.2, and initial placement of atoms being simulated
is discussed in Appendix A.3. The mode-elimination (ME) step of the AITEM, discussed in 
Secs.~\ref{sec_3} and \ref{sec_4}, is shown to consistently 
improve convergence of the 	optimally preconditioned ITEM
by at least an order of magnitude, and often by much more. 
Values of the ME parameter $s$, which controls the amount of modes that are being
eliminated, are discussed in Secs.~\ref{sec_3B} and \ref{sec_4B}. For HF
equations, we found optimal $s$ values to be significantly lower than those
found in earlier studies for just one or two equations (which also did not have
hard-core-type potentials).

The developments listed in the previous paragraph pertain to {\em optimizing}
the form of the AITEM, which was proposed in earlier studies. This work also
presents a {\em novel}
 method to accelerate the AITEM: the multiple-mode elimination (mME).
Its derivation and Algorithm are found in Sec.~\ref{sec_4A}, and implementation
details are discussed at the end of Sec.~\ref{sec_4B} and in Appendix D. 
As we demonstrate in Secs.~\ref{sec_4B} and \ref{sec_5}, the mME
even with a small number $K$ of eliminated modes provides an improvement 
of around 30\% (on average) over the single-mode elimination (sME)
in those cases where the iteration error obtained by the sME
fluctuates moderately (see the thick line in Fig.~\ref{fig_7}). 
There, the mME also makes the number of iterations required for convergence 
slightly less dependent on the initial conditions and on values of parameter $s$.
We also found that when, on the contrary, error oscillations in the AITEM with sME
are large, as illustrated by the thin line in Fig.~\ref{fig_7} 
(see also the discussion at the end of Sec.~\ref{sec_3C}),
the mME does not further improve convergence.

Understanding the mechanism behind error oscillations remains an open problem. 
In the Supplemental Material we listed several approaches that we tried to
reduce or remove them. 
Here we present an argument that even if a ``cure" for those oscillations is
not found in the future, the problem of waiting for the iterations to converge
to a prescribed low tolerance can be circumvented. This argument is based on
two facts:\vspace{-0.2cm}
\begin{itemize}
\item
	The reason for setting a low tolerance for convergence is not that such
	accuracy of solutions is needed in practice. 
	Rather, it is merely to be assured that	the iterations would {\em eventually}
	reach a (local) minimum of the energy landscape rather than 
	a saddle point. (In the latter case, the iterations would eventually diverge
	after converging initially; this is the type of the behavior
	seen at points 1 and 3 in Fig.~\ref{fig_7}.)\vspace{-0.2cm}
\item
	By comparing solutions obtained at points 1--3 in Fig.~\ref{fig_7} with
	the final solution, we observed that 
	$\left\| \vec{\phi}_{\rm pt} - \vec{\phi}_{\rm final}\right\|$ 
	decreases {\em monotonically} when `pt' increases from 1 to 2 to 3. 
	That is, even though the error of the {\em equations} does not decrease
	monotonically, 	the error of the {\em solution} does. (This is similar to
	the discussion about Fig.~\ref{fig_6} found in Sec.~\ref{sec_3C} about
	the AITEM without ME.)\vspace{-0.2cm}
\end{itemize}
Therefore, a meaningful way to stop iteration in a case where the equation
error has large oscillations could be when one observes that the error
begins to grow the {\em second} time, as at point 3 in Fig.~\ref{fig_7}.
Indeed, by waiting past that point one would obtain a slightly
more accurate solution, but this will be achieved at the expense of several
hundred more iterations.

Both single- and multiple-mode versions of ME were shown, in Sec.~\ref{sec_4C},
to perform significantly better than the well-known Anderson Acceleration.
From a philosophical perspective, this result may appear not surprising given
that the ME uses (and requires) more information of the iteration operator ---
namely, its Hermitianness, --- than the AA, which would work for any iteration
operator. 

We believe that the acceleration techniques proposed in this work for the ITEM
will also be useful for other iteration methods.

As a byproduct, our 2D simulations revealed three interesting
periodic patterns that
helium atoms can form over graphene (or graphite): $f\!f=5/12$, $1/2$,
and $2/3$ 
(surface coverages of 0.0796, 0.0955, and 0.127 \AA$^{-2}$, respectively);
 see Sec.~\ref{sec_5}. 
As explained in Sec.~\ref{sec_2C}, 
the latter pattern can form only in a 2D system, or in 3D when 
the motion of helium atoms in the direction perpendicular to graphene's surface 
is prohibited/restricted.
The periodicity of the patterns was verified by taking their 2D Fourier transform.
We acknowledge that a periodic pattern for $f\!f=31/75\approx 0.4133$, which 
is very close to $f\!f=5/12\approx 0.4167$, was mentioned in 
Refs.~\cite{2009_GordilloBoronat,2014_Gordillo}, but no picture of it was provided.
Also, a phase with $f\!f=13/24\approx 1/2$ was shown in Fig.~5(c) of 
Ref.~\cite{2012_Kwon}; it appears similar to the $f\!f= 1/2$ pattern
shown in our Fig.~\ref{fig_10}, but was termed `incommensurate' in \cite{2012_Kwon}.

Finally, we comment on some extensions of our method and future research.
In general, the AITEM  developed in this work can be used to simulate hard-core bosons and
fermions in various settings.
For example, it can be used to compute the nearest- and 
next-nearest-neighbor
interaction strength in a variety of lattice Bose--Hubbard models (see, e.g., 
\cite{WesselTroyer, 2012_PRB_longrange, 2013_PRB_anisotropic, 2015_PRB_spinhalf}),
as two of the present authors did in Ref.~\cite{PRB}, which motivated this research.
The same method can also be used to study distribution of atoms (helium and others),
including their forming various periodic patterns 
over surfaces periodic at the atomic level 
\cite{2012_GHandGF, 2019_6612graphene, 2016_JLTP_Rus, 2019_O2onG}.

When spatial orientation of the adsorbed atom or molecule is essential
(as, e.g., in \cite{2019_O2onG}), the method would need to be extended to include
the rotational degrees of freedom in the Hamiltonian and the third spatial 
dimension, while using the same underlying concepts. A 3D extension of the method
can also  be used to study formation of a small number of layers of adsorbed atoms
over surfaces \cite{1998_Whitlock}. 


Simulations of the full $N$-body wavefunction by
ab initio methods such as path integral quantum Monte Carlo produce more 
physically accurate results for adsorbed phases of atoms than simulations of the 
HF equations. However, the latter simulations by the numerical method
 presented in this work are substantially less  computationally costly.  
Therefore, the AITEM could be a useful tool in high-throughput schemes \cite{2023_Barik} 
that would employ HF equations to identify promising scenarios for realizing exotic 
low-dimensional 
adsorbed superfluid phases. This may include searching for optimal materials as
substrates or for optimal tunable parameters of known substrates, as, e.g., in
\cite{2024_Kim}.

\acknowledgments

This work was supported by NASA grant number 80NSSC19M0143.  
A.D. acknowledges partial support from the National Science Foundation 
Materials Research Science and Engineering Center program through the 
UT Knoxville Center for Advanced Materials and Manufacturing (DMR-2309083).

\appendix


\section*{Appendix A: \ DFT for evaluation of terms in (\ref{e2_04}) and for  
	initial placement of helium atoms}
\label{app_A}

We use Discrete Fourier transform (DFT) and its inverse defined as:
\be
\sum_{\bf{n} =
	\bf{0}}^{\bf{\max}} 
e^{-i\,{\bfr}_{\bf{n}} \cdot \bf{k}_{\bf{m}}}\, \phi_{\bf{n}} \equiv
	 F[\phi]_{\bf{m}}, 
\qquad
\phi_{\bf{n}} = \frac1{M_x M_y} \sum_{\bf{m}=\bf{0}}^{\bf{\max}} 
e^{i\,{\bfr}_{\bf{n}} \cdot  \bf{k}_{\bf{m}} }\,
 F[\phi]_{\bf{m}}  \equiv
  F^{-1}\left[ \,F[\phi]\, \right]_{\bf{n}}  ,
\label{A_01}
\ee
where: $M_x,M_y$ are the numbers of grid points along $x$
and $y$; all vectors have two components pertaining to $x$ and $y$,
e.g.,: ${\bf n}=(n_x,n_y)$, ${\bfr}_{\bf{n}} = (x_{n_x}, y_{n_y})$,
${\bf\max}=(M_x-1,M_y-1)$, 
${\bf k}_{\bf{m}} = \big( (k_x)_{m_x}, (k_y)_{m_y}\big)$, etc. 
Here $x_{n_x}=(n_x-1)\D x$, $(k_x)_{m_x}=(m_x-1)\D k_x$, etc., where
$\D x$ and $\D k_x$ are the respective mesh sizes. 
Note that in this Appendix, $\bf{n}$ refers to the index of grid points in the
physical space, while in the main text $n$ referred to the iteration number.

The Laplacian in \eqref{e2_04} is computed in the standard way:
\be
-\nabla^2 \phi = F^{-1} \big[ \, \| ({\bf k})_{\rm shifted} \|^2
\, F[\phi] \, \big], 
\label{A_02}
\ee
where $\|\ldots\|$ stands for the Euclidean norm, 
\be
\big((k_x)_{\rm shifted}\big)_{m_x} =  \left\{
\ba{ll}
(k_x)_{m_x}, & m_x \in [0,\, M_x/2-1]; \\
- (k_x)_{M_x-m_x}, & m_x \in [M_x/2, \, M_x-1],
\ea 
\right. 
\label{A_03}
\ee
and similarly for $(k_y)_{\rm shifted}$. 

\subsection*{A.1: Computing the nonlocal term in \eqref{e2_04}}
\label{app_A1}

\begin{figure}[!ht]
	\centering
	\includegraphics[width=8cm,angle=0]{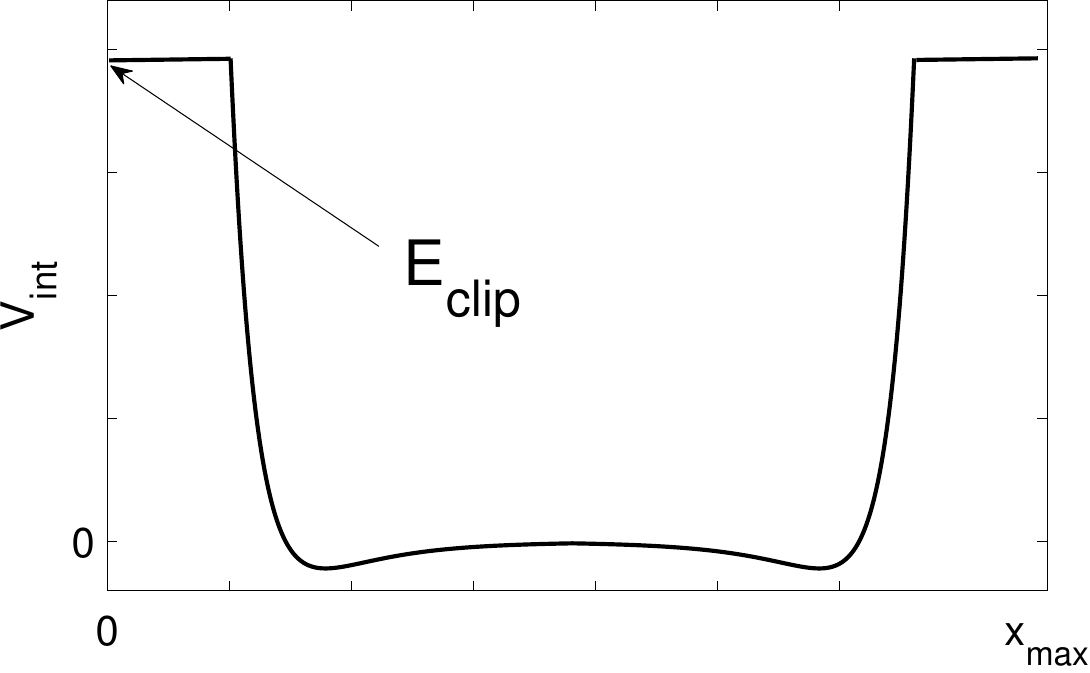}
	\caption{Schematics of the interaction potential between helium atoms, set
		 up as explained before Eq.~\eqref{A_05}. Details along the $x$-axis are not to scale: 
		 the attracting ``tail" past the minimum of $\vint$ at $x\approx 3$\AA\ actually extends
		 much longer towards $x_{\max}/2$. 
	}
	\label{fig_A1}
\end{figure}

In this paragraph, we treat the case of one spatial dimension for the sake of clarity;
its 2D generalization will be given later. 
We begin by setting up 
$\vint$ to be periodic on the computational grid and also clip it removing
the ``singularity'' at ${\bfr}=\bf{0}$ at some empirically chosen value
\be
\eclip \gg 1,
\label{A_04}
\ee
as illustrated in Fig.~\ref{fig_A1}. 
Note that, given the definition of $x_{n_x}$ above, 
array $\vint(x_{n_x})$ is defined so as to reflect the fact that the
interaction is the strongest at zero separation between helium atoms.
With this setup, for any function $\psi({\bfr}')$, one has:
\be
\int \vint({\bfr} -{\bfr}') \, \psi({\bfr}')\,d{\bfr}' 
 = F^{-1} \left[ \, 
 F[\vint] \, F[\psi] \, \right]. 
\label{A_05}
\ee

We verified that a derivative discontinuity introduced by the clipping of $\vint$ does not 
appear to introduce numerical artifacts in $F[\vint]$. This is because the very high gradient
of this potential at sufficiently small distances already produces a long ``tail" in its
Fourier spectrum.

The code snippet below shows how $\vint$, originally defined in \cite{1995_Aziz}
as a 1D array over a 1D vector \verb+rint+, is interpolated on a 2D grid in Matlab:
\begin{verbatim}
[X,Y] = meshgrid(x,y); % create 2D grid along x and y
rint_arr = sqrt(X.^2 + Y.^2); % distance from center of grid
rint_vec = reshape(rint_arr,[1,M_x*M_y]); % make it 1 x (Mx*My) vector
Vint_vec = interp1(rint,Vint,rint_vec); % interpolate Vint from rint to rint_vec
Vint_arr = reshape(Vint_vec,[Mx,My]); % put it over Mx x My grid
\end{verbatim}

The resulting 2D array \verb+Vint_arr+ peaks at the center of the numerical grid and decays 
towards its boundaries, which is {\em not} how $\vint$ is set up in
 Fig.~\ref{fig_A1} and 
upon which assumption Eq.~\eqref{A_05} is written. 
To adjust for this difference, an 
\verb+fftshift+ command must be placed either around the 
entire r.h.s. of that equation or around the $F[\vint]$ term.

\smallskip

We now show how all the $N(N+1)/2$ Hartree terms in Eqs.~\eqref{e2_04} can
be computed in $O(N)$ steps: \vspace{-0.2cm}
\begin{itemize}
	\item 
	Compute $F[\phi_i^2]$ for all $i$ and then define $F[S]\equiv \sum_{i=1}^N F[\phi_i^2]$. 
	\item 
	Use \eqref{A_05} with $\psi = S - \phi_j^2$ 
	to compute the Hartree sum in \eqref{e2_04} for each $j$. 
\end{itemize}

\subsection*{A.2: Verifying validity of clipping $\vint$}
\label{app_A2}

The clipping here refers to the procedure mentioned before Eq.~\eqref{A_04}. 
Specifically, we need to verify that a somewhat arbitrarily chosen value  $\eclip$, which we 
took to be  $\eclip=1000$ K and $2000$ K for 1D and 2D simulations,  
respectively, does not significantly alter the simulation results. 
(For the original $\vint$ from \cite{1995_Aziz}, 
$ (\vint)_{\rm not\; clipped}(0) \approx 2\cdot 10^6$ K.)
To that end, one can monitor the difference between quantities
\be
(E_{{\rm int},\,j})_{\rm clipped} = 
\langle \, \sum_{i=1,\, i\neq j}^N \phi^2_i | (\vint)_{\rm clipped} | \phi^2_j \, \rangle 
\quad {\rm and} \quad
(E_{{\rm int},\,j})_{\rm not\;clipped} =
\langle \, \sum_{i=1,\, i\neq j}^N \phi^2_i | (\vint)_{\rm not\; clipped} | \phi^2_j \, \rangle\,,
\label{A_06}
\ee
and check whether it exceeds a specified amount (see below), for all atoms $j$.
Here all the solutions $\phi_i$ in \eqref{A_06} are computed with 
$(\vint)_{\rm clipped}$. 
Choosing $\eclip$ too low, obviously, increases the difference 
in \eqref{A_06}. On the other hand,
choosing $\eclip$ too high leads to the need to decrease $\dt$ so as to avoid
extreme sensitivity of the solution, whereby tiny changes of $\phi_i$ or $\phi_j$ at the
locations of its neighbors would lead to very large and unphysical changes in
the entire system (examine the first term in $[\cdots]$ in \eqref{e2_05a}).
In turn, too small a $\dt$ slows down convergence of the iterations.

Now, since the simulated equations \eqref{e2_05} ``do not know" that the true 
helium--helium interaction potential is $ (\vint)_{\rm not\; clipped}$ rather than
$ (\vint)_{\rm clipped}$, they have no way of correcting a discrepancy between the
two quantities in \eqref{A_06} if such arises. 
Such a discrepancy did occur in many of our simulations and was due, usually, to
$\phi_j$ ``developing too big of a piece" (usually, of a size less than 1\% of its peak value)
at the location of one of its neighbors. 
An external correcting procedure was needed when that occurred. 
Through extensive testing, we found the following one to work.

{\em Procedure for bounding \ 
	$\delta E_j=|(E_{{\rm int},\,j})_{\rm clipped} -  (E_{{\rm int},\,j})_{\rm not\;clipped}|$}
\vspace{-0.2cm}
\begin{itemize}
	\item
	Compute $\delta E_j$ every $n_{\rm monitor}$ iterations.
	We used $n_{\rm monitor}=200$. Doing so too often will slow down the code,
	while doing so too infrequently may extend a transient period which is required 
	for helium atoms move sufficiently close to their equilibrium locations.\\
	Find those $j$ for which $\delta E_j$ exceeds some threshold. We used the
	criterion \ $\delta E_j > 5\cdot \min_j \delta E_j$, which amounted to 
	roughly 1K. 
	\item
	Estimate the width of any one helium atom whose $\delta E_j$ is below the threshold.
	Replace (see \eqref{A_07} below) those $\phi_j$ whose $\delta E_j$ exceeds the threshold
	with a Gaussian of the above width.
	(As a simpler alternative, which may slightly increase the length of the 
	calculations, use the initial Gaussian guess.)
	\item
	Apply the Gram--Schmidt orthogonalization to enforce conditions \eqref{e2_02}.
\end{itemize}

\subsection*{A.3: Initial placement of helium atoms}
\label{app_A3}

We first outline the two procedures that can be used to create an initial filling fraction
$p/q$ of helium atoms and then how they were implemented using DFT.

{\em First initial placement procedure}: \  
We first distributed $2(p/q)\,N_{{\rm cells},\;x}$ atoms over two consecutive rows of cells, 
placing them equidistantly along each row and horizontally shifting atoms in the second
row relative to
those in the previous row so as to maximize the distance between any two atoms in the two
rows. We then added to coordinates
of each atom a random perturbation of size $\sim 20\%$ of the cell's side (i.e., of $d_0/\sqrt{3}$). 

{\em Second initial placement procedure}: \  We first found integer factors $N_{{\rm He},\;x}$ 
and $N_{{\rm He},\;y}$
of the total number of helium atoms $N$; the code does so by going through various combinations
of these factors starting with an initial guess $N_{{\rm He},\;x} = \sqrt{f\!f}\,N_{{\rm cells},\;x}$
rounded to the nearest integer. Since, as mentioned in Sec.~\ref{sec_2C}, 
$N_{{\rm cells},\;y}$ must be even, so
must be the selected factor $N_{{\rm He},\;y}$. Second, we placed the atoms so that any
two consecutive atoms within one row (i.e., along $x$) are  $\delta_{{\rm He},\;x} = (M_x/N_{{\rm He},\;x})\dx$ apart, with the
rows being spaced by $\delta_{{\rm He},\;y} = (M_y/N_{{\rm He},\;y})\D y$ along $y$. 
Any two consecutive rows of
atoms are shifted horizontally by $\delta_{{\rm He},\;x}/2$ so as to maximize 
the distance between any two atoms, as in the first placement procedure. 
Finally, to those locations we add random perturbations of the size of $\sim 20\%$ of
$(\delta_{{\rm He},\;x} - d_0)/2$ and $(\delta_{{\rm He},\;y} - d_0)/2$ along the $x$- and 
$y$-directions, respectively. 

Of these two procedures, the second one usually yields a more uniform initial placement
for smaller $ff$. 

Finally, the implementation of the initial placement of helium atoms at specified locations 
$\vec{\bfr}_c\equiv (x_c,y_c)$ of the
lattice is straightforward with \eqref{A_01}. Namely, one first creates a
localized (typically, Gaussian) function $\phi_0(\bfr)$ centered at $\bfr=0$
and then shifts it to the desired location via:
\be
\phi(\bfr-\bfr_c) = F^{-1}\left[ F[\phi_0] \,e^{-i\,\bfr_c\cdot \bf{k}} \,\right].
\label{A_07}
\ee
%


\section*{Appendix B: \ Discussion of the $\vint$-term in Sec.~\ref{sec_3A}}
\label{app_B}

For the purpose of the estimate in \eqref{e3_05} and the text below it,
\be
\mbox{`$\vint$-term'} \equiv 
\left\|
\sum_{i\neq j} \langle\phi_i|\vint|\phi_i\rangle \, f_j + 
2 \sum_{i\neq j} \langle\phi_i|\vint|f_i\rangle \, \phi_j \, \right\|
\,/\, \| \vec{f}\|,
\label{B_01}
\ee
where $f_j$ is the component of eigenfunction $\vec{f}$ of $P^{-1}L$
corresponding to helium atom $j$. 
Our goal here is to show that it is feasible that this term can be 
$O(E_{\rm clip})$ for some of the eigenfunctions, even though it may not
be obvious at first sight.

Both terms in the numerator above depend on the overlap 
between $\phi_i$ and $f_j$ for $i\neq j$. 
Note that the magnitude of neither term can be inferred from the
$\vint$-term in the original, non-linearized equation, \eqref{e2_03} or
\eqref{e3_03}, i.e., from $\left\| \sum_{i\neq j} \langle\phi_i|\vint|\phi_i\rangle \, \phi_j\,\right\|$.
The latter term is on the 
order of $\vint(d)$, where $d$ is the distance between neighboring helium atoms. When
$d=d_0$, the distance between centers of adjacent graphene cells, $O(\vint(d_0))=10\ldots100$ K $\ll E_{\rm clip}$ \cite{PRB}. 
(For $d>d_0$, $\vint(d)$ is even smaller.)
This occurs because the overlap between (the wavefunctions of) 
helium atoms even in adjacent wells of $\vext$, i.e., 
$\langle \phi_i | \phi_j \rangle$, $i\neq j$, is quite small.

\begin{figure}[!ht]
	\vspace*{0.5cm}
	\centering
	\includegraphics[width=8cm,angle=0]{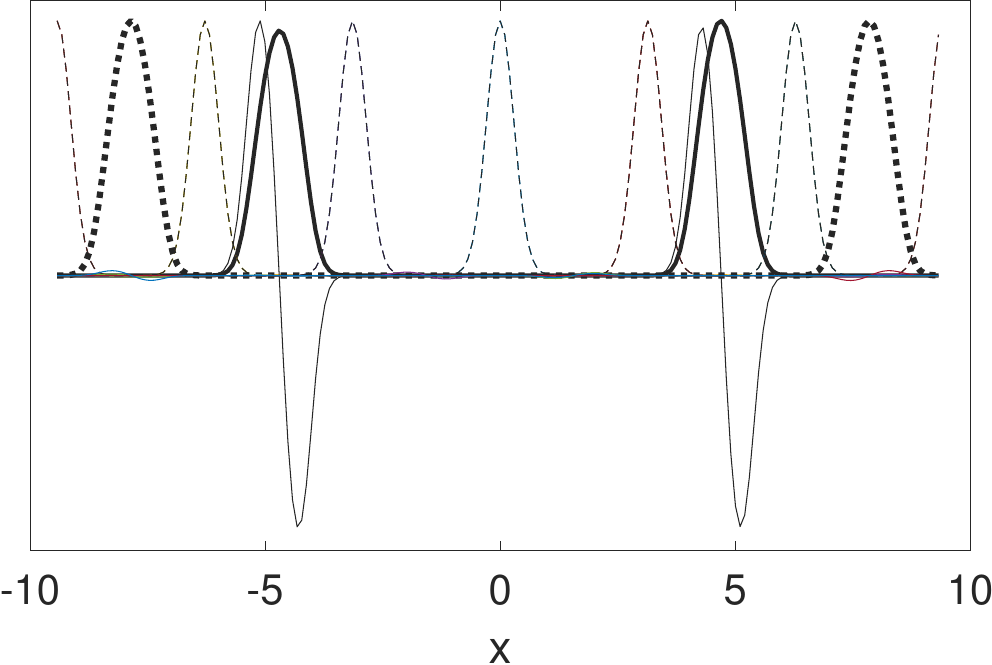}
	\caption{Eigenfunctions of $P^{-1}L$ with $c=\eclip=100$ for 
		$N=6$ helium atoms placed in $N$ adjacent wells of $\vext$,
		corresponding to
		some of its largest eigenvalues: $\lambda=1.777=\lambda_{\max}$ (thick solid),
		$\lambda=1.772$ (thick dotted), $\lambda=1.702$ (thin solid).
		Thin dashed lines show the locations of the helium atoms.
		Vertical scale is in arbitrary units.
	}
	\label{fig_B1}
\end{figure}

However, for certain eigenfunctions of $P^{-1}L$, 
the overlap $\phi_i$ and $f_j$, $i\neq j$, can be much greater 
than that between helium atoms in the adjacent wells of $\vext$.  
This is illustrated in Fig.~\ref{fig_B1}. Due to this greater
overlap, the size of the $\vint$-term as defined in \eqref{B_01}
is much greater than $O(\vint(d_0))$. In fact, from the estimate
\eqref{e3_06} extrapolated to $k=O(1)$ and the fact that using
$c=E_{\rm clip}$ yields $max(\lambda)=O(1)$ in this case, one
can infer that  the $\vint$-term for these eigenfunctions is
$O(E_{\rm clip})$.

%


\section*{Appendix C: \ On the lowest eigenvalue of of $P^{-1}L$ and its eigenfunction}

\subsection*{C.1: \ The lowest-$\lambda$ eigenfunction corresponds to
	a common shift of all atoms}

In this subsection, we will justify its title.
Suppose that $\vext\equiv 0$. 
Then any periodic array of atoms can be shifted as a whole without changing its
energy. In this case, the common shift mode has $\lambda=0$. Other modes that 
involve unequal shifts or breathing of atoms would increase the total energy via the term
$\langle \phi_i\,|\vint | \,\phi_j \rangle$, due to the rapidly-increasing repulsion
between the atoms with their decreasing separation. Therefore, such modes
will have $\lambda>0$. 
Now, when $\vext \,\cancel{\equiv}\, 0$, a common shift of all atoms would raise
the array's energy due to the atoms shifting away from $\vext$'s minima; hence
the eigenvalue of that mode becomes positive. The reason that it is small is that 
the change in the external potential energy due to such a shift is still much smaller
than a change that would occur had the distance between adjacent atoms changed. In
other words, $\vext$ acts as a small perturbation in a system where the total potential
energy is dominated by $\vint$.

For the commensurate filling fraction $f\!f=1/2$ the low-$\lambda$ part of the 
spectrum of $P^{-1}L$ looks quite different than that for $f\!f=1$, shown in Fig.~\ref{fig_5}.
For the same parameters as used in Fig.~\ref{fig_5} except that now $N=3$, the common
shift mode is still the lowest with $\lambda_{\min}=0.020$; however, that value is 
considerably greater than $\lambda_{\min}=0.0053$ for $f\!f=1$. This explains the faster
convergence of the AITEM without ME for $f\!f=1/2$ (given that $\lambda_{\max}\approx 1$),
as noted at the end of Sec.~\ref{sec_3B}. 
Moreover, the next few modes have eigenvalues that are much closer to $\lambda_{\min}$
than in the $f\!f=1$ case: $\lambda_=0.023,\, 0.034,\, 0.048, \, 0.051\ldots$.
 They correspond to combinations of shifts, breathing, and other shape deformations
 of the atoms. The reason that eigenvalues of these latter modes are much lower for
 $f\!f=1/2$ than for $f\!f=1$ is that the role of the helium--helium interaction is
 much reduced for atoms spaced farther apart, thereby ``penalizing" their independent
 motions less.

\subsection*{C.2: \ Estimation of the lowest eigenvalue of $P^{-1}L$ from error evolution plots}

Consider the error evolution curve corresponding to $N=24$ in Fig.~\ref{fig_6}.
From point `2' on, the error decays monotonically by 7 orders of magnitude (i.e., a
factor of about $\exp[-16]$), and therefore one can estimate $\lambda_{\min}$ of
the linearized iteration operator from (see \eqref{e3_add_01}):
\be
\left( 1 - \dt\,(\lambda_{P^{-1}L})_{\min} \right) ^{120,000} \approx e^{-16},
\label{C_01}
\ee
where the exponent on the l.h.s. is the number of iteration from point `2' to convergence. 
This yields $(\lambda_{P^{-1}L})_{\min} \approx 4\cdot 10^{-4}$.

A similar estimate for the $N=23$ case, where it takes over 115 million iterations
for the error to decrease from $10^{-8}$ to $10^{-10}$, yields 
$(\lambda_{P^{-1}L})_{\min} \approx 7\cdot 10^{-8}$.

The fact that $P^{-1}L$, and hence $L$, has very small eigenvalues
means that the ``energy landscape" of the considered configuration of atoms 
is very shallow near the energy minimum. The following order-of-magnitude 
estimate illustrates this statement. 
First, for smooth function, which the lowest-$\lambda$ eigenfunctions are,
$P^{-1}$ reduces eigenvalues of $L$ by the factor $O(1/c)=O(10^{-2})$; 
see Sec.~\ref{sec_3A}. Hence 
\be
(\lambda_{L})_{\min} \sim 10^2 (\lambda_{P^{-1}L})_{\min}
\label{C_02}
\ee
for $c=O(10^2)$. 
%
%
Next, the error norm, defined in the caption to Fig.~\ref{fig_3}, can be related to 
the (minimum) eigenvalue by:
\bsube
\label{C_03}
\be
\| L_0^{\rm 1D,H}\phi_n\| = \frac1N 
\sqrt{ \int dx \sum_{j=1}^N | L_0^{\rm 1D,H}(\phi_j(x))_n |^2 } \sim 
\sqrt{\frac{\dx}{N}} \, (\lambda_{L})_{\min} \left| (\tilde{\phi_j})_n \right|,
\label{C_03a}
\ee
where: $(\tilde{\phi_j})_n$ stands for any of the components of the iteration error
$(\widetilde{\vec{\phi}})_n$ (see \eqref{e2_08}); 
 $\dx$ is the numerical mesh size defined in Appendix A; and 
we have used \eqref{e2_08} and the fact that the observed width of $\tilde{\phi}_j$ is
$O(1)$. 
In the calculations considered here, $\dx\approx 0.1$, $N\gtrsim 20$; whence
\eqref{C_03a} yields:
\be
\| L_0^{\rm 1D,H}\phi_n\| \sim 10^{-1}\, (\lambda_{L})_{\min} \,
 \left| (\tilde{\phi_j})_n \right|.
\label{C_03b}
\ee
\esube
Together, estimates \eqref{C_02}, \eqref{C_03b}, and the above estimates of 
$(\lambda_{P^{-1}L})_{\min}$ imply that when $\| L_0^{\rm 1D,H}\phi_n\| $
reaches $10^{-10}$, the difference $(\tilde{\phi_j})_n$ 
between the final and exact solution is $O(10^{-8}\ldots 10^{-7})$ for
the case $N=24$ and only $O(10^{-4})$ for the case $N=23$. 
In other words, when $(\lambda_{L})_{\min}\ll 1$, relatively large deviations 
from the exact solutions still lead
to the error (of the HF equations) reaching a prescribed small tolerance.


\section*{Appendix D: \ Implementation of the mME Algorithm of Sec.~\ref{sec_4}}

\subsection*{D.1: \ Implementation issues of the mME algorithm}

The Algorithm presented in Sec.~\ref{sec_4A} assumes infinite numerical
precision and the relation between the middle and right sides of 
Eq.~\eqref{e4_03} holding exactly rather than approximately. To account for
deviation from these conditions, the following measures need to be taken
when implementing the Algorithm. 
\begin{enumerate}
\item
	Matrix $\wD$ is composed of nearly linearly dependent rows because
	the (smooth part of the) solution changes only little 
	from one iteration to the next. Therefore, 
	the smallest eigenvalues of $P^{1/2}\wD\,(P^{1/2}\wD)^T$, used in 
	\eqref{e4_10}, are very small and can even be 
	comparable with the numerical precision of Matlab, $O(10^{-16})$, 
	for $K \gtrsim 10$. In such a case, Matlab would not be able to 
	compute the Cholesky factorization \eqref{e4_10}. To overcome this
	problem, we ensured that the matrix on the l.h.s. of \eqref{e4_10} is
	positive definite, and yet only minimally different from the original
	matrix, by adding to it a matrix proportional to the identity
	with diagonal entries of the size
	\be
	2\,\min(\,|\lambda_{\rm lhs}|\,) + 10^{-12}\,\max(\lambda_{\rm lhs}),
	\label{D_01}
	\ee
	where $\lambda_{\rm lhs}$ are the eigenvalues of the original matrix
	on the l.h.s. of \eqref{e4_10} (the smallest of which could be negative
	and on the order of machine precision, as noted above). \\
	A version of the Algorithm that leads to a significantly smaller
	condition number of $P^{1/2}\wD\,(P^{1/2}\wD)^T$, although not
	improving performance of the mME, is presented in the Supplemental
	Material. 
\item
	Due to the reason stated in the first paragraph of this Appendix, matrix
	$\bB$ defined by the second relation in \eqref{e4_12} and computed from the middle side of \eqref{e4_03} deviates from
	being symmetric. Then, for the
	calculations in the Algorithm one uses the symmetric matrix
	\bsube
	\be
	\bB_{\rm sym} = \frac12 \left( L \wD\,\wD^T + (L \wD\,\wD^T )^T \right)
	\label{D_02a}
	\ee
	while monitoring that the anti-symmetric part
	\be
	\bB_{\rm anti-sym} = \frac12 \left( L \wD\,\wD^T - (L \wD\,\wD^T )^T \right)
	\label{D_02b}
	\ee
	\label{D_02}
	\esube
	is sufficiently small. We indeed found that to be the case, with the
	anti-symmetric part being on the order of $10^{-5}$ of the symmetric
	one when the iterations are converging. 
\item
	We also monitored the difference between the l.h.s. and r.h.s. of
	\eqref{e4_08} and verified that the matrix on the l.h.s. is within
	$O(10^{-10})$ from being orthogonal. 
\item	
	Finally, we verified that with the exceptions of those situations 
	where the error increases (as in the case shown in 
	Fig.~\ref{fig_7} for $N=23$), the relative difference between 
	$\alpha_i$'s computed from \eqref{e4_14} and from \eqref{e3_09b}
	(given the modes $\vec{u}_i$ found by the Algorithm) is as small as
	$O(10^{-4})$. \\
	We clarify that for computational efficiency, we computed 
	$\alpha_i$'s from \eqref{e4_14}. 
\end{enumerate}

\subsection*{D.2: \ Collecting data for matrices $\wD$ and $L\wD$, and efficient computation of related matrix products}

Here we will give a high-level structure of our AITEM code
focusing on the order in which it collects data for the mME 
Algorithm. We will also demonstrate that the computational 
(but not storage) cost
associated with the Algorithm grows very mildly with the number
$K$ of the modes being eliminated (under the practical
 assumption that $K$ is much smaller than the number of points
 in the numerical grid). 
 
%
\begin{enumerate}
\item
 	Save the solution $(\vec{\phi})_n$ 
 	at the current iteration $n$.\\
 	(For $n\le K$, store $(\vec{\phi})_n$ in a variable 
 	$\overrightarrow{\{\phi\}}[1]$. Just prior to that, stored solutions from
 	 previous iterations must be reassigned: 
 	$\overrightarrow{\{\phi\}}[k-1] \mapsto \overrightarrow{\{\phi\}}[k]$
 	following the order $k=K,\ldots, 2$.) 
\item
	Compute $\big(L^{(0)}\vec{\phi}\big)_n$ via \eqref{e2_05b}.
\item
	a) \ For $n>K+1$, reassign entries in \eqref{e4_03} available from the 
	previous	iterations: \ 
	$P^{-1/2}\,\overrightarrow{\{\D L\}}[k-1] \mapsto 
	 P^{-1/2} \overrightarrow{\{\D L\}}[k]$ following the order $k=K,\ldots, 2$.
	 Here the notation 
	 $\overrightarrow{\{\D L\}}$ is defined similarly to that in `b)' below.\\
	 (For $n = K+1$, compute $P^{-1/2} \overrightarrow{\{\D L\}}[k]$ using
	 $\big(L^{(0)}\vec{\phi}\big)_{n-k+1}$ which are stored in Step 6a
	 (below) at the previous iterations.)\\
	b) \ Compute $P^{-1/2}\,\overrightarrow{\{\D L\}}[1] \equiv 
	P^{-1/2} L \overrightarrow{\D\phi}_{n-1}$, where the r.h.s. is found as the top entry
    in the middle part of \eqref{e4_03} and from \eqref{e4_04}.  
\item
	Perform the mME as per the Algorithm in Sec.~\ref{sec_4A}.
\item
	Update the solution as in \eqref{e2_05a}, with the $\Gamma$-term being computed as
	explained in the last two steps in Sec.~\ref{sec_4A}. Apply the Gram--Schmidt
	orthonormalization step as needed. Note that at this point, we have the 
	solution $(\vec{\phi})_{n+1}$.
\item
    a) \ Reassign  $\overrightarrow{\{L^{(0)}\phi\}}[k-1] \mapsto 
    \overrightarrow{\{L^{(0)}\phi\}}[k]$ following the order $k=K,\ldots, 2$.
    Here the notation $\overrightarrow{\{L^{(0)}\phi\}}$ 
    is defined similarly to that in `b)' below.\\
	b) \ Save $\big(L^{(0)}\vec{\phi}\big)_n$ as
     $\overrightarrow{\{L^{(0)}\phi\}}[1]$; it will be used in Step 3b above
	at the next iteration. 
\item
	a) \ For $n > K$, reassign entries in \eqref{e4_01} available from the previous	iterations: \ 
	$\overrightarrow{P^{1/2}\{\D \phi\}}[k-1] \mapsto 
	\overrightarrow{P^{1/2}\{\D \phi\}}[k]$, following the order $k=K,\ldots, 2$.
	Here the notation $\overrightarrow{P^{1/2}\{\D \phi\}}$ 
	is defined similarly to that in `b)' below.\\
	(For $n=K$, compute $\overrightarrow{P^{1/2}\{\D \phi\}}[k]$ using the
	solutions at the previous iterations stored in Step 1 above.)\\
 	b) \ Compute $\overrightarrow{P^{1/2}\{\D \phi\}}[1] = 
 	P^{1/2} \overrightarrow{\D\phi}_{n-1}$, where the r.h.s. is found as the top
 	 entry 	in \eqref{e4_01} and from \eqref{e4_02}.  	
 \end{enumerate}

Note that at iterations $n>K$, the operational cost of all Steps above except
Step 4 is clearly independent of $K$. The cost of Step 4 can potentially be
$O(K^2\,N\,M_xM_y)$ due to the need to compute the $K\times K$ matrix
 $P^{1/2}\wD \, (P^{1/2}\wD)^T$, each term of which requires $O(N\,M_xM_y)$ 
 operations. However, Step 4 can be implemented in only $O(K\,N\,M_xM_y)$ 
 operations once one notices that the upper-left 
 $(K-1)\times(K-1)$ block of $P^{1/2}\wD\, (P^{1/2}\wD)^T$ computed at 
 iteration $n$ can be reassigned as the lower-right $(K-1)\times(K-1)$ 
 block of that matrix at iteration $(n+1)$. Then only $(2K-1)$ entries of
 that matrix would need to be computed anew at each iteration $n>K$. 
 The same estimate also applies to the computation of $L\wD\,\wD$. 
 When $K<5$, the resulting cost of $4K\,O(N\,M_xM_y)$ operations adds a
 relatively small amount to the baseline computational cost of the method.


\nocite{apsrev42Control}
\bibliographystyle{apsrev4-2}
\bibliography{sub_refs}

\begin{thebibliography}{42}%
\makeatletter
\providecommand \@ifxundefined [1]{%
 \@ifx{#1\undefined}
}%
\providecommand \@ifnum [1]{%
 \ifnum #1\expandafter \@firstoftwo
 \else \expandafter \@secondoftwo
 \fi
}%
\providecommand \@ifx [1]{%
 \ifx #1\expandafter \@firstoftwo
 \else \expandafter \@secondoftwo
 \fi
}%
\providecommand \natexlab [1]{#1}%
\providecommand \enquote  [1]{``#1''}%
\providecommand \bibnamefont  [1]{#1}%
\providecommand \bibfnamefont [1]{#1}%
\providecommand \citenamefont [1]{#1}%
\providecommand \href@noop [0]{\@secondoftwo}%
\providecommand \href [0]{\begingroup \@sanitize@url \@href}%
\providecommand \@href[1]{\@@startlink{#1}\@@href}%
\providecommand \@@href[1]{\endgroup#1\@@endlink}%
\providecommand \@sanitize@url [0]{\catcode `\\12\catcode `\$12\catcode
  `\&12\catcode `\#12\catcode `\^12\catcode `\_12\catcode `\%12\relax}%
\providecommand \@@startlink[1]{}%
\providecommand \@@endlink[0]{}%
\providecommand \url  [0]{\begingroup\@sanitize@url \@url }%
\providecommand \@url [1]{\endgroup\@href {#1}{\urlprefix }}%
\providecommand \urlprefix  [0]{URL }%
\providecommand \Eprint [0]{\href }%
\providecommand \doibase [0]{https://doi.org/}%
\providecommand \selectlanguage [0]{\@gobble}%
\providecommand \bibinfo  [0]{\@secondoftwo}%
\providecommand \bibfield  [0]{\@secondoftwo}%
\providecommand \translation [1]{[#1]}%
\providecommand \BibitemOpen [0]{}%
\providecommand \bibitemStop [0]{}%
\providecommand \bibitemNoStop [0]{.\EOS\space}%
\providecommand \EOS [0]{\spacefactor3000\relax}%
\providecommand \BibitemShut  [1]{\csname bibitem#1\endcsname}%
\let\auto@bib@innerbib\@empty
\bibitem [{\citenamefont {Nakamura}\ \emph {et~al.}(2016)\citenamefont
  {Nakamura}, \citenamefont {Matsui}, \citenamefont {Matsui},\ and\
  \citenamefont {Fukuyama}}]{2016_Nakamura}%
  \BibitemOpen
  \bibfield  {author} {\bibinfo {author} {\bibfnamefont {S.}~\bibnamefont
  {Nakamura}}, \bibinfo {author} {\bibfnamefont {K.}~\bibnamefont {Matsui}},
  \bibinfo {author} {\bibfnamefont {T.}~\bibnamefont {Matsui}},\ and\ \bibinfo
  {author} {\bibfnamefont {H.}~\bibnamefont {Fukuyama}},\ }\bibfield  {title}
  {\bibinfo {title} {{P}ossible quantum liquid crystal phases of helium
  monolayers},\ }\href {https://doi.org/10.1103/physrevb.94.180501} {\bibfield
  {journal} {\bibinfo  {journal} {Phys. Rev. B}\ }\textbf {\bibinfo {volume}
  {94}},\ \bibinfo {pages} {180501} (\bibinfo {year} {2016})}\BibitemShut
  {NoStop}%
\bibitem [{\citenamefont {Ny{\'e}ki}\ \emph {et~al.}(2017)\citenamefont
  {Ny{\'e}ki}, \citenamefont {Phillis}, \citenamefont {Ho}, \citenamefont
  {Lee}, \citenamefont {Coleman}, \citenamefont {Parpia}, \citenamefont
  {Cowan},\ and\ \citenamefont {Saunders}}]{2017_Nyeki}%
  \BibitemOpen
  \bibfield  {author} {\bibinfo {author} {\bibfnamefont {J.}~\bibnamefont
  {Ny{\'e}ki}}, \bibinfo {author} {\bibfnamefont {A.}~\bibnamefont {Phillis}},
  \bibinfo {author} {\bibfnamefont {A.}~\bibnamefont {Ho}}, \bibinfo {author}
  {\bibfnamefont {D.}~\bibnamefont {Lee}}, \bibinfo {author} {\bibfnamefont
  {P.}~\bibnamefont {Coleman}}, \bibinfo {author} {\bibfnamefont
  {J.}~\bibnamefont {Parpia}}, \bibinfo {author} {\bibfnamefont
  {B.}~\bibnamefont {Cowan}},\ and\ \bibinfo {author} {\bibfnamefont
  {J.}~\bibnamefont {Saunders}},\ }\bibfield  {title} {\bibinfo {title}
  {{I}ntertwined superfluid and density wave order in two-dimensional
  $^4${H}e},\ }\href {https://doi.org/10.1038/nphys4023} {\bibfield  {journal}
  {\bibinfo  {journal} {Nature Phys.}\ }\textbf {\bibinfo {volume} {13}},\
  \bibinfo {pages} {455} (\bibinfo {year} {2017})}\BibitemShut {NoStop}%
\bibitem [{\citenamefont {Choi}\ \emph {et~al.}(2021)\citenamefont {Choi},
  \citenamefont {Zadorozhko}, \citenamefont {Choi},\ and\ \citenamefont
  {Kim}}]{2021_Choi}%
  \BibitemOpen
  \bibfield  {author} {\bibinfo {author} {\bibfnamefont {J.}~\bibnamefont
  {Choi}}, \bibinfo {author} {\bibfnamefont {A.~A.}\ \bibnamefont
  {Zadorozhko}}, \bibinfo {author} {\bibfnamefont {J.}~\bibnamefont {Choi}},\
  and\ \bibinfo {author} {\bibfnamefont {E.}~\bibnamefont {Kim}},\ }\bibfield
  {title} {\bibinfo {title} {Spatially modulated superfluid state in
  two-dimensional $^{4}$he films},\ }\href
  {https://doi.org/10.1103/PhysRevLett.127.135301} {\bibfield  {journal}
  {\bibinfo  {journal} {Phys. Rev. Lett.}\ }\textbf {\bibinfo {volume} {127}},\
  \bibinfo {pages} {135301} (\bibinfo {year} {2021})}\BibitemShut {NoStop}%
\bibitem [{\citenamefont {Gordillo}\ and\ \citenamefont
  {Boronat}(2012)}]{2012_Gordillo}%
  \BibitemOpen
  \bibfield  {author} {\bibinfo {author} {\bibfnamefont {M.}~\bibnamefont
  {Gordillo}}\ and\ \bibinfo {author} {\bibfnamefont {J.}~\bibnamefont
  {Boronat}},\ }\bibfield  {title} {\bibinfo {title} {Zero-temperature phase
  diagram of the second layer of $^4${H}e adsorbed on graphene},\ }\href@noop
  {} {\bibfield  {journal} {\bibinfo  {journal} {Phys. Rev. B}\ }\textbf
  {\bibinfo {volume} {85}},\ \bibinfo {pages} {195457} (\bibinfo {year}
  {2012})}\BibitemShut {NoStop}%
\bibitem [{\citenamefont {Gordillo}\ and\ \citenamefont
  {Boronat}(2020)}]{2020_Gordillo}%
  \BibitemOpen
  \bibfield  {author} {\bibinfo {author} {\bibfnamefont {M.~C.}\ \bibnamefont
  {Gordillo}}\ and\ \bibinfo {author} {\bibfnamefont {J.}~\bibnamefont
  {Boronat}},\ }\bibfield  {title} {\bibinfo {title} {{S}uperfluid and
  {S}upersolid {P}hases of $^4$he on the {S}econd {L}ayer of {G}raphite},\
  }\href {https://doi.org/10.1103/physrevlett.124.205301} {\bibfield  {journal}
  {\bibinfo  {journal} {Phys. Rev. Lett.}\ }\textbf {\bibinfo {volume} {124}},\
  \bibinfo {pages} {205301} (\bibinfo {year} {2020})}\BibitemShut {NoStop}%
\bibitem [{\citenamefont {Corboz}\ \emph {et~al.}(2008)\citenamefont {Corboz},
  \citenamefont {Boninsegni}, \citenamefont {Pollet},\ and\ \citenamefont
  {Troyer}}]{2008_Corboz}%
  \BibitemOpen
  \bibfield  {author} {\bibinfo {author} {\bibfnamefont {P.}~\bibnamefont
  {Corboz}}, \bibinfo {author} {\bibfnamefont {M.}~\bibnamefont {Boninsegni}},
  \bibinfo {author} {\bibfnamefont {L.}~\bibnamefont {Pollet}},\ and\ \bibinfo
  {author} {\bibfnamefont {M.}~\bibnamefont {Troyer}},\ }\bibfield  {title}
  {\bibinfo {title} {Phase diagram of $^4${H}e adsorbed on graphite},\
  }\href@noop {} {\bibfield  {journal} {\bibinfo  {journal} {Phys. Rev. B}\
  }\textbf {\bibinfo {volume} {78}},\ \bibinfo {pages} {245414} (\bibinfo
  {year} {2008})}\BibitemShut {NoStop}%
\bibitem [{\citenamefont {Steele}(1973)}]{1973_Steele}%
  \BibitemOpen
  \bibfield  {author} {\bibinfo {author} {\bibfnamefont {W.}~\bibnamefont
  {Steele}},\ }\bibfield  {title} {\bibinfo {title} {The physical interaction
  of gases with crystalline solids},\ }\href@noop {} {\bibfield  {journal}
  {\bibinfo  {journal} {Surf. Sci.}\ }\textbf {\bibinfo {volume} {36}},\
  \bibinfo {pages} {317} (\bibinfo {year} {1973})}\BibitemShut {NoStop}%
\bibitem [{\citenamefont {Bruch}\ \emph {et~al.}(2007)\citenamefont {Bruch},
  \citenamefont {Cole},\ and\ \citenamefont {Zaremba}}]{2007_Bruch}%
  \BibitemOpen
  \bibfield  {author} {\bibinfo {author} {\bibfnamefont {L.}~\bibnamefont
  {Bruch}}, \bibinfo {author} {\bibfnamefont {M.}~\bibnamefont {Cole}},\ and\
  \bibinfo {author} {\bibfnamefont {E.}~\bibnamefont {Zaremba}},\ }\href
  {https://books.google.com/books?id=qZiRDQAAQBAJ} {\emph {\bibinfo {title}
  {Physical Adsorption: Forces and Phenomena}}},\ Dover Books on Physics\
  (\bibinfo  {publisher} {Dover Publications},\ \bibinfo {address} {Mineola,
  New York, USA},\ \bibinfo {year} {2007})\BibitemShut {NoStop}%
\bibitem [{\citenamefont {Yu}\ \emph {et~al.}(2021)\citenamefont {Yu},
  \citenamefont {Lauricella}, \citenamefont {Elsayed}, \citenamefont
  {Shepherd}, \citenamefont {Nichols}, \citenamefont {Lombardi}, \citenamefont
  {Kim}, \citenamefont {Wexler}, \citenamefont {Vanegas}, \citenamefont
  {Lakoba}, \citenamefont {Kotov},\ and\ \citenamefont {Maestro}}]{PRB}%
  \BibitemOpen
  \bibfield  {author} {\bibinfo {author} {\bibfnamefont {J.}~\bibnamefont
  {Yu}}, \bibinfo {author} {\bibfnamefont {E.}~\bibnamefont {Lauricella}},
  \bibinfo {author} {\bibfnamefont {M.}~\bibnamefont {Elsayed}}, \bibinfo
  {author} {\bibfnamefont {K.}~\bibnamefont {Shepherd}}, \bibinfo {author}
  {\bibfnamefont {N.~S.}\ \bibnamefont {Nichols}}, \bibinfo {author}
  {\bibfnamefont {T.}~\bibnamefont {Lombardi}}, \bibinfo {author}
  {\bibfnamefont {S.~W.}\ \bibnamefont {Kim}}, \bibinfo {author} {\bibfnamefont
  {C.}~\bibnamefont {Wexler}}, \bibinfo {author} {\bibfnamefont {J.~M.}\
  \bibnamefont {Vanegas}}, \bibinfo {author} {\bibfnamefont {T.}~\bibnamefont
  {Lakoba}}, \bibinfo {author} {\bibfnamefont {V.~N.}\ \bibnamefont {Kotov}},\
  and\ \bibinfo {author} {\bibfnamefont {A.~D.}\ \bibnamefont {Maestro}},\
  }\bibfield  {title} {\bibinfo {title} {Two-dimensional {Bose}-{Hubbard} model
  for helium on graphene},\ }\href@noop {} {\bibfield  {journal} {\bibinfo
  {journal} {Phys. Rev. B}\ }\textbf {\bibinfo {volume} {103}},\ \bibinfo
  {pages} {235414} (\bibinfo {year} {2021})}\BibitemShut {NoStop}%
\bibitem [{\citenamefont {Shukla}\ \emph {et~al.}(1998)\citenamefont {Shukla},
  \citenamefont {Dolg}, \citenamefont {Fulde},\ and\ \citenamefont
  {Stoll}}]{1998_goodHFpaper}%
  \BibitemOpen
  \bibfield  {author} {\bibinfo {author} {\bibfnamefont {A.}~\bibnamefont
  {Shukla}}, \bibinfo {author} {\bibfnamefont {M.}~\bibnamefont {Dolg}},
  \bibinfo {author} {\bibfnamefont {P.}~\bibnamefont {Fulde}},\ and\ \bibinfo
  {author} {\bibfnamefont {H.}~\bibnamefont {Stoll}},\ }\bibfield  {title}
  {\bibinfo {title} {Obtaining {Wannier} functions of a crystalline insulator
  within a {Hartree}-{Fock} approach: Applications to {LiF} and {LiCl}},\
  }\href@noop {} {\bibfield  {journal} {\bibinfo  {journal} {Phys. Rev. B}\
  }\textbf {\bibinfo {volume} {57}},\ \bibinfo {pages} {1471} (\bibinfo {year}
  {1998})}\BibitemShut {NoStop}%
\bibitem [{\citenamefont {Goedecker}(1999)}]{1999_Goedecker_Review}%
  \BibitemOpen
  \bibfield  {author} {\bibinfo {author} {\bibfnamefont {S.}~\bibnamefont
  {Goedecker}},\ }\bibfield  {title} {\bibinfo {title} {Linear scaling
  electronic structure methods},\ }\href@noop {} {\bibfield  {journal}
  {\bibinfo  {journal} {Rev. Mod. Phys.}\ }\textbf {\bibinfo {volume} {71}},\
  \bibinfo {pages} {1085} (\bibinfo {year} {1999})}\BibitemShut {NoStop}%
\bibitem [{\citenamefont {Talman}(2010)}]{2010_NumSolHF_molorbs}%
  \BibitemOpen
  \bibfield  {author} {\bibinfo {author} {\bibfnamefont {J.~D.}\ \bibnamefont
  {Talman}},\ }\bibfield  {title} {\bibinfo {title} {Numerical solution of the
  {H}artree-{F}ock equation in molecular geometries},\ }\href
  {https://doi.org/10.1103/PhysRevA.82.052518} {\bibfield  {journal} {\bibinfo
  {journal} {Phys. Rev. A}\ }\textbf {\bibinfo {volume} {82}},\ \bibinfo
  {pages} {052518} (\bibinfo {year} {2010})}\BibitemShut {NoStop}%
\bibitem [{\citenamefont {Dziedzic}\ \emph {et~al.}(2013)\citenamefont
  {Dziedzic}, \citenamefont {Hill},\ and\ \citenamefont
  {Skylaris}}]{2013_nonorthogWannier}%
  \BibitemOpen
  \bibfield  {author} {\bibinfo {author} {\bibfnamefont {J.}~\bibnamefont
  {Dziedzic}}, \bibinfo {author} {\bibfnamefont {Q.}~\bibnamefont {Hill}},\
  and\ \bibinfo {author} {\bibfnamefont {C.-K.}\ \bibnamefont {Skylaris}},\
  }\bibfield  {title} {\bibinfo {title} {Linear-scaling calculation of
  {Hartree}-{Fock} exchange energy with non-orthogonal generalised {Wannier}
  functions},\ }\href@noop {} {\bibfield  {journal} {\bibinfo  {journal} {J.
  Chem. Phys.}\ }\textbf {\bibinfo {volume} {139}},\ \bibinfo {pages} {214103}
  (\bibinfo {year} {2013})}\BibitemShut {NoStop}%
\bibitem [{\citenamefont {Anderson}(1965)}]{1965_AA}%
  \BibitemOpen
  \bibfield  {author} {\bibinfo {author} {\bibfnamefont {D.}~\bibnamefont
  {Anderson}},\ }\bibfield  {title} {\bibinfo {title} {Iterative procedures for
  nonlinear integral equations},\ }\href@noop {} {\bibfield  {journal}
  {\bibinfo  {journal} {J. ACM}\ }\textbf {\bibinfo {volume} {12}},\ \bibinfo
  {pages} {547} (\bibinfo {year} {1965})}\BibitemShut {NoStop}%
\bibitem [{\citenamefont {Duminil}\ and\ \citenamefont
  {Sadok}(2011)}]{2011_RedRankExtrap}%
  \BibitemOpen
  \bibfield  {author} {\bibinfo {author} {\bibfnamefont {S.}~\bibnamefont
  {Duminil}}\ and\ \bibinfo {author} {\bibfnamefont {H.}~\bibnamefont
  {Sadok}},\ }\bibfield  {title} {\bibinfo {title} {Reduced rank extrapolation
  applied to electronic structure computations},\ }\href@noop {} {\bibfield
  {journal} {\bibinfo  {journal} {Electron. Trans. Numer. Anal.}\ }\textbf
  {\bibinfo {volume} {38}},\ \bibinfo {pages} {347} (\bibinfo {year}
  {2011})}\BibitemShut {NoStop}%
\bibitem [{\citenamefont {Walker}\ and\ \citenamefont
  {Ni}(2011)}]{2011_WalkerNi}%
  \BibitemOpen
  \bibfield  {author} {\bibinfo {author} {\bibfnamefont {H.~F.}\ \bibnamefont
  {Walker}}\ and\ \bibinfo {author} {\bibfnamefont {P.}~\bibnamefont {Ni}},\
  }\bibfield  {title} {\bibinfo {title} {Anderson acceleration for fixed-point
  iterations},\ }\href {https://doi.org/10.1137/10078356X} {\bibfield
  {journal} {\bibinfo  {journal} {SIAM J. Numer. Anal.}\ }\textbf {\bibinfo
  {volume} {49}},\ \bibinfo {pages} {1715} (\bibinfo {year}
  {2011})}\BibitemShut {NoStop}%
\bibitem [{\citenamefont {Ryssens}\ \emph {et~al.}(2019)\citenamefont
  {Ryssens}, \citenamefont {Bender},\ and\ \citenamefont
  {Heenen}}]{2019_Nuclear}%
  \BibitemOpen
  \bibfield  {author} {\bibinfo {author} {\bibfnamefont {W.}~\bibnamefont
  {Ryssens}}, \bibinfo {author} {\bibfnamefont {M.}~\bibnamefont {Bender}},\
  and\ \bibinfo {author} {\bibfnamefont {P.-H.}\ \bibnamefont {Heenen}},\
  }\bibfield  {title} {\bibinfo {title} {Iterative approaches to the
  self-consistent nuclear energy density functional problem: Heavy ball
  dynamics and potential preconditioning},\ }\href@noop {} {\bibfield
  {journal} {\bibinfo  {journal} {Eur. Phys. J. A}\ }\textbf {\bibinfo {volume}
  {55}},\ \bibinfo {pages} {93} (\bibinfo {year} {2019})}\BibitemShut {NoStop}%
\bibitem [{\citenamefont {Marder}(2015)}]{book_Marder}%
  \BibitemOpen
  \bibfield  {author} {\bibinfo {author} {\bibfnamefont {M.~P.}\ \bibnamefont
  {Marder}},\ }\bibinfo {title} {Condensed matter physics}\ (\bibinfo
  {publisher} {Wiley},\ \bibinfo {year} {2015})\ Chap.\ \bibinfo {chapter}
  {9.2}\BibitemShut {NoStop}%
\bibitem [{\citenamefont {Masiello}\ \emph {et~al.}(2005)\citenamefont
  {Masiello}, \citenamefont {McKagan},\ and\ \citenamefont
  {Reinhardt}}]{2005_HFforBosons}%
  \BibitemOpen
  \bibfield  {author} {\bibinfo {author} {\bibfnamefont {D.}~\bibnamefont
  {Masiello}}, \bibinfo {author} {\bibfnamefont {S.~B.}\ \bibnamefont
  {McKagan}},\ and\ \bibinfo {author} {\bibfnamefont {W.~P.}\ \bibnamefont
  {Reinhardt}},\ }\bibfield  {title} {\bibinfo {title} {Multiconfigurational
  {Hartree}-{Fock} theory for identical bosons in a double well},\ }\href@noop
  {} {\bibfield  {journal} {\bibinfo  {journal} {Phys. Rev. A}\ }\textbf
  {\bibinfo {volume} {72}},\ \bibinfo {pages} {063624} (\bibinfo {year}
  {2005})}\BibitemShut {NoStop}%
\bibitem [{\citenamefont {Aziz}\ \emph {et~al.}(1995)\citenamefont {Aziz},
  \citenamefont {Janzen},\ and\ \citenamefont {Moldover}}]{1995_Aziz}%
  \BibitemOpen
  \bibfield  {author} {\bibinfo {author} {\bibfnamefont {R.~A.}\ \bibnamefont
  {Aziz}}, \bibinfo {author} {\bibfnamefont {A.~R.}\ \bibnamefont {Janzen}},\
  and\ \bibinfo {author} {\bibfnamefont {M.~R.}\ \bibnamefont {Moldover}},\
  }\bibfield  {title} {\bibinfo {title} {Ab initio calculations for helium: A
  standard for transport property measurements},\ }\href@noop {} {\bibfield
  {journal} {\bibinfo  {journal} {Phys. Rev. Lett.}\ }\textbf {\bibinfo
  {volume} {74}},\ \bibinfo {pages} {1586} (\bibinfo {year}
  {1995})}\BibitemShut {NoStop}%
\bibitem [{\citenamefont {Lakoba}(2011)}]{multicomp}%
  \BibitemOpen
  \bibfield  {author} {\bibinfo {author} {\bibfnamefont {T.~I.}\ \bibnamefont
  {Lakoba}},\ }\bibfield  {title} {\bibinfo {title} {Convergence conditions for
  iterative methods seeking multi-component solitary waves with prescribed
  quadratic conserved quantities},\ }\href@noop {} {\bibfield  {journal}
  {\bibinfo  {journal} {Math. Comput. Simul.}\ }\textbf {\bibinfo {volume}
  {81}},\ \bibinfo {pages} {1572} (\bibinfo {year} {2011})}\BibitemShut
  {NoStop}%
\bibitem [{\citenamefont {Lakoba}\ and\ \citenamefont {Yang}(2007)}]{ME}%
  \BibitemOpen
  \bibfield  {author} {\bibinfo {author} {\bibfnamefont {T.}~\bibnamefont
  {Lakoba}}\ and\ \bibinfo {author} {\bibfnamefont {J.}~\bibnamefont {Yang}},\
  }\bibfield  {title} {\bibinfo {title} {A mode elimination technique to
  improve convergence of iteration methods for finding solitary waves},\ }\href
  {https://doi.org/https://doi.org/10.1016/j.jcp.2007.06.010} {\bibfield
  {journal} {\bibinfo  {journal} {J. Comput. Phys.}\ }\textbf {\bibinfo
  {volume} {226}},\ \bibinfo {pages} {1693} (\bibinfo {year}
  {2007})}\BibitemShut {NoStop}%
\bibitem [{\citenamefont {Yang}\ and\ \citenamefont {Lakoba}(2008)}]{AITEM}%
  \BibitemOpen
  \bibfield  {author} {\bibinfo {author} {\bibfnamefont {J.}~\bibnamefont
  {Yang}}\ and\ \bibinfo {author} {\bibfnamefont {T.}~\bibnamefont {Lakoba}},\
  }\bibfield  {title} {\bibinfo {title} {Accelerated imaginary-time evolution
  methods for the computation of solitary waves},\ }\href@noop {} {\bibfield
  {journal} {\bibinfo  {journal} {Stud. Appl. Math.}\ }\textbf {\bibinfo
  {volume} {120}},\ \bibinfo {pages} {265} (\bibinfo {year}
  {2008})}\BibitemShut {NoStop}%
\bibitem [{\citenamefont {Tuckerman}\ \emph {et~al.}(1992)\citenamefont
  {Tuckerman}, \citenamefont {Berne},\ and\ \citenamefont
  {Martyna}}]{1992_Molly}%
  \BibitemOpen
  \bibfield  {author} {\bibinfo {author} {\bibfnamefont {M.}~\bibnamefont
  {Tuckerman}}, \bibinfo {author} {\bibfnamefont {B.}~\bibnamefont {Berne}},\
  and\ \bibinfo {author} {\bibfnamefont {G.}~\bibnamefont {Martyna}},\
  }\bibfield  {title} {\bibinfo {title} {Reversible multiple time scale
  molecular dynamics},\ }\href@noop {} {\bibfield  {journal} {\bibinfo
  {journal} {J. Chem. Phys.}\ }\textbf {\bibinfo {volume} {97}},\ \bibinfo
  {pages} {1990} (\bibinfo {year} {1992})}\BibitemShut {NoStop}%
\bibitem [{\citenamefont {Izaguirre}\ \emph {et~al.}(1999)\citenamefont
  {Izaguirre}, \citenamefont {Reich},\ and\ \citenamefont
  {Skeel}}]{1999_Molly}%
  \BibitemOpen
  \bibfield  {author} {\bibinfo {author} {\bibfnamefont {J.}~\bibnamefont
  {Izaguirre}}, \bibinfo {author} {\bibfnamefont {S.}~\bibnamefont {Reich}},\
  and\ \bibinfo {author} {\bibfnamefont {R.}~\bibnamefont {Skeel}},\ }\bibfield
   {title} {\bibinfo {title} {Longer time steps for molecular dynamics},\
  }\href@noop {} {\bibfield  {journal} {\bibinfo  {journal} {J. Chem. Phys.}\
  }\textbf {\bibinfo {volume} {110}},\ \bibinfo {pages} {9853} (\bibinfo {year}
  {1999})}\BibitemShut {NoStop}%
\bibitem [{\citenamefont {Lakoba}(2009)}]{2009_CGM}%
  \BibitemOpen
  \bibfield  {author} {\bibinfo {author} {\bibfnamefont {T.~I.}\ \bibnamefont
  {Lakoba}},\ }\bibfield  {title} {\bibinfo {title} {Conjugate gradient method
  for finding fundamental solitary waves},\ }\href@noop {} {\bibfield
  {journal} {\bibinfo  {journal} {Physica D: Nonlinear Phenomena}\ }\textbf
  {\bibinfo {volume} {238}},\ \bibinfo {pages} {2308} (\bibinfo {year}
  {2009})}\BibitemShut {NoStop}%
\bibitem [{\citenamefont {Shewchuk}(1994)}]{1994_nopainCGM}%
  \BibitemOpen
  \bibfield  {author} {\bibinfo {author} {\bibfnamefont {J.~R.}\ \bibnamefont
  {Shewchuk}},\ }\href@noop {} {\emph {\bibinfo {title} {An Introduction to the
  Conjugate Gradient Method Without the Agonizing Pain}}},\ \bibinfo {type}
  {Tech. Rep.}\ (\bibinfo {address} {Pittsburgh, PA},\ \bibinfo {year}
  {1994})\BibitemShut {NoStop}%
\bibitem [{\citenamefont {Whitlock}\ \emph {et~al.}(1998)\citenamefont
  {Whitlock}, \citenamefont {Chester},\ and\ \citenamefont
  {Krishnamachari}}]{1998_Whitlock}%
  \BibitemOpen
  \bibfield  {author} {\bibinfo {author} {\bibfnamefont {P.~A.}\ \bibnamefont
  {Whitlock}}, \bibinfo {author} {\bibfnamefont {G.~V.}\ \bibnamefont
  {Chester}},\ and\ \bibinfo {author} {\bibfnamefont {B.}~\bibnamefont
  {Krishnamachari}},\ }\bibfield  {title} {\bibinfo {title} {Monte {C}arlo
  simulation of a helium film on graphite},\ }\href
  {https://doi.org/10.1103/PhysRevB.58.8704} {\bibfield  {journal} {\bibinfo
  {journal} {Phys. Rev. B}\ }\textbf {\bibinfo {volume} {58}},\ \bibinfo
  {pages} {8704} (\bibinfo {year} {1998})}\BibitemShut {NoStop}%
\bibitem [{\citenamefont {Kwon}\ and\ \citenamefont
  {Ceperley}(2012)}]{2012_Kwon}%
  \BibitemOpen
  \bibfield  {author} {\bibinfo {author} {\bibfnamefont {Y.}~\bibnamefont
  {Kwon}}\ and\ \bibinfo {author} {\bibfnamefont {D.}~\bibnamefont
  {Ceperley}},\ }\bibfield  {title} {\bibinfo {title} {$^4${H}e adsorption on a
  single graphene sheet: {P}ath-integral {M}onte {C}arlo study},\ }\href@noop
  {} {\bibfield  {journal} {\bibinfo  {journal} {Phys. Rev. B}\ }\textbf
  {\bibinfo {volume} {85}},\ \bibinfo {pages} {224501} (\bibinfo {year}
  {2012})}\BibitemShut {NoStop}%
\bibitem [{\citenamefont {Skorobogatiy}\ and\ \citenamefont
  {Yang}(2009)}]{book_PC_Jianke}%
  \BibitemOpen
  \bibfield  {author} {\bibinfo {author} {\bibfnamefont {M.}~\bibnamefont
  {Skorobogatiy}}\ and\ \bibinfo {author} {\bibfnamefont {J.}~\bibnamefont
  {Yang}},\ }\bibinfo {title} {Fundamentals of photonic crystal guiding}\
  (\bibinfo  {publisher} {Cambridge University Press},\ \bibinfo {year}
  {2009})\ Chap.\ \bibinfo {chapter} {7.2}\BibitemShut {NoStop}%
\bibitem [{\citenamefont {Gordillo}\ and\ \citenamefont
  {Boronat}(2009)}]{2009_GordilloBoronat}%
  \BibitemOpen
  \bibfield  {author} {\bibinfo {author} {\bibfnamefont {M.~C.}\ \bibnamefont
  {Gordillo}}\ and\ \bibinfo {author} {\bibfnamefont {J.}~\bibnamefont
  {Boronat}},\ }\bibfield  {title} {\bibinfo {title} {{$^{4}\mathrm{He}$} on a
  single graphene sheet},\ }\href@noop {} {\bibfield  {journal} {\bibinfo
  {journal} {Phys. Rev. Lett.}\ }\textbf {\bibinfo {volume} {102}},\ \bibinfo
  {pages} {085303} (\bibinfo {year} {2009})}\BibitemShut {NoStop}%
\bibitem [{\citenamefont {Gordillo}(2014)}]{2014_Gordillo}%
  \BibitemOpen
  \bibfield  {author} {\bibinfo {author} {\bibfnamefont {M.}~\bibnamefont
  {Gordillo}},\ }\bibfield  {title} {\bibinfo {title} {Diffusion {M}onte
  {C}arlo calculations of the phase diagram of $^{4}\mathrm{He}$ on corrugated
  graphene},\ }\href@noop {} {\bibfield  {journal} {\bibinfo  {journal} {Phys.
  Rev. B}\ }\textbf {\bibinfo {volume} {89}},\ \bibinfo {pages} {155401}
  (\bibinfo {year} {2014})}\BibitemShut {NoStop}%
\bibitem [{\citenamefont {Wessel}\ and\ \citenamefont
  {Troyer}(2005)}]{WesselTroyer}%
  \BibitemOpen
  \bibfield  {author} {\bibinfo {author} {\bibfnamefont {S.}~\bibnamefont
  {Wessel}}\ and\ \bibinfo {author} {\bibfnamefont {M.}~\bibnamefont
  {Troyer}},\ }\bibfield  {title} {\bibinfo {title} {Supersolid hard-core
  bosons on the triangular lattice},\ }\href@noop {} {\bibfield  {journal}
  {\bibinfo  {journal} {Phys. Rev. Lett.}\ }\textbf {\bibinfo {volume} {95}},\
  \bibinfo {pages} {127205} (\bibinfo {year} {2005})}\BibitemShut {NoStop}%
\bibitem [{\citenamefont {Yamamoto}\ \emph {et~al.}(2012)\citenamefont
  {Yamamoto}, \citenamefont {Masaki},\ and\ \citenamefont
  {Danshita}}]{2012_PRB_longrange}%
  \BibitemOpen
  \bibfield  {author} {\bibinfo {author} {\bibfnamefont {D.}~\bibnamefont
  {Yamamoto}}, \bibinfo {author} {\bibfnamefont {A.}~\bibnamefont {Masaki}},\
  and\ \bibinfo {author} {\bibfnamefont {I.}~\bibnamefont {Danshita}},\
  }\bibfield  {title} {\bibinfo {title} {Quantum phases of hardcore bosons with
  long-range interactions on a square lattice},\ }\href@noop {} {\bibfield
  {journal} {\bibinfo  {journal} {Phys. Rev. B}\ }\textbf {\bibinfo {volume}
  {86}},\ \bibinfo {pages} {054516} (\bibinfo {year} {2012})}\BibitemShut
  {NoStop}%
\bibitem [{\citenamefont {Ying}\ \emph {et~al.}(2013)\citenamefont {Ying},
  \citenamefont {Batrouni}, \citenamefont {Rousseau}, \citenamefont {Jarrell},
  \citenamefont {Moreno}, \citenamefont {Sun},\ and\ \citenamefont
  {Scalettar}}]{2013_PRB_anisotropic}%
  \BibitemOpen
  \bibfield  {author} {\bibinfo {author} {\bibfnamefont {T.}~\bibnamefont
  {Ying}}, \bibinfo {author} {\bibfnamefont {G.~G.}\ \bibnamefont {Batrouni}},
  \bibinfo {author} {\bibfnamefont {V.~G.}\ \bibnamefont {Rousseau}}, \bibinfo
  {author} {\bibfnamefont {M.}~\bibnamefont {Jarrell}}, \bibinfo {author}
  {\bibfnamefont {J.}~\bibnamefont {Moreno}}, \bibinfo {author} {\bibfnamefont
  {X.~D.}\ \bibnamefont {Sun}},\ and\ \bibinfo {author} {\bibfnamefont {R.~T.}\
  \bibnamefont {Scalettar}},\ }\bibfield  {title} {\bibinfo {title} {Phase
  stability in the two-dimensional anisotropic boson {Hubbard} {Hamiltonian}},\
  }\href@noop {} {\bibfield  {journal} {\bibinfo  {journal} {Phys. Rev. B}\
  }\textbf {\bibinfo {volume} {87}},\ \bibinfo {pages} {195142} (\bibinfo
  {year} {2013})}\BibitemShut {NoStop}%
\bibitem [{\citenamefont {Streib}\ and\ \citenamefont
  {Kopietz}(2015)}]{2015_PRB_spinhalf}%
  \BibitemOpen
  \bibfield  {author} {\bibinfo {author} {\bibfnamefont {S.}~\bibnamefont
  {Streib}}\ and\ \bibinfo {author} {\bibfnamefont {P.}~\bibnamefont
  {Kopietz}},\ }\bibfield  {title} {\bibinfo {title} {Hard-core boson approach
  to the $\text{spin}\ensuremath{-}\frac{1}{2}$ triangular-lattice
  antiferromagnet {${\mathrm{Cs}}_{2}{\mathrm{CuCl}}_{4}$} at finite
  temperatures in magnetic fields higher than the saturation field},\
  }\href@noop {} {\bibfield  {journal} {\bibinfo  {journal} {Phys. Rev. B}\
  }\textbf {\bibinfo {volume} {92}},\ \bibinfo {pages} {094442} (\bibinfo
  {year} {2015})}\BibitemShut {NoStop}%
\bibitem [{\citenamefont {Reatto}\ \emph {et~al.}(2012)\citenamefont {Reatto},
  \citenamefont {Nava}, \citenamefont {Galli}, \citenamefont {Billman},
  \citenamefont {Sofo},\ and\ \citenamefont {Cole}}]{2012_GHandGF}%
  \BibitemOpen
  \bibfield  {author} {\bibinfo {author} {\bibfnamefont {L.}~\bibnamefont
  {Reatto}}, \bibinfo {author} {\bibfnamefont {M.}~\bibnamefont {Nava}},
  \bibinfo {author} {\bibfnamefont {D.~E.}\ \bibnamefont {Galli}}, \bibinfo
  {author} {\bibfnamefont {C.}~\bibnamefont {Billman}}, \bibinfo {author}
  {\bibfnamefont {J.~O.}\ \bibnamefont {Sofo}},\ and\ \bibinfo {author}
  {\bibfnamefont {M.~W.}\ \bibnamefont {Cole}},\ }\bibfield  {title} {\bibinfo
  {title} {Novel substrates for helium adsorption: Graphane and
  graphene{\textemdash}fluoride},\ }\href@noop {} {\bibfield  {journal}
  {\bibinfo  {journal} {J. Phys. Conf. Ser.}\ }\textbf {\bibinfo {volume}
  {400}},\ \bibinfo {pages} {012010} (\bibinfo {year} {2012})}\BibitemShut
  {NoStop}%
\bibitem [{\citenamefont {Ahn}\ \emph {et~al.}(2019)\citenamefont {Ahn},
  \citenamefont {You}, \citenamefont {Lee}, \citenamefont {Volkoff},\ and\
  \citenamefont {Kwon}}]{2019_6612graphene}%
  \BibitemOpen
  \bibfield  {author} {\bibinfo {author} {\bibfnamefont {J.}~\bibnamefont
  {Ahn}}, \bibinfo {author} {\bibfnamefont {M.}~\bibnamefont {You}}, \bibinfo
  {author} {\bibfnamefont {G.}~\bibnamefont {Lee}}, \bibinfo {author}
  {\bibfnamefont {T.}~\bibnamefont {Volkoff}},\ and\ \bibinfo {author}
  {\bibfnamefont {Y.}~\bibnamefont {Kwon}},\ }\bibfield  {title} {\bibinfo
  {title} {Symmetry-changing commensurate-incommensurate solid transition in
  the $^{4}\mathrm{He}$ monolayer on 6,6,12-graphyne},\ }\href
  {https://doi.org/10.1103/PhysRevB.99.024113} {\bibfield  {journal} {\bibinfo
  {journal} {Phys. Rev. B}\ }\textbf {\bibinfo {volume} {99}},\ \bibinfo
  {pages} {024113} (\bibinfo {year} {2019})}\BibitemShut {NoStop}%
\bibitem [{\citenamefont {Burganova}\ \emph {et~al.}(2016)\citenamefont
  {Burganova}, \citenamefont {Lysogorskiy}, \citenamefont {Nedopekin},\ and\
  \citenamefont {Tayurskii}}]{2016_JLTP_Rus}%
  \BibitemOpen
  \bibfield  {author} {\bibinfo {author} {\bibfnamefont {R.}~\bibnamefont
  {Burganova}}, \bibinfo {author} {\bibfnamefont {Y.}~\bibnamefont
  {Lysogorskiy}}, \bibinfo {author} {\bibfnamefont {O.}~\bibnamefont
  {Nedopekin}},\ and\ \bibinfo {author} {\bibfnamefont {D.}~\bibnamefont
  {Tayurskii}},\ }\bibfield  {title} {\bibinfo {title} {Adsorption of helium
  atoms on two-dimensional substrates},\ }\href@noop {} {\bibfield  {journal}
  {\bibinfo  {journal} {J. Low Temp. Phys.}\ }\textbf {\bibinfo {volume}
  {185}},\ \bibinfo {pages} {392} (\bibinfo {year} {2016})}\BibitemShut
  {NoStop}%
\bibitem [{\citenamefont {Shin}\ \emph {et~al.}(2019)\citenamefont {Shin},
  \citenamefont {Luo}, \citenamefont {Benali},\ and\ \citenamefont
  {Kwon}}]{2019_O2onG}%
  \BibitemOpen
  \bibfield  {author} {\bibinfo {author} {\bibfnamefont {H.}~\bibnamefont
  {Shin}}, \bibinfo {author} {\bibfnamefont {Y.}~\bibnamefont {Luo}}, \bibinfo
  {author} {\bibfnamefont {A.}~\bibnamefont {Benali}},\ and\ \bibinfo {author}
  {\bibfnamefont {Y.}~\bibnamefont {Kwon}},\ }\bibfield  {title} {\bibinfo
  {title} {Diffusion {Monte} {Carlo} study of {${\mathrm{O}}_{2}$} adsorption
  on single layer graphene},\ }\href@noop {} {\bibfield  {journal} {\bibinfo
  {journal} {Phys. Rev. B}\ }\textbf {\bibinfo {volume} {100}},\ \bibinfo
  {pages} {075430} (\bibinfo {year} {2019})}\BibitemShut {NoStop}%
\bibitem [{\citenamefont {Barik}\ and\ \citenamefont
  {Woods}(2023)}]{2023_Barik}%
  \BibitemOpen
  \bibfield  {author} {\bibinfo {author} {\bibfnamefont {R.~K.}\ \bibnamefont
  {Barik}}\ and\ \bibinfo {author} {\bibfnamefont {L.~M.}\ \bibnamefont
  {Woods}},\ }\bibfield  {title} {\bibinfo {title} {{H}igh throughput
  calculations for a dataset of bilayer materials},\ }\href
  {https://doi.org/10.1038/s41597-023-02146-7} {\bibfield  {journal} {\bibinfo
  {journal} {Scientific Data}\ }\textbf {\bibinfo {volume} {10}},\ \bibinfo
  {pages} {1038} (\bibinfo {year} {2023})}\BibitemShut {NoStop}%
\bibitem [{\citenamefont {Kim}\ \emph {et~al.}(2024)\citenamefont {Kim},
  \citenamefont {Elsayed}, \citenamefont {Nichols}, \citenamefont {Lakoba},
  \citenamefont {Vanegas}, \citenamefont {Wexler}, \citenamefont {Kotov},\ and\
  \citenamefont {Del~Maestro}}]{2024_Kim}%
  \BibitemOpen
  \bibfield  {author} {\bibinfo {author} {\bibfnamefont {S.~W.}\ \bibnamefont
  {Kim}}, \bibinfo {author} {\bibfnamefont {M.~M.}\ \bibnamefont {Elsayed}},
  \bibinfo {author} {\bibfnamefont {N.~S.}\ \bibnamefont {Nichols}}, \bibinfo
  {author} {\bibfnamefont {T.}~\bibnamefont {Lakoba}}, \bibinfo {author}
  {\bibfnamefont {J.}~\bibnamefont {Vanegas}}, \bibinfo {author} {\bibfnamefont
  {C.}~\bibnamefont {Wexler}}, \bibinfo {author} {\bibfnamefont {V.~N.}\
  \bibnamefont {Kotov}},\ and\ \bibinfo {author} {\bibfnamefont
  {A.}~\bibnamefont {Del~Maestro}},\ }\bibfield  {title} {\bibinfo {title}
  {Atomically thin superfluid and solid phases for atoms on strained
  graphene},\ }\href {https://doi.org/10.1103/PhysRevB.109.064512} {\bibfield
  {journal} {\bibinfo  {journal} {Phys. Rev. B}\ }\textbf {\bibinfo {volume}
  {109}},\ \bibinfo {pages} {064512} (\bibinfo {year} {2024})}\BibitemShut
  {NoStop}%
\end{thebibliography}%

\newpage


\section*{Supplementary material}

Supplemental Material for

"Efficient simulations of Hartree–Fock equations by an accelerated gradient descent method"

by  Y. Ohno, A. Del Maestro, and T.I. Lakoba

## I.  DERIVATION OF THE SECOND TERM IN EQ. (6b)

In Sec. 2 of [S1] it was shown that when minimizing a functional $\mathcal{H}$ such that $\delta H/\delta\phi_j^* = L_{00}\phi_j$ subject to constraints given by equations $Q^{(k)} = 0$, $k = 1, \ldots, K$, the second term in Eq. (6b)[S2] is given by:

$$\Phi\,\langle P^{-1}\Phi\,|\,\Phi\rangle^{-1}\,\langle P^{-1}\Phi\,|\,\overrightarrow{L_{00}\phi}^T\rangle. \tag{S1}$$

Here the $(j,k)$-th component of the $N \times K$ matrix $\Phi$ is:

$$\Phi_{j,k} = \delta Q^{(k)}/\delta\phi_j, \tag{S2}$$

$$\overrightarrow{L_{00}\phi} = [L_{00}\phi_1, \ldots, L_{00}\phi_N], \tag{S3}$$

the rest of the notations — here and below — are defined in Sec. II A of the main text, and we have omitted iteration's index $n$.

For the constraints $Q$ given by the $K = N^2$ Eqs. (2), definition (S2) yields:

$$\Phi = \vec{\phi} \otimes I, \tag{S4}$$

where $\otimes$ stands for the Kronecker (tensor) product and $I$ is the identity matrix of appropriate dimension ($N \times N$ in this case). Then $P^{-1}\Phi = P^{-1}\vec{\phi} \otimes I$. Next, using the identities (for matrices $A, B, C, D$ of appropriate dimensions):

$$(A \otimes B)\,(C \otimes D) = (AC) \otimes (BD), \qquad (A \otimes B)^{-1} = A^{-1} \otimes B^{-1}, \tag{S5}$$

one has

$$\langle P^{-1}\Phi\,|\,\Phi\rangle^{-1} = \langle P^{-1}\vec{\phi}\,|\,\vec{\phi}\rangle^{-1} \otimes I \equiv \begin{pmatrix} a_{11} \cdot I & \cdots & a_{1N} \cdot I \\ \vdots & \vdots & \vdots \\ a_{N1} \cdot I & \cdots & a_{NN} \cdot I \end{pmatrix}, \tag{S6}$$

with $a_{ij}$ denoting entries of the inverse matrix in the middle of the above line. To proceed, one needs to

write out an explicit form of the last factor in (S1):

$$\langle P^{-1}\Phi \,|\, \overrightarrow{L_{00}\phi}^{\,T}\rangle = \begin{pmatrix} \langle P^{-1}\phi_1 \,|\, L_{00}\phi_1\rangle \\ \vdots \\ \langle P^{-1}\phi_1 \,|\, L_{00}\phi_N\rangle \\ \hline \vdots \\ \hline \langle P^{-1}\phi_N \,|\, L_{00}\phi_1\rangle \\ \vdots \\ \langle P^{-1}\phi_N \,|\, L_{00}\phi_N\rangle \end{pmatrix}. \tag{S7}$$

Multiplying the second and third factors in (S1) and using (S6), (S7) yields:

$$\langle P^{-1}\Phi \,|\, \Phi\rangle^{-1}\,\langle P^{-1}\Phi \,|\, \overrightarrow{L_{00}\phi}^{\,T}\rangle = \begin{pmatrix} \sum_{k=1}^{N} a_{1k}\langle P^{-1}\phi_k \,|\, L_{00}\phi_1\rangle \\ \vdots \\ \sum_{k=1}^{N} a_{1k}\langle P^{-1}\phi_k \,|\, L_{00}\phi_N\rangle \\ \hline \vdots \\ \hline \sum_{k=1}^{N} a_{Nk}\langle P^{-1}\phi_k \,|\, L_{00}\phi_1\rangle \\ \vdots \\ \sum_{k=1}^{N} a_{Nk}\langle P^{-1}\phi_k \,|\, L_{00}\phi_N\rangle \end{pmatrix}. \tag{S8}$$

Finally, using (S4), one can see by inspection that $j$-th row of the entire quantity in (S1) is given by:

$$\sum_{i=1}^{N} \phi_i \sum_{k=1}^{N} a_{ik}\langle P^{-1}\phi_k \,|\, L_{00}\phi_j\rangle, \tag{S9}$$

which is indeed the second term on the r.h.s. of (6b).

## II.  EQUATIONS $L_0\phi_j = 0$ AND $L^{(0)}\phi_j = 0$ ARE EQUIVALENT

The operator in the second equation in the section title is defined in (6b). Now, the former equation, i.e., (3), can be written as

$$L_{00}\phi_j = \vec{\phi}\,\langle \vec{\phi}\,|\, L_{00}\phi_j\rangle. \tag{S10}$$

Similarly, the latter equation is written as:

$$L_{00}\phi_j = \vec{\phi}\,\langle P^{-1}\vec{\phi}\,|\,\vec{\phi}\rangle^{-1}\,\langle P^{-1}\vec{\phi}\,|\, L_{00}\phi_j\rangle. \tag{S11}$$

In addition, constraints (2) can be written as

$$\langle \vec{\phi}\,|\,\vec{\phi}\rangle = I. \tag{S12}$$

Let us denote $\langle P^{-1}\vec{\phi}\,|\,\vec{\phi}\rangle \equiv M$ (an $N \times N$ matrix).

The proof consists of showing that (S10) implies (S11) and vice versa. To show the former step, substitute (S10) into the r.h.s. of (S11); this yields:

$$\vec{\phi}\,M^{-1}\,\langle P^{-1}\vec{\phi}\,|\,\vec{\phi}\,\langle\vec{\phi}\,|\, L_{00}\phi_j\rangle\,\rangle = \vec{\phi}\,M^{-1}\,M\,\langle\vec{\phi}\,|\, L_{00}\phi_j\rangle, \tag{S13}$$

which, per (S10), equals the l.h.s. of (S11).

To show that (S11) along with (S12) implies (S10), substitute the r.h.s. of (S11) into the r.h.s. of (S10) while multiplying the left (right) part of the inner product by $P^{-1}$ ($P$) and using the fact that $P$ is Hermitian. Then the r.h.s. of (S10) becomes:

$$\vec{\phi}\,\langle P^{-1}\vec{\phi}\,|\,P\,\vec{\phi}\,M^{-1}\langle P^{-1}\vec{\phi}\,|\,L_{00}\phi_j\rangle\,\rangle = \vec{\phi}\,\langle P^{-1}\vec{\phi}\,|\,P\,\vec{\phi}\,\rangle\,M^{-1}\langle P^{-1}\vec{\phi}\,|\,L_{00}\phi_j\rangle = \vec{\phi}\,M^{-1}\langle P^{-1}\vec{\phi}\,|\,L_{00}\phi_j\rangle, \quad \text{(S14)}$$

where we have used (S12). Per (S11), this equals the l.h.s. of (S10).

## III. SUMMARY OF UNSUCCESSFUL ATTEMPTS TO IMPROVE CONVERGENCE OF THE ME-ACCELERATED AITEM BY REDUCING ERROR OSCILLATIONS

- We tried to restart ME (i.e., stop it and return to it after various numbers of no-ME AITEM iterations), in analogy to restarting certain other methods [S3]. We varied the stopping criterion for ME (e.g., did it when the error would grow by a specified factor between 2 and 10, or ran ME only for a specified number of iterations). The oscillations did not, in general, reduce, and in some cases would stall at a higher error than the tolerance.

- When the error would continue to grow either for a specified number of iterations or by a given factor, we would flip the sign of parameter $s$ to negative and kept it so for a (typically small) prescribed number of iterations. The error would decay in the few initial iterations but then would grow or stall.

- As explained in Sec. II B, the idea behind ME is that $Lu_{\text{slow},j}$ is aligned with $L\phi_j$ so that $\text{corr}[Lu_{\text{slow},j}L\phi_j] \equiv \langle Lu_{\text{slow},j}|L\phi_j\rangle/(\,\|Lu_{\text{slow},j}\|\,\|L\phi_j\|\,) \approx -1$. Therefore, we monitored $\text{corr}[Lu_{\text{slow},j}L\phi_j]$ and flipped the sign of $\gamma$ to make it positive when we detected that $\text{corr}[Lu_{\text{slow},j}L\phi_j] > 0$. This led to the same results as in the previous item.

- From Fig. 8(c) of the main text, one can see that the error grows when the smallest of the computed $\alpha$'s decreases. The decrease of $\alpha$ implies, via (8b), that $\gamma$ becomes more negative. To arrest this behavior, we tried to set the lower bound for $\gamma$ in (37) at a less negative number, e.g., $-10^{-5}$. However, the error would continue to grow, but slower, and the "bulge" in the error curve would simply take many more iterations.

- Guided by the same observation about the correlation between the evolutions of $\alpha$ and the error, we tried to limit values of $\alpha$ to a range as follows: Once we computed $\alpha \equiv \alpha_{\text{first ME}}$ at the first ME iteration, we restricted $\alpha$ to a range $[\alpha_{\text{first ME}}/\text{sf}, [\alpha_{\text{first ME}} \cdot \text{sf}]$. Here sf is some factor, with whose value we also experimented. This did not consistently reduce either the duration or the magnitude of the error oscillations.

- In a slight variation to the idea of the previous item, we modulated the ME parameter $s$ depending on the value of $\alpha$ at a given iteration; we explored various dependencies of $s$ on $\alpha$. The result was the same as in the previous item.

- As a variation of the previous idea, we modulated $s$ according to the error rather than $\alpha$. We tried various modification of:

$$s_{n+1} = s_n \, \frac{\|\mathrm{error}_{n-1}\|}{\|\mathrm{error}_n\|}.$$

This sometimes reduced the duration and/or magnitude of the oscillation of the error evolution, but never by more than 20 % and not in all cases.

- In a different line of thought, once we detected that the error has grown by a specified factor, we would add a spatially smooth random perturbation to each atom, the size of the perturbation being on the order of the error at the given iteration. The hypothesis behind this was that when the error begins to grow, the iterations could be in the vicinity of some metastable state, and adding noise to the solution would push it away from that state and could restore convergence of the iterations. This did not work, and the error would typically stall.

- We also tried a number of variations of the stochastic gradient descent method. For example, we updated, by the AITEM equation (6a), only a subset of $\phi_j$'s. Such a subset, different for different iterations, was chosen either randomly or based on the size of the error of individual atoms (estimated either from $\|(\phi_j)_{n+1} - (\phi_j)_n\|$ or from $\left\| \left( L^{(0)} \phi_j \right)_n \right\|$. While the error of the updated atoms would diminish, it would grow back when a different subset of atoms would get updated later, and the overall result was that the average error would stall.

- Finally, we tried to include memory in the definitions of $\alpha_n$ and $\phi_{\mathrm{slow},\,n}$, by replacing $\alpha_n$ with $(1 - \mathrm{mf})\,\alpha_n + \mathrm{mf}\,\alpha_{n-1}$, where $\alpha_n$ was computed by (20b) and $\mathrm{mf}$ was some 'memory factor' that we varied between 0 and 1. A similar update was used for $u_{\mathrm{slow},\,n}$. This trick would occasionally help by a small amount, around 10%, and only for small $\mathrm{mf} \lesssim 0.1$.

## IV.   PATTERNS FOR $ff = 7/16$ AND $7/12$

These patterns were reported by previous researchers, specifically: for $ff = 7/16$ — in Figs. 2(c) of [S4], 5(b) of [S5], and 2 of [S6]; for $ff = 7/12$ — in Figs. 5 of [S4], 7 of [S5], and 2(a) of [S7]. Our patterns, shown in Fig. S1, agree with those up to, possibly, a 90° rotation of the computational window.

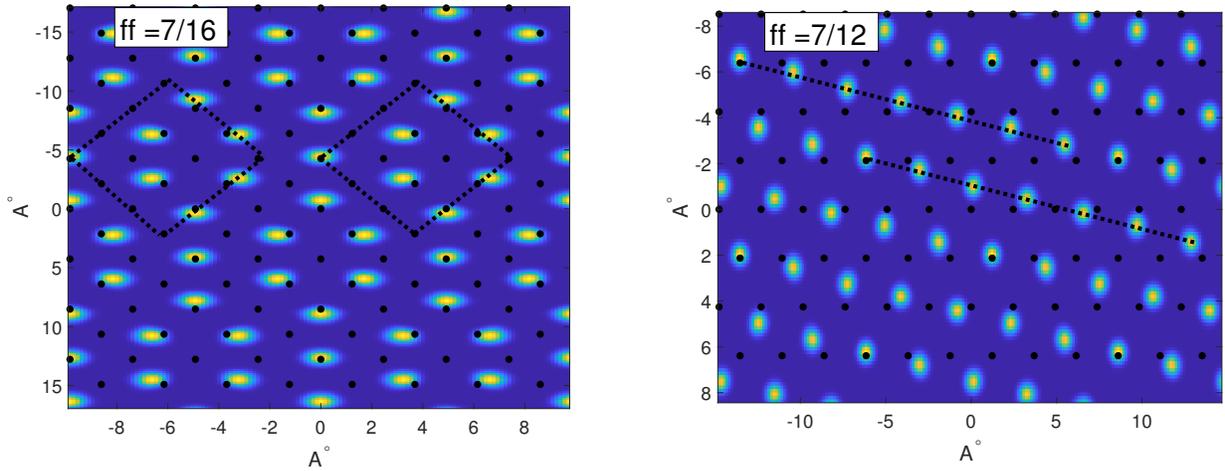

FIG. S1. Patterns for $ff = 7/16$ and $7/12$. See caption of Fig. 10 of the main text for explanations.

## V. QR-VERSION OF THE MME ALGORITHM

As noted in Appendix D, the condition number of matrix

$$P^{1/2}\widehat{\Delta}\left(P^{1/2}\widehat{\Delta}\right)^T \tag{S15}$$

becomes very large for the number of eliminated modes $K \gtrsim 10$, which compromises the numerical calculation of its Cholesky decomposition (31) and hence the accuracy of the computed eigenmodes of $P^{-1}L$. One can avoid the computation of (S15) by first computing the row-counterpart of the QR-factorization of $P^{1/2}\widehat{\Delta}$:

$$P^{1/2}\widehat{\Delta} = \mathbf{R}\,\widehat{Q}, \tag{S16}$$

where $\mathbf{R}$ is a lower-triangular $K \times K$ matrix and $\widehat{Q}$ is the matrix of the same dimensions as $\widehat{\Delta}$ whose *rows* are orthonormal. Their computation is described below. Once such $\widehat{Q}$ and $L\widehat{Q}$ (defined below) are found, they can be used in place of $P^{1/2}\widehat{\Delta}$ and $P^{-1/2}L\widehat{\Delta}$ in the Algorithm of Sec. IV A. The condition number of $\widehat{Q}\widehat{Q}^T$ is 1 by design. The ill-conditioning of $\widehat{\Delta}$ is inherited by $\mathbf{R}$, which will need to be inverted (see below). However, cond$\mathbf{R} \sim$ cond$\widehat{\Delta} = \sqrt{\text{cond}\,\widehat{\Delta}\,\widehat{\Delta}^T}$, and thus the calculations involving $\widehat{Q}$ are much less ill-conditioned than those involving (S15).

We will now outline how (S16) is found. First, the rows of $P^{1/2}\widehat{\Delta}$ are orthonormalized by the modified Gram–Schmidt algorithm. The latter is used instead of the classical Gram–Schmidt because the rows of $\widehat{\Delta}$ are almost linearly dependent. The modified Gram–Schmidt algorithm of orthonormalizing $K$ vectors $\vec{d_i}$ is the following (see, e.g., [S8]):

- Copy all $\vec{d_i}$ to $\vec{v_i}$.

- for $i = 1 : K$
    $r_{ii} = \|\vec{v_i}\|_2; \qquad \vec{q_i} = \vec{v_i}/r_{ii}$
    for $j = (i+1) : K$
        $r_{ij} = \vec{q_i}^T\vec{v_j}; \qquad \vec{v_j} = \vec{v_j} - r_{ij}\vec{q_i}$
    end

end

The resulting vectors $\vec{q}_i$ are orthonormal. They form the rows of $\widehat{Q}$. The matrix $\mathbf{R}$ is constructed next.

One begins with constructing matrices

$$\widetilde{\mathbf{R}}_i = \mathbf{I}_i + \begin{pmatrix} & \mathcal{O}_{(i-1),K} & \\ \mathcal{O}_{1,K-i-1} & r_{ii} & \ldots r_{iK} \\ & \mathcal{O}_{(K-i),K} & \end{pmatrix}, \tag{S17}$$

where $\mathbf{I}_i$ is the $K \times K$ identity matrix whose $i$th diagonal entry is replaced with 0, $\mathcal{O}_{j,l}$ is a zero matrix of size $j \times l$, and $r_{ij}$ have been computed above. Then [S8]

$$\mathbf{R} = \left( \widetilde{\mathbf{R}}_K \, \widetilde{\mathbf{R}}_{K-1} \cdots \widetilde{\mathbf{R}}_1 \right)^T. \tag{S18}$$

Finally, to accomplish step (33) of the mME Algorithm, we need a counterpart of $L\widehat{\Delta}\,\widehat{\Delta}^T$ in terms of $\widehat{Q}$ and $L\widehat{Q}$. More specifically,

$$L\widehat{\Delta}\,\widehat{\Delta}^T \equiv P^{-1/2}LP^{-1/2}\,P^{1/2}\widehat{\Delta}\,(P^{1/2}\widehat{\Delta})^T \tag{S19}$$

(since all operators in the physical (i.e., $(x,y)$-) space commute with the boldfaced $K \times K$ matrices that act on rows of the $K \times (NM_xM_y)$ matrices with a hat). Therefore, we seek $P^{-1/2}LP^{-1/2}\,P^{1/2}\widehat{\Delta}$, which from (S16) becomes:

$$P^{-1/2}LP^{-1/2}\,P^{1/2}\widehat{\Delta} = \mathbf{R}\,P^{-1/2}LP^{-1/2}\,\widehat{Q}, \tag{S20a}$$

whence

$$P^{-1/2}LP^{-1/2}\,\widehat{Q} = \mathbf{R}^{-1}\,P^{-1/2}L\widehat{\Delta}. \tag{S20b}$$

To summarize: In this version of the mME Algorithm, one uses $\widehat{Q}$ instead of $P^{1/2}\widehat{\Delta}$, whence the r.h.s. of (31) is the identity matrix and hence $\mathbf{C} = \mathbf{I}$. In (33), $\mathbf{B}$ is computed as $P^{-1/2}LP^{-1/2}\,\widehat{Q}\,\widehat{Q}^T$.

The drawback of this version, as presented above, is that one would have to perform the (row-version of) QR factorization (S16) at every iteration, which has a relatively high operational cost of $2K^2\,O(NM_xM_y)$ [S8]. As an alternative, one may be able to design a variation of the modified Gram–Schmidt process where one of the vectors in the set (corresponding to the last row of $\widehat{\Delta}$) is discarded and a new row is made orthogonal to all remaining $(K-1)$ rows. We did not explore this option since the more straightforward QR-version of the mME algorithm described above did not yield any consistent improvement in the iteration number compared to the version presented in the main text.

---

[S1] T. I. Lakoba, Math. Comput. Simul. **81**, 1572 (2011).
[S2] Here and below equations numbered without prefix 'S' refer to the main text, unless stated otherwise.

[S3] B. O'Donoghue and E. Candès, Found. Comput. Math. **15**, 715 (2015).

[S4] P. Corboz, M. Boninsegni, L. Pollet, and M. Troyer, Phys. Rev. B **78**, 245414 (2008).

[S5] Y. Kwon and D. Ceperley, Phys. Rev. B **85**, 224501 (2012).

[S6] J. Happacher, P. Corboz, M. Boninsegni, and L. Pollet, Phys. Rev. B **87**, 094514 (2013).

[S7] J. Ahn, H. Lee, and Y. Kwon, Phys. Rev. B **93**, 064511 (2016).

[S8] L. Trefethen and D. Bau, Numerical linear algebra (SIAM, 1997) Chap. 8.



\end{document}